\documentclass[12pt]{iopart}
\expandafter\let\csname equation*\endcsname\relax
\expandafter\let\csname endequation*\endcsname\relax
\usepackage{amssymb}
\usepackage{amsmath}
\usepackage{geometry}
\usepackage{cite}
\usepackage{epstopdf, epsfig}
\usepackage{url}

\renewcommand\vec\mathbf

\begin{document}

\title{Characteristics of monotonic sheaths near a wall with grazing magnetic incidence}
\author{A. Geraldini$^{1}$, R. J. Ewart$^{2}$, S. Brunner$^{1}$, F. I. Parra$^{3}$}
\address{$^1$ Swiss Plasma Center, \'Ecole Polytechnique F\'ed\'erale de Lausanne (EPFL), CH-1015 Lausanne, Switzerland}
\address{$^2$ Rudolf Peierls Centre for Theoretical Physics, University of Oxford, Oxford, OX1 3NP, UK}
\address{$^3$  Princeton Plasma Physics Laboratory, Princeton, NJ 08540, USA }
\ead{alessandro.geraldini@epfl.ch}

\begin{abstract}
We consider a magnetised plasma in contact with an absorbing planar wall, where the angle $\alpha$ between the magnetic field and the wall is small, $\alpha \ll 1$ (in radians) and the system is symmetric tangential to the wall.
The finite ratio $\gamma$ of the characteristic electron gyroradius $\rho_{\rm e}$ to the Debye length $\lambda_{\rm D}$, $\gamma = \rho_{\rm e} / \lambda_{\rm D}$, is retained via a grazing-incidence ($\alpha \ll 1$) gyrokinetic treatment \cite{Geraldini-2017, Geraldini-2018}.
Building on a previously developed iterative scheme \cite{Geraldini-2018, Ewart-2021} to solve for the steady-state electrostatic potential in the quasineutral magnetic presheath of width $\sim \rho_{\rm S}$, we developed a scheme that simultaneously solves for both the presheath and the non-neutral Debye sheath of width $\sim \lambda_{\rm D}$ in the limit $\lambda_{\rm D} / \rho_{\rm S} \rightarrow 0$.
The code, called \url{GYRAZE}, thus provides the energy-angle distribution of ions at the wall and the velocity distributions of electrons reflected by the wall for different values of wall potential.
A monotonic electrostatic potential profile, assumed in this work, can only exist for magnetic field angles larger than a critical value \cite{Ewart-2021}.
While the critical angle is shown here to significantly increase with $\gamma$, it is still typically smaller than the magnetic field angle at divertor targets of a fusion device.
\end{abstract}

\section{Introduction} \label{sec-intro}

Research in the area of magnetised plasma-wall interaction is important in many areas of plasma physics, such as fusion devices \cite{Stangeby-book}, plasma thrusters \cite{Martinez-1998}, magnetic filters \cite{Anders-1995-filters}, and probes \cite{Hutchinson-book}.
Focusing on a case particularly relevant to fusion devices, we consider a magnetic field at a shallow angle $\alpha$ with a solid target, 
\begin{align} \label{ordering-angle}
 \alpha \ll 1  \text{ (in radians).}
\end{align} 

When a plasma is in contact with a solid target, or wall, a potential difference develops in a very thin positively charged region next to the wall called a Debye sheath.
The size of this region is a few Debye lengths $\lambda_{\rm D} = \sqrt{\varepsilon_0 T_{\rm e} / n_{\rm e, ref} e^2}$.
Here, $\varepsilon_0$ is the permittivity of free space, $T_{\rm e}$ is the electron temperature close to the target, $n_{\rm e, ref}$ is a reference electron density close to the target, and $e$ is the proton charge.
The wall potential is usually negative relative to the plasma one in order to repel the lighter---and thus more mobile---electrons from the wall so that no net charge leaves the plasma. 
The Debye sheath is positively charged in order to shield the bulk plasma from the negative wall potential.

With a magnetic field, the sheath structure is strongly affected.
By considering the magnetic field to be at an oblique angle with the wall, Chodura \cite{Chodura-1982} was the first to show that the magnetised plasma sheath exhibits potential variation on two distinct length scales.
A fraction of the potential drop between the plasma and the wall occurs over a quasineutral region, known as the magnetic presheath or Chodura sheath, whose characteristic size is the ion sound gyroradius $\rho_{\rm S} \equiv \sqrt{m_{\rm i} (Z T_{\rm e} + T_{\rm i}) }/ (ZeB)$,  where $T_{\rm i}$ is the ion temperature, $m_{\rm i}$ is the ion mass, $Z$ is the charge state of the ion and $B$ is the magnetic field strength.
The remaining potential drop occurs over the typically much smaller Debye sheath scale, $ \lambda_{\rm D}$.
The magnetised sheath is thus characterised by a small dimensionless parameter equal to the ratio of the characteristic size of the two subregions comprising it,
\begin{align} \label{ordering-length}
\epsilon_{\rm ms} \equiv \frac{ \lambda_{\rm D}}{\rho_{\rm S}} \ll 1 \rm . 
\end{align}
The two subregions can be analysed separately from the bulk plasma of length scale $L$ provided the ordering
\begin{align} \label{rhostar}
\rho_{\star} \equiv \frac{\rho_{\rm S}}{L} \ll 1 \rm 
\end{align}
is satisfied.
Using (\ref{rhostar}), we can consider the magnetised sheath as a boundary layer covering the walls surrounding the bulk plasma.
This can be formalised in the asymptotic limit $\rho_{\star} \rightarrow 0$, where the magnetic presheath entrance is simultaneously a point infinitely far from the wall on the magnetic presheath scale, $x/\rho_{\rm S} \rightarrow \infty$, with $x$ the distance from the wall, and infinitely close to the wall on the plasma scale, $x/L \rightarrow 0$ \cite{Geraldini-2024-Chodura}.
Also using (\ref{ordering-length}) we can consider the Debye sheath as a boundary layer at the wall side of the magnetic presheath.
In the asymptotic limit $\epsilon_{\rm ms} \rightarrow 0$, the Debye sheath entrance is simultaneously a point infinitely far from the wall on the Debye sheath scale, $x/\lambda_{\rm D} \rightarrow \infty$, and infinitely close to the wall on the magnetic presheath scale, $x/\rho_{\rm S} \rightarrow 0$.
We denote the wall potential, at $x/\lambda_{\rm D} = 0$, as $\phi_{\rm w}$, and we choose the potential at the magnetised sheath entrance, $x/\rho_{\rm S} \rightarrow \infty$, to be zero. 
We further assume that $\phi_{\rm w} < 0$, such that the magnetised sheath repels electrons from the wall, and we refer to $\phi_{\rm w}$ as the potential drop across the magnetised sheath.

An important observation made by Chodura by examining numerical solutions is that the floating-wall potential drop across the magnetised sheath, denoted $\phi_{\text{w,fl}}$, only depends weakly on the angle and strength of the magnetic field, satisfying the scaling \cite{Stangeby-book, Stangeby-2012, Coulette-Manfredi-2016}
\begin{align} \label{ordering-phiw-float}
\exp \left( \frac{e \phi_{\text{w,fl}}}{T_{\rm e}} \right)  \sim  \left( \frac{1+\tau}{M} \right)^{1/2} \ll 1 \rm .
\end{align}
In (\ref{ordering-phiw-float}), $M$ is the mass ratio,
\begin{align} \label{ordering-massratio}
M \equiv \frac{m_{\rm i}}{m_{\rm e}} \gg 1 \rm ,
\end{align}
and $\tau$ is the temperature ratio,
\begin{align} \label{ordering-tau}
\tau \equiv \frac{T_{\rm i}}{ZT_{\rm e}}  \rm .
\end{align}
Chodura found that a larger potential drop occurs across the magnetic presheath for shallow magnetic field angles, but also that this is almost exactly balanced by a smaller potential drop across the Debye sheath.

The floating wall potential in (\ref{ordering-phiw-float}) corresponds to when the electron and ion current densities in the direction $x$ perpendicular to the wall, respectively $J_{x, \rm e}$ and $J_{x, \rm i} $, are equal and opposite.
This situation is a conventional one to consider, but it is a special case, since the preservation of plasma quasineutrality only imposes that the overall outflow of positive and negative charge be equal, ``global ambipolarity'', and not that the current density be zero everywhere, ``local ambipolarity''.
In practice, a conducting wall can be biased such that a current flows across the plasma, and there may be current loops closing through the wall.
The ion current towards the wall is constrained to scale linearly with $\alpha$ at the magnetic presheath entrance, since it is determined by the projection in the direction normal to the target of the ion current parallel to the magnetic field.
The projection gives a factor $\sin \alpha \simeq \alpha $ for $\alpha \ll 1$.
When $T_{\rm i} \ll T_{\rm e}$, the Chodura condition \cite{Chodura-1982}, necessary for a monotonic and electron-repelling electric field at the magnetic presheath entrance \cite{Riemann-1994, Geraldini-2024-Chodura}, further ensures a parallel ion flow into the wall at the Bohm speed $v_{\rm B} = \sqrt{ZT_{\rm e}/m_{\rm i}}$, so that in general (accounting for the ion temperature)
\begin{align} \label{Ji-mpe}
|J_{x, \rm i} |\sim  \alpha e n_{\rm e, mp}(\infty) c_{\rm S} \rm ,
\end{align}
with $c_{\rm S} = \sqrt{(ZT_{\rm e} + T_{\rm i})/m_{\rm i}}$ denoting the sound speed.
Here, $n_{\rm e, mp}(\infty)$ is the electron density at the entrance of the magnetised sheath.
The electron current satisfies the scaling
\begin{align} \label{Je-mpe}
J_{x, \rm e} \sim J_{x, \rm e, Max,0} = \exp \left( \frac{e\phi_{\rm w} }{ T_{\rm e} } \right) \frac{1}{\sqrt{2\pi}} \alpha e n_{\rm e, mp}(\infty) v_{\rm t, e},
\end{align}
where $J_{x, \rm e, Max,0}$ denotes the analytical prediction for the electron current that is obtained from a Maxwellian electron distribution function by considering that electrons reach the wall (instead of being reflected) if their velocity component directed along the magnetic field is larger than $\sqrt{-2e\phi_{\rm w}/m_{\rm e}}$.
Here, $v_{\rm t, e}$ is the electron thermal speed, where the thermal speed of species $s$ is defined through $v_{\text{t}, s} = \sqrt{T_{s}/m_{s}}$.
The floating-wall potential drop scaling (\ref{ordering-phiw-float}) follows by equating (\ref{Ji-mpe}) and (\ref{Je-mpe}).
The ordering (\ref{Je-mpe}) assumes that the distribution of energetic electrons is still well-described by a Maxwellian with temperature $T_{\rm e}$, although it has been found that the tail of the electron distribution function can have a significantly higher temperature than the bulk in some cases \cite{Tskhakaya-2017}.

The magnetic field strength $B$ can be parameterised by a dimensionless parameter equal to the ratio of the electron gyroradius to the Debye length, denoted $\gamma$,
\begin{align} \label{gamma-def}
\gamma \equiv \frac{  \rho_{\rm e}   }{ \lambda_{\rm D}  } = \frac{1}{B} \sqrt{\frac{ m_{\rm e} n_{\rm e, ref} }{ \varepsilon_0} } \rm .
\end{align} 
Here, we have used the definition of the thermal electron gyroradius $\rho_{\rm e} = \sqrt{m_{\rm e}T_{\rm e}}/eB$. 
If $ \gamma = \rho_{\rm e} / \lambda_{\rm D}$ is negligibly small, such that $\gamma = 0$ can be taken, the electron motion is effectively one-dimensional because the electric field only accelerates the electron in the direction parallel to the magnetic field.
Therefore, if an electron enters the magnetic presheath with a large enough velocity component parallel to the magnetic field, it overcomes the potential barrier of the magnetised sheath and reaches the target.
Otherwise, it reflects within the magnetised sheath and exits the system with equal and opposite parallel velocity (to within a small correction, not considered here, coming from gradients tangential to the target \cite{Cohen-Ryutov-2004-sheath-boundary-conditions}).
Hence, the flux of electrons parallel to the magnetic field at the magnetic presheath entrance only depends on the distribution of parallel electron velocities and on the total electrostatic potential drop, $\phi_{\rm w}$.
The component of the parallel-to-$\vec{B}$ electron flux directed normal to the target is thus proportional to $\alpha$ (see (\ref{Je-mpe})), just like the ion flux (see (\ref{Ji-mpe})), but additionally depends on $\phi_{\rm w}$.
Hence, the value of $\phi_{\rm w, fl}$ corresponding to $J_{x, \rm e} = - J_{x, \rm i}$ is entirely independent of $\alpha$ for $ \gamma \rightarrow 0$.

The situation changes if $ \rho_{\rm e} \sim \lambda_D $, with a weak dependence on $\alpha$ emerging.
The parallel electron velocity makes the electron gyro-orbit move, \emph{on average}, towards or away from the wall.
However, it is the electron motion perpendicular to the magnetic field that contributes to how the electron finally reaches the wall. 
Reflecting an electron from the wall requires reversing its parallel velocity before its gyro-orbit touches the wall.
For an electron that is an electron gyroradius away from the wall, the gyroaveraged electrostatic potential is larger than the wall potential, which implies that the capability of the Debye sheath to repel the electron before reaching the wall is reduced.
This effect is amplified for electrons with a larger magnetic moment, which have a larger gyro-orbit and thus a larger gyroaveraged potential when touching the wall.
The parallel electron flux is thus not only controlled by the value of $\phi_{\rm w}$, but also by the value of $\gamma$ and by the Debye sheath potential variation $\phi_{\rm ds}(x)$ on the scale $x \sim \rho_{\rm e} = \gamma \lambda_{\rm D}$.
Moreover, since the potential profile across the Debye sheath strongly depends on the magnetic field angle $\alpha$, the parallel electron flux depends also on $\alpha$, as well as on the potential drop $\phi_{\rm w}$.
Consequently, the component of the parallel electron flux directed normal to the target is not exactly proportional to $\alpha$ as is $J_{x, \rm e, Max,0}$ in (\ref{Je-mpe}), and the ambipolar potential drop $\phi_{\rm w,fl}$ thus depends on $\alpha$, albeit still satisfying the ordering (\ref{ordering-phiw-float}).

In this paper, numerical solutions of the magnetised sheath for a particular choice of incoming ion and electron distribution functions (corresponding to $T_{\rm i} \sim T_{\rm e}$) and for a range of values of $\alpha \ll 1$, $\gamma \lesssim 1$ and $M \gg 1$ are presented.
The dependence of various characteristics of the magnetised sheath on our chosen parameters is studied.
The ambipolar potential drop across the magnetised sheath is found to slightly increase with $\gamma$ and $\alpha$, consistent with results in the literature.
This increase occurs almost entirely in the Debye sheath, with no change in the magnetic presheath potential drop. 
For finite values of $\gamma$, the nontrivial relation between the magnetised sheath potential drop and the electron fluxes is obtained. We stress that our numerical method is much faster than a time-dependent simulation of the entire magnetised sheath.
Its speed makes it an attractive tool also to obtain the self-consistent ion energy-angle distribution at the target \cite{Geraldini-2021}, useful for accurate sputtering predictions.

The numerical method used in this work presents a substantial advancement relative to previous work which focused on the magnetic presheath for $\alpha \ll 1$ \cite{Geraldini-2018, Ewart-2021}.
Previously, only the quasineutrality equation was solved. 
This was efficiently carried out using an iterative approach.
In this work, an iterative scheme to obtain the steady-state solution of Poisson's equation is developed. 
The already existing iterative scheme to solve for the quasineutrality equation on length scales of $\rho_{\rm i}$ \cite{Geraldini-2018} is modified to solve both Poisson's equation and the quasineutrality equation interchangeably.
Notably, this opens up the possibility of iteratively solving the steady state of the full magnetised sheath using an adaptive non-equidistant grid which resolves both length scales.
In this work, however, the magnetic presheath and Debye sheath equations are solved iteratively and separately by using the scale separation $\epsilon_{\rm ms} \rightarrow 0$.
At each iteration, the outputs and inputs of both the Debye sheath and magnetic presheath are appropriately matched with one another.
A third iterative scheme is also carried out to simultaneously solve for the wall potential when the current density to the wall is prescribed (e.g. the floating wall potential is recovered by prescribing zero current density). 
The code, which builds onto the code developed in references \cite{Geraldini-2018, Ewart-2021}, has been made available at \url{https://github.com/alessandrogeraldini/GYRAZE}.

A central result obtained here is that no convergence to an ambipolar monotonic electron-repelling electrostatic potential solution in the Debye sheath is obtained numerically for $\alpha < \alpha_{\star}$, where the critical angle $\alpha_{\star}$ is a function of $\gamma$ and $M$ while also depending on the incoming ion and electron distribution functions.
We argue that the analytical constraints posed by a monotonic and electron-repelling solution for the magnetised sheath potential profile for $\epsilon_{\rm ms} = \lambda_{\rm D} / \rho_{\rm S} \rightarrow 0$ imply that such a solution does not exist below the critical angle, which explains why our scheme cannot find it.
The critical angle arises because for smaller magnetic field angle one would expect the Debye sheath to have a smaller potential drop, yet monotonicity imposes a minimum Debye sheath potential drop, as we proceed to explain.
For negligible electron gyro-orbit size, $\gamma = 0$, the Debye sheath can collapse such that the potential profile is flat and the minimum Debye sheath potential drop is zero.
Ewart \emph{et al.} \cite{Ewart-2021} demonstrated that the critical angle at which this happens is $\alpha_{\star} \sim  M^{-1}$ for $\tau \sim 1$, numerically evaluated to $\approx 0.3^{\circ}$ for a deuterium plasma, which is much smaller than the previously predicted angle $\alpha_{\star} \sim  M^{-1/2}$ \cite{Stangeby-2012}.
In this paper, we show both numerically and analytically that the critical angle required to preserve monotonicity of the electron-repelling magnetised sheath potential is significantly enhanced at finite values of $\gamma$ due to the impossibility of a Debye sheath with finite electron gyro-orbits to sustain a monotonic profile with a potential drop smaller than a certain critical value.
For a deuterium plasma, the value $\alpha_{\star} \approx M^{-1/2} \approx 1^{\circ}$ predicted in reference \cite{Stangeby-2012} fortuitiously coincides with our numerically predicted value in fusion-relevant regimes. 

The rest of this paper is organized as follows. 
In section~\ref{sec-orderings}, the assumptions and orderings of this paper are summarised and discussed.
Particle trajectories are analysed and particle densities are derived in section~\ref{sec-particle}.
In section~\ref{sec-analytical}, the equations for the self-consistent electrostatic potential in the stationary magnetic presheath and Debye sheath are given, and some analytical constraints arising from the requirement of a monotonic electron-repelling potential profile are derived.
The numerical method used to calculate monotonic and electron-repelling electrostatic potential solutions and the numerical results are presented in section~\ref{sec-num}.
Section~\ref{sec-conc} concludes with a summary and a discussion of the main results.

\section{Assumptions and orderings} \label{sec-orderings}

\begin{figure} 
\centering
\includegraphics[width=0.6\textwidth]{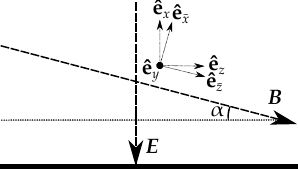} 
\caption{Geometry of the magnetised sheath, with the direction of the unit vectors $\hat{\vec{e}}_k$ shown with respect to that of the magnetic field $\vec{B}$ and electric field $\vec{E}$. The angle between the magnetic field and the target (full horizontal line) is $\alpha$.}
\label{fig-geom}
\end{figure}

We consider a magnetic field $\vec{B}$ impinging on the target at an angle $\alpha$ such that 
\begin{align}
    \vec{B} = - B \sin \alpha ~ \hat{\vec{e}}_x + B \cos \alpha ~ \hat{\vec{e}}_z = B \hat{\vec{e}}_{\tilde{z}} \rm .
\end{align}
The unit vectors $\hat{\vec{e}}_x$, $\hat{\vec{e}}_y$ and $\hat{\vec{e}}_z$ are aligned with the $x$, $y$ and $z$ coordinates, respectively, as shown in figure \ref{fig-geom}.
Since $x$ is the distance from the wall (also referred to as target in fusion devices), the vector $\hat{\vec{e}}_x$ is the unit vector normal to the target, directed away from the target.
The unit vector $\hat{\vec{e}}_{\tilde{z}}$ is directed along the magnetic field towards the target, such that $\hat{\vec{e}}_{\tilde{z}} \cdot \hat{\vec{e}}_{x} = - \sin \alpha$.
The unit vector $\hat{\vec{e}}_z$ is, instead, in the direction of the component of the magnetic field which is tangential to the wall, $\vec{B} - (\hat{\vec{e}}_x \cdot \vec{B}) \hat{\vec{e}}_x$, such that $\hat{\vec{e}}_{\tilde{z}} \cdot \hat{\vec{e}}_{z} = \cos \alpha$.
The electric field is assumed to be electrostatic, directed normal to the wall and only varying with the distance from the wall, such that
\begin{align}
\vec{E} = - \nabla \phi = - \hat{\vec{e}}_x \phi'(x) \rm ,
\end{align}
where $\phi$ is an electrostatic potential and, in general, a prime denotes differentiation with respect to the argument of a function.
The direction $\hat{\vec{e}}_y$ is parallel to the $\vec{E} \times \vec{B}$ drift, $\hat{\vec{e}}_y = \hat{\vec{e}}_z \times \hat{\vec{e}}_x$.
The unit vector $\hat{\vec{e}}_{\tilde{x}}$ is defined to be parallel to $\hat{\vec{e}}_y \times \vec{B} $, such that $\vec{\hat{e}}_x \cdot \vec{\hat{e}}_{\tilde{x}} = \cos \alpha$.
The unit vectors are shown relative to the magnetic field and the wall in figure~\ref{fig-geom}.

We assume an electrostatic plasma: this assumption has been justified in reference \cite{Geraldini-2017} based on the typically very small ratio of plasma thermal pressure to magnetic pressure in fusion devices.
While our treatment of the magnetised sheath is electrostatic, electromagnetic effects may be important in turbulence at the larger length scale $L$ of the Scrape-Off Layer \cite{Mandell-2022, Chang-2024}.
Although spatial gradients tangential to the target may influence ion transport to the wall at shallow magnetic field angles \cite{Stangeby-Chankin-1995, Loizu-2012, Geraldini-2024-Chodura}, in this work we assume that the system is invariant upon translations tangential to the target.
We consequently also neglect surface roughness, although it is expected to be important in larger fusion devices such as ITER \cite{Tskhakaya-2017}.
The system is effectively 1-dimensional in physical space, but 3-dimensional in velocity space.

The magnetic field angle is assumed to be small by (\ref{ordering-angle}), as is typical at divertor or limiter targets.
Collisions in the magnetised sheath are neglected by the ordering $\rho_{\rm S} \ll \alpha \lambda_{\text{c},s}$ \cite{Geraldini-2017}, where $\lambda_{\text{c},s}$ is the collisional mean free path of ions ($s=\rm i$) or electrons ($s=\rm e$).
The length scales of the subregions comprising the magnetised sheath are typically different in size, so that the ratio $\epsilon_{\rm ms}$ between the Debye length and the ion sound gyroradius satisfies (\ref{ordering-length}).  
This ordering allows the solution of two separate systems, the Debye sheath and the magnetic presheath, at two asymptotically separate length scales (the Debye length $\lambda_{\rm D}$ and the ion gyroradius $\rho_{\rm S}$) to approximate well the exact solution.
We consider a temperature ratio of order unity, $\tau \sim 1$, consistent with the Scrape-Off Layer \cite{Mosetto-2015}, which implies that the the magnetic presheath size is comparable to the thermal ion gyroradius, $\rho_{\rm i}  \equiv \sqrt{m_{\rm i} T_{\rm i}  }/ (ZeB)$.
For a deuterium ($M = 3671$) or a hydrogen ($M = 1836$) plasma, the physical ion to electron mass ratio is obviously large as stated by equation (\ref{ordering-massratio}). 

The Debye sheath is characterised by the ordering $ x \sim \lambda_{\rm D}$ and $x/\rho_{\rm i} \ll 1$, and the magnetic presheath is characterised by the ordering $x \sim \rho_{\rm i}$ and $\lambda_{\rm D} / x \ll 1$.
From (\ref{ordering-length}), we take the asymptotic limit $\epsilon_{\rm ms} \rightarrow 0$, $x/\rho_{\rm i} \rightarrow 0$ in the Debye sheath and $ \lambda_{\rm D} / x \rightarrow 0$ in the magnetic presheath.
The Debye sheath and magnetic presheath are thus treated as two distinct regions connected by the Debye sheath entrance, where $x/\rho_{\rm i} \rightarrow 0$ and $ \lambda_{\rm D} / x \rightarrow 0$ hold simultaneously.
The electrostatic potential in the magnetised sheath is thus approximated by two functions: the magnetic presheath potential, $\phi_{\rm mp}(x)$, which satisfies $\phi_{\rm mp}(\infty) = 0$ and $\phi_{\rm mp} (\lambda_{\rm D}) \simeq \phi_{\rm mp}(0)$; and the Debye sheath potential, $\phi_{\rm ds}(x)$, which satisfies $\phi_{\rm ds}(\rho_{\rm i}) \simeq \phi_{\rm ds}(\infty) = 0$.
The wall potential is defined to be
\begin{align}
\phi_{\rm w} = \phi_{\rm mp} (0) + \phi_{\rm ds}(0) \rm .
\end{align} 
The two functions, $ \phi_{\rm mp} $ and $ \phi_{\rm ds}$, are obtained by solving separate equations for the magnetic presheath and Debye sheath, respectively.
A crucial assumption of this work is that the electric field repels electrons from the wall and that it varies monotonically in the magnetised sheath\footnote{To be precise, our calculation of the ion density in the magnetic presheath makes a further assumption of monotonicity of the gradient of the electric field, as explained in section~\ref{subsec-IMPEDS} and reference \cite{Geraldini-2018}.}.
More specifically, we emphasise that the signs of the electrostatic potential values and first two derivatives are as follows: $\phi_{\rm ds}(x) < 0$ and $\phi_{\rm mp}(x) < 0$ for all $x$; $\phi_{\rm mp}'(x) >0$ and $\phi_{\rm ds}'(x) >0$ for all $x$; $\phi_{\rm ds}''(x) < 0$ and $\phi_{\rm mp}''(x) < 0$ for all $x$.
Taking $x=0$, $\phi_{\rm ds}(0) < 0$, $\phi_{\rm mp}(0) < 0$ and consequently $\phi_{\rm w} < 0$.

The ion and electron velocities in the magnetised sheath are ordered as follows. 
Their kinetic energy is determined by a combination of their thermal energy and the energy gained (if any) from the electric field.
Since the magnetised sheath electric field's purpose is to repel electrons, one must have $e|\phi_{\rm w}| \sim T_{\rm e}$, which is consistent with the ordering (\ref{ordering-phiw-float}).
The energy gained by accelerating ions is therefore of order $Ze |\phi_{\rm w}| \sim ZT_{\rm e}$.
Furthermore, ions have a scatter of kinetic energy determined by the ion velocity distribution and parameterised by the ion temperature $T_{\rm i}$.
Thus, the typical kinetic energy of an ion in the magnetised sheath is $ZT_{\rm e} + T_{\rm i}$, and its typical speed is therefore the ion sound speed, $c_{\rm S} = \sqrt{\left(ZT_{\rm e} + T_{\rm i}\right)/m_{\rm i}}$.
If the electron temperature is small, $\tau \gtrsim 1$, $c_{\rm S}$ is similar to the ion thermal velocity $v_{\rm t,i} = \sqrt{ T_{\rm i}/m_{\rm i}}$;
otherwise, it is comparable to the Bohm speed, $v_{\rm B} = \sqrt{ZT_{\rm e}/m_{\rm i}}$.
The typical speed of an electron is the electron thermal velocity, $v_{\rm t, e} = \sqrt{T_{\rm e} / m_{\rm e}}$.

The magnetised sheath is considered to be in a steady state, such that it adapts immediately to changes in the external plasma.
The equilibration time of the magnetised sheath is set by the slowest timescale of the magnetised sheath: the time taken by ions to reach the target,  $\sim \rho_{\rm S} / (c_{\rm S} \sin \alpha) \sim 1/(\alpha \Omega_{\rm i})$.
Here, we introduced the signed gyrofrequency of charged particles,
\begin{align} \label{Omegas}
\Omega_s = \frac{q_s B}{m_s} \rm ,
\end{align}
where $q_{\rm i} = Ze$, $Z$ is the ion charge state, and $q_{\rm e} = -e$.
Note that for the electrons the signed gyrofrequency is negative, $\Omega_{\rm e} < 0$, and the gyrofrequency is thus strictly its absolute value $|\Omega_{\rm e}|$.
The assumption that the magnetised sheath is in steady state requires that the timescale $( \alpha  \Omega_{\rm i} )^{-1}$ be much faster than any bulk plasma timescale of interest \cite{Geraldini-2017}.

\section{Particle densities} \label{sec-particle}

In this section we analyse particle trajectories in a prescribed electrostatic field, and obtain expressions for the particle densities.
The scale of particle gyromotion in the magnetised sheath can range from much larger to much smaller than the length scale of variation of the electrostatic potential, depending on whether electrons or ions in the magnetic presheath or Debye sheath are considered.
Therefore, we analyse the possible cases separately: the motion of particles of each species $s \in \{\text{i},\text{e}\}$ in each region $R\in \{\text{ds}, \text{mp}\}$ with length scale $l_R\in \{\lambda_{\rm D},\rho_{\rm i}\}$.
In section~\ref{subsec-electronsMP}, we analyse the motion of electrons in the magnetic presheath, where their gyromotion is much smaller than the length scale of the electrostatic potential and electrons are thus essentially tied to the magnetic field line.
Then, in section~\ref{subsec-IMPEDS} we analyse the motion of ions in the magnetic presheath, $R = \rm mp$, and electrons in the Debye sheath, $R = \rm ds$. 
In both cases the length scale of electrostatic potential variation is comparable to the size of the gyro-orbits.
The analysis of these complicated trajectories is simplified by invoking $\alpha \ll 1$.
Finally, in section~\ref{subsec-ionsDS} we analyse the motion of ions in the Debye sheath, where the gyromotion is much larger than the length scale of the electrostatic potential variation and so the magnetic force can be neglected altogether.

\subsection{Electrons in the magnetic presheath}  \label{subsec-electronsMP}

In the magnetic presheath, $\rho_{\rm e} \ll x  \sim l_{\rm mp} \sim \rho_{\rm S}$ is satisfied.
When the motion of a particle of species $s$ in a system with characteristic length scale $l_R$ satisfies the ordering $\rho_s \ll l_R$, the gyration velocity of the particle about the magnetic field can be expressed as a function of the magnetic moment,
\begin{align} \label{mu-tilde}
\tilde{\mu} = \frac{v_{\tilde{x}}^2 + v_{y}^2}{2|\Omega_s|}  \rm ,
\end{align}
which is approximately conserved.
Here, $\vec v$ is the velocity of a particle, while $v_k$ is the $k$-component $\vec v \cdot \hat{\vec{e}}_k$.
The velocity component $v_{\tilde{x}}$ is related to $v_x$ and $v_z$ via
\begin{align} \label{vxtil-def}
v_{\tilde{x}} = v_x \cos \alpha + v_z \sin \alpha  \simeq v_x  +  \alpha v_z   \rm .
\end{align}
The phase $\varphi$ of the electron gyration, or gyrophase, is defined by
\begin{align} \label{varphi-def}
\tan \varphi = \frac{-v_{\tilde{x}}}{v_{y}} \rm .
\end{align}
The gyrophase is defined such that $d\varphi / dt = |\Omega_{\rm e}|$.
In equation (\ref{mu-tilde}) we have neglected the $\vec{E} \times \vec{B}$ drift of the particle because it is small relative to its velocity \cite{Ewart-2021}. 

The total energy (per unit mass) of an electron in the magnetic presheath is
\begin{align}
U_{\rm mp} =  \frac{1}{2} v_{\tilde{x}}^2 + \frac{1}{2} v_{y}^2 + \frac{1}{2} v_{\parallel}^2 + \frac{\Omega_{\rm e}\phi_{\rm mp}(x)}{B} \rm . 
\end{align}
Using the magnetic moment $\tilde{\mu}_{\rm e}$, the energy $U_{\rm mp}$ and the gyrophase angle $\varphi$, the velocity components in the $\tilde{x}, ~y,$ and $ \tilde{z}~(\parallel)$ directions are
\begin{align} \label{vxtil-varphi}
    v_{\tilde{x}} = \sqrt{2 |\Omega_e| \tilde{\mu}} \sin \varphi \rm ,
\end{align}
\begin{align} \label{vytil-varphi}
    v_{y} = - \sqrt{2 |\Omega_e| \tilde{\mu}} \cos \varphi \rm ,
\end{align}
\begin{align} \label{vztil-U-mu}
    v_{\parallel} = \sigma_{\parallel} \sqrt{2\left( U_{\rm mp} - |\Omega_e| \tilde{\mu} - \frac{ \Omega_{\rm e} \phi_{\rm mp} (x) }{ B } \right)}  \rm ,
\end{align}
where $\sigma_{\parallel} = v_{\parallel} / |v_{\parallel}|$.
The electron velocity component in the $\tilde{z}$ ($\parallel$) direction is, in terms of $v_{z} $ and $v_{x}$,
\begin{align}
v_{\parallel} = v_{z} \cos \alpha - v_{x} \sin \alpha \simeq v_{z}  - \alpha v_{x}  \rm .
\end{align}

Electrons moving through the magnetic presheath conserve $\tilde{\mu}$ and $U_{\rm mp}$ \cite{Ewart-2021}. 
While the conservation of $U_{\rm mp}$ is exact in steady state, the conservation of the magnetic moment $\tilde{\mu}$ relies on $\rho_{\rm e} \ll l_{\rm mp} \sim \rho_{\rm S}$, and is connected with the gyromotion of electrons being a very fast timescale.
This makes the electron distribution function independent of the gyrophase angle $\varphi$ to lowest order in $\rho_{\rm e} / \rho_{\rm S} \sim M^{-1/2}$.
If the distribution function of electrons at $x / \rho_{\rm S} \rightarrow \infty$ travelling into the magnetic presheath, that is, with $v_\parallel > 0$ and thus $\sigma_{\parallel} =+1$, is expressed as a function only of $\tilde{\mu}$ and $U_{\rm mp}$, $F_{\rm e\infty}(\tilde{\mu}, U_{\rm mp})$, then the distribution throughout the magnetic presheath is given by $f_{\rm e, mp}(x, \tilde \mu, v_{\parallel}) = F_{\rm e\infty}(\tilde{\mu}, U_{\rm mp})$ for $v_\parallel > 0$.
This follows from the conservation of $\tilde{\mu}$ and $U_{\rm mp}$. 
By convention, we have denoted the stationary distribution function expressed in terms of conserved variables using an upper case $F$, while denoting the $x$-dependent velocity distribution expressed as a function of $v_\parallel$ and $\tilde \mu$ using a lower case $f$.
Usually, most of the electrons entering the magnetic presheath do so with values of $\tilde{\mu}$ and $U_{\rm mp}$ such that they are reflected within the magnetic presheath or the Debye sheath and come back out with $\sigma_{\parallel} = -1$.
We will see in subsection~\ref{subsubsec-reflection} that this happens for $U_{\rm mp} < U_{\rm cut,ds} ( \tilde{\mu} ) + \Omega_{\rm e} \phi_{\rm mp}(0) / B$, where $U_{\rm cut,ds} ( \tilde{\mu} )$ is an energy cutoff \cite{Castillo-2024} for electron reflection within the Debye sheath.
The overall distribution of magnetic moment and energy, including particles streaming in both directions along the magnetic field line, is denoted with an overbar and can be written as
\begin{align} \label{Fe-mp}
\bar{F}_{\rm e, mp} & (\tilde{\mu}, U_{\rm mp}, \sigma_{\parallel}) \nonumber \\
& = \begin{cases} F_{\text{e}\infty} (\tilde{\mu}, U_{\rm mp}) & \text{ for } \sigma_{\parallel}  = +1 \rm , \\
 F_{\text{e}\infty} (\tilde{\mu}, U_{\rm mp}) \Theta \left( U_{\rm cut, ds} (\tilde{\mu}) + \frac{\Omega_{\rm e} \phi_{\rm mp}(0)}{B} - U_{\rm mp} \right) & \text{ for } \sigma_{\parallel} = -1 \rm ,
\end{cases}
\end{align}
where $\Theta$ is the Heaviside step function defined by
\begin{align} \label{Heaviside-def}
\Theta(U) = \begin{cases} 1 \text{ for } U \geqslant 0 \rm , \\
0 \text{ for } U < 0 \rm . \end{cases}
\end{align}
To obtain the numerical results of this paper, we will assume a Maxwellian distribution $F_{\rm e \infty}$ (see section~\ref{sec-num}), although the numerical scheme is general enough that other distributions could be used.
An important point concerning (\ref{Fe-mp}) is that the reflected part ($\sigma_{\parallel} = -1$) of the electron distribution function $\bar{F}_{\rm e, mp}$ is not known until $\phi_{\rm mp}(0)$ and the cutoff function $U_{\rm cut, ds}(\tilde\mu)$ are known.
Yet these depend on the solution profiles $\phi_{\rm mp}(x)$ and $\phi_{\rm ds}(x)$ which will be derived iteratively.
This also implies that the precise value of the electron density at the entrance of the magnetised sheath depends on the potential profile of the magnetised sheath.
This dependence is weak provided that most electrons are reflected by the magnetised sheath, as is the case when $\phi_{\rm w}$ satisfies the ordering (\ref{ordering-phiw-float}).
Since in this paper we solve the magnetised sheath independently of the rest of the plasma, we consider the overall size of the electron distribution function at the magnetised sheath entrance, given by the electron density at the magnetised sheath entrance, to be specified \emph{a posteriori} from the magnetised sheath solution in such a way that quasineutrality is exactly satisfied there.
That is, for a given magnetised sheath potential profile, we provide $F_{\rm e \infty}$ up to a constant multiplier, and this constant is determined by requiring that quasineutrality be satisfied at the magnetised sheath entrance.

The distribution function of electrons in the magnetic presheath expressed in terms of the parallel velocity $v_{\parallel}$, magnetic moment $\tilde{\mu}$ and position $x$ is
\begin{align} \label{fe-mp}
f_{\rm e, mp} (x, \tilde{\mu}, v_{\parallel}) = & F_{\text{e}\infty} \left( \tilde{\mu}, |\Omega_{\rm e}| \tilde{\mu} + \frac{1}{2} v_{\parallel}^2 + \frac{\Omega_{\rm e} \phi_{\rm mp}(x)}{B} \right)   \Theta \left( v_{\parallel} + v_{\rm cut, mp} (x, \tilde{\mu}) \right)   \rm ,
\end{align}
with the cutoff parallel velocity function $v_{\rm cut, mp} (x, \tilde{\mu})$ defined by 
\begin{align} \label{v-cut-mp}
v_{\rm cut, mp} (x, \tilde{\mu}) = \sqrt{2\left(  U_{\rm cut, ds} (\tilde{\mu}) - |\Omega_e| \tilde{\mu} + \frac{ |\Omega_{\rm e}| \left( \phi_{\rm mp} (x) - \phi_{\rm mp}(0)\right) }{ B } \right)} \rm .
\end{align}
The electron density is an integral of the two branches, incoming and reflected, of the magnetic presheath distribution function (\ref{Fe-mp}) in velocity space,
\begin{align} \label{ne-mp}
n_{\rm e, mp}(x) = 2\pi \sum_{\sigma_{\parallel} = \pm 1} \int_0^{\infty} |\Omega_{\rm e}| d\tilde{\mu} \int_{|\Omega_{\rm e}| \tilde{\mu} + \Omega_e \phi_{\rm mp} (x) / B}^{\infty} dU_{\rm mp} \nonumber \\
\times \frac{\bar{F}_{\rm e, mp} ( \tilde{\mu}, U_{\rm mp}, \sigma_{\parallel})}{\sqrt{2\left( U_{\rm mp} - |\Omega_e| \tilde{\mu} - \frac{ \Omega_{\rm e} \phi_{\rm mp} (x)}{B} \right) }} \rm .
\end{align}
To write (\ref{ne-mp}), the standard Jacobian $|\partial (v_{\tilde{x}}, v_y, v_{\parallel}) / \partial (\tilde{\mu}, \varphi, U_{\rm mp}) | = |\Omega_{\rm e}| / |v_{\parallel}|$ was used and the integral over the gyrophase $\varphi$ was evaluated to $2\pi$.
Furthermore, we have used that electrons with $U_{\rm mp} < |\Omega_e| \tilde{\mu} + \Omega_{\rm e} \phi_{\rm mp} (x) / B $ must have been reflected before reaching the position $x$, and are thus absent (see equation (\ref{vztil-U-mu}) for $v_{\parallel}$).
The density can also be more conveniently expressed as
\begin{align} \label{ne-mpre}
n_{\rm e, mp} (x) = &  2\pi \int_0^{\infty} |\Omega_{\rm e}| d\tilde{\mu}  \int_{-v_{\rm cut, mp} (x, \tilde{\mu})}^{\infty} F_{\rm e \infty} \left( \tilde{\mu}, |\Omega_{\rm e}| \tilde{\mu} + \frac{1}{2} v_{\parallel}^2 + \frac{\Omega_{\rm e} \phi_{\rm mp}(x)}{B} \right)  dv_{\parallel} \rm .
\end{align}
The electron density at the entrance of the magnetic presheath is $n_{\rm e, mp}(\infty) = n_{\rm e, \infty}$, and as we explained above, it is forced to be equal to the ion density at the same location by modifying a free constant multiplier in $F_{\rm e \infty}$.

The electron current density to the target can be calculated from the fluxes along the magnetic field at the magnetic presheath entrance.
Since the magnetised sheath is assumed to be collisionless, implying also that ionization is neglected, the particle flux across it is conserved.
The electron flux in the $\tilde z$ direction can be calculated by multiplying the integrand in (\ref{ne-mp}) by $v_{\parallel} = \sigma_{\parallel} \sqrt{2( U_{\rm mp} - |\Omega_e| \tilde{\mu} - \Omega_{\rm e} \phi_{\rm mp} (x) / B ) }$,
\begin{align} \label{Phipare-mp}
\Phi_{\parallel, \rm e} = 2\pi \sum_{\sigma_{\parallel} = \pm 1} \sigma_{\parallel} \int_0^{\infty} |\Omega_{\rm e}| d\tilde{\mu} \int_{|\Omega_{\rm e}| \tilde \mu + \frac{\Omega_{\rm e} \phi_{\rm mp}(x)}{B}}^{\infty} \bar F_{\rm e, mp} (\tilde{\mu}, U_{\rm mp}, \sigma_{\parallel}) dU_{\rm mp}  \rm .
\end{align}
Since the electrons are strongly tied to the magnetic field line at the entrance of the magnetised sheath, the flux perpendicular to the magnetic field is zero, $\Phi_{\tilde{x}, e} \simeq 0$, and the electron current density in the direction normal to the wall is $J_{x, \rm e} = -e\left( \Phi_{\tilde{x}, \rm e} \cos \alpha - \Phi_{\parallel, \rm e} \sin \alpha \right) \simeq e \alpha \Phi_{\parallel, \rm e}  $.
Hence, and upon inserting (\ref{Fe-mp}) in (\ref{Phipare-mp}), we obtain
\begin{align} \label{Je-mp}
J_{x, \rm e} = 2\pi e \alpha \int_0^{\infty} |\Omega_{\rm e}| d\tilde{\mu} \int_{\frac{\Omega_{\rm e} \phi_{\rm mp}(0)}{B} + U_{\rm cut, ds} (\tilde{\mu})}^{\infty} F_{\rm e \infty} (\tilde{\mu}, U_{\rm mp}) dU_{\rm mp}  \rm .
\end{align}

\subsection{Ions in the magnetic presheath and electrons in the Debye sheath} \label{subsec-IMPEDS}

This subsection is a summary of the grazing-angle gyrokinetic treatment of particles in a strong inhomogenous electric field approximately perpendicular to a constant magnetic field which is discussed in detail in references \cite{Geraldini-2017} and \cite{Geraldini-2018}. 
Some crucial elements of this treatment had previously been developed in references \cite{Gerver-Parker-Theilhaber-1990} and \cite{Cohen-Ryutov-1998}.
Within the small-angle ($\alpha \ll 1$) framework, the effect of finite electron Larmor orbits on the electron reflection in the Debye sheath had only been studied in the limit $\rho_{\rm e} / \lambda_{\rm D} \rightarrow \infty$ \cite{Cohen-Ryutov-1995-spreading}. 
Here, we treat electron reflection in the Debye sheath for $\rho_{\rm e} \sim \lambda_{\rm D}$.
We present the equations in a mostly self-contained manner, while not going into any of the more advanced derivations which had been previously presented.
For more details, the reader is referred to the aforementioned papers.

Particles moving in the magnetised sheath experience a strong electric field normal to the target.
Owing to the shallow magnetic field angle at the target, the electric field is approximately in the plane of the gyro-orbit.
When a particle gets close enough to the target, this electric field is strong and inhomogeneous on the length scale of the gyro-orbit, and thus strongly distorts the gyro-motion away from its usual circular shape.
For ions, this happens at distances from the target comparable to the ion sound gyroradius, $\rho_{\rm S}$, in the magnetic presheath.
Hence, the piece of the electrostatic potential responsible for distorting ion gyro-orbits is $\phi_{\rm mp}$.
For electrons, when $\gamma \sim 1$, this happens in the Debye sheath at distances $ x \sim \rho_e = \gamma \lambda_D $.
Hence, the piece of the electrostatic potential responsible for distorting electron gyro-orbits is $\phi_{\rm ds}$.

With $\alpha \ll 1$, particle motion in distorted gyro-orbits can be approximately split into ``closed'' orbits (section \ref{subsubsec-closed}), conserving an adiabatic invariant (section \ref{subsubsec-adiabatic}), and ``open'' orbits (section \ref{subsubsec-open}), hitting the wall. 
If $\Omega_s \phi_{R} >0$, such as for electrons in the Debye sheath, reflected particles with $v_z < 0$ can be present (section \ref{subsubsec-reflection}). 
The density contribution from closed and open orbits can be calculated and combined to obtain the total particle density for $\alpha \ll 1$ (section \ref{subsubsec-dens}).

\subsubsection{Closed orbits.} \label{subsubsec-closed}

The equations of motion for electrons in the Debye sheath, $R = $ ds, and for ions in the magnetic presheath, $R = $ mp, are
\begin{align}
\centering
\dot{x} & = v_{x} \text{,} \label{x-EOM-exact}
\end{align}
\begin{align}
\dot{v}_{x} & = -\frac{ \Omega_s \phi_{R}'(x) }{B} + \Omega_s v_{y}\cos\alpha \text{,} \label{vx-EOM-exact} 
\end{align}
\begin{align}
\dot{v}_{y} & =  - \Omega_s v_{x}\cos\alpha - \Omega_s v_{z}\sin\alpha \text{,} \label{vy-EOM-exact} 
\end{align}
\begin{align}
\dot{v}_{z} &  = \Omega_s v_{y}\sin\alpha \label{vz-EOM-exact} \text{.}
\end{align}
Note that, from (\ref{Omegas}), for ions $\Omega_{\rm i} = ZeB/m_{\rm i}$ is positive while for electrons $\Omega_{\rm e} = -eB/m_e$ is negative.
Retaining only terms small in $\alpha \ll 1$ to first order, we obtain
\begin{align}
\dot{v}_{x} & \simeq -\frac{\Omega_s \phi_{R}'(x) }{B} + \Omega_s v_{y} \text{,} \label{vx-EOM-smallalpha} 
\end{align}
\begin{align}
\dot{v}_{y} & \simeq  - \Omega_s v_{x} - \alpha \Omega_s v_{z} \text{,} \label{vy-EOM-smallalpha} 
\end{align}
\begin{align}
\dot{v}_{z} &  \simeq \alpha \Omega_s v_{y} \label{vz-EOM-smallalpha} \text{.}
\end{align}

Equations (\ref{vx-EOM-exact})-(\ref{vz-EOM-exact}) have been analysed extensively \cite{Gerver-Parker-Theilhaber-1990, Cohen-Ryutov-1998, Geraldini-2017, Geraldini-2018} for ions in the magnetic presheath.
The characteristic frequency of the motion is $|\Omega_s|$.
Setting $\alpha = 0$ gives a motion consisting of parallel streaming in the $z$ direction at constant $v_z$ and an $\vec{E} \times \vec{B}$ drift in the $y$ direction, superimposed on closed (i.e., periodic), generally non-circular, gyromotion in the $xy$ plane.
The gyromotion is exactly circular, with period $2\pi / |\Omega_s|$, only if $\phi_{R}'' = 0$ everywhere. 
The orbit parameters
\begin{align} \label{xbar-def}
\bar{x} & = x + \frac{v_{y}}{\Omega_s} \text{,}
\end{align}
\begin{align}
U_{\perp} & =  \frac{1}{2} v_{x}^2 + \frac{1}{2} v_{y}^2 + \frac{\Omega_s \phi_{R} (x)}{B}  \text{,} \label{Uperp-def} 
\end{align}
\begin{align}
U_{R} & =  \frac{1}{2} v_{x}^2 + \frac{1}{2} v_{y}^2 + \frac{1}{2} v_{z}^2 + \frac{\Omega_s \phi_{R} (x)}{B}  \text{,}  \label{U-def}
\end{align}
are constants of the motion when $\alpha = 0$.
When $\alpha \neq 0$, the orbit position $\bar{x}$ and the perpendicular energy $U_{\perp}$ are no longer constant, and for $0 < \alpha \ll 1$ one has
\begin{align} \label{xbardot}
    \dot{\bar{x}} = v_{x} + \frac{\dot{v}_{y}}{\Omega_{s}} \simeq - \alpha v_{z} \rm ,
\end{align}
and
\begin{align}
    \dot{U}_{\perp} = v_{x} \dot{v}_{x} + v_{y} \dot{v}_{y} + \frac{\Omega_{s} v_{x} \phi_{R}'(x)}{B} \simeq - \alpha \Omega_{s} v_{y} v_{z} \rm ,
\end{align}
while the total energy $U_{R}$ remains constant,
\begin{align}
    \dot{U}_{R} = 0 \rm .
\end{align}
The characteristic frequency of variation of $\bar{x}$ and $U_{\perp}$ is $\alpha |\Omega_s|$, much slower than the characteristic frequency of the lowest order gyromotion.
Thus, although not exactly constant, the orbit parameters can be considered to be constant over the fast timescale $2\pi / |\Omega_s|$.
The particle therefore undergoes a quasi-periodic motion. 

Using the orbit parameters, the particle velocities can be re-expressed as
\begin{align} \label{vx-GK}
v_{x} & = \sigma_{x} \sqrt{2\left(U_{\perp} - \chi_R (x, \bar{x}) \right)} \rm , 
\end{align}
\begin{align} \label{vy-GK}
v_{y} & = \Omega_s \left( \bar{x} - x \right) \rm , 
\end{align}
\begin{align} 
v_{z} & = \sigma_{z} \sqrt{2\left(U_{R} - U_{\perp} \right)}  \label{vz-U-Uperp } \rm ,
\end{align}
where we have introduced an effective potential
\begin{align} \label{chis}
\chi_R (x, \bar{x}) = \frac{1}{2} \Omega_{s}^2 \left( x - \bar{x} \right)^2 + \frac{\Omega_{s} \phi_{R} (x)}{B} \rm ,
\end{align}
the sign of $v_{x}$, $\sigma_{x} = v_{x} / |v_{x}|$, and the sign of $v_{z}$, $\sigma_{z} = v_{z}/|v_{z}|$.
When $\phi_{R}(x)$ is constant for all values of $x$, the effective potential (\ref{chis}) is parabolic and the lowest order charged particle gyromotion is circular in the $xy$ plane.
In general, however, $\phi_{R}(x)$ is strongly inhomogeneous on the scale of the gyro-orbit size.
The presence of a local minimum in the effective potential is necessary to have quasi-periodic motion.

The condition on $\bar{x}$ for $\chi_{R}$ to have a minimum of the effective potential is obtained by analysing the first derivative of  $\chi_{R}$ with respect to $x$ at fixed $\bar{x}$.
Extremas of $\chi_R$ occur at values of $x$ satisfying $\partial \chi (x, \bar{x}) / \partial x = \Omega_{s}^2 (x - \bar{x} ) + \Omega_{s} \phi_{R}'(x) / B = 0$.
Therefore, the function
\begin{align} \label{xi}
\xi_R (x) = x + \frac{\phi_{R}'(x)}{\Omega_{s} B} \rm 
\end{align}
returns, for a given value of $x$, the value of $\bar{x}$ for which $x$ is a stationary point of $\chi_R (x, \bar{x})$.
The position $x_{\text{m}} \geqslant 0$ is a local minimum of the function $\chi_{R} (x, \bar{x})$ if 
\begin{align}
\bar{x} = \xi_R ( x_{\text{m}} ) & \text{ and } \xi_R' (x_{\text{m}}) > 0  \rm ,
\end{align}
since $\xi_R'(x) = \Omega_{s}^{-2} \partial^2 \chi_R (x, \bar {x}) / \partial x^2$.
The value of the effective potential at the minimum is denoted
\begin{align}
\chi_{R, \rm m} (\bar{x}) \equiv \chi_R (x_{\text{m}}, \bar{x}) \rm .
\end{align}
As was done in reference \cite{Geraldini-2018}, we search for electrostatic potential solutions such that at most one minimum of the effective potential exists for each $\bar x$.
This is guaranteed if $\xi_R'(x)$ is monotonic, which is the case if $\phi_R''(x)$ is monotonic.\footnote{A monotonic $\xi_{R}'$ can have at most one zero. At that zero, $\xi_R$ has a minimum, or maximum, and around that extremum, the same value of $\xi_R = \bar x$ has two corresponding $x$ values: that is, for some $\bar x$, there are two extrema of $\chi_R$ with respect to $x$. Given the change of sign of $\xi_{R}'$, one of these extrema of $\chi_R$ is a maximum, and the other is a minimum.}
Then, the smallest value of $\bar{x}$, denoted $\bar{x}_{\star}$, for which a minimum of the effective potential exists is defined by the equation
\begin{align} \label{xbarstar}
\bar{x}_{\star} = \begin{cases}  \xi_R ( x_{\star} ) & \text{ if a point } x_{\star} \text{ with } \xi_R' (x_{\star}) = 0 \text{ exists,}  \\
\xi_R ( 0 ) & \text{ otherwise.} 
\end{cases}
\end{align}
Equation (\ref{xbarstar}) also defines the critical point $x_{\star}$ which corresponds to the inflexion point of the effective potential curve when there exists an effective potential curve with such an inflexion point. When there are no effective potential curves with an inflexion point (e.g. for electrons in the Debye sheath), we define $x_{\star}$ to be $0$, $x_{\star} = 0$.
Since $\phi_{R}''(x) < 0$, for electrons in the Debye sheath $\xi_{\rm ds}'(x) > 0$ everywhere and the stationary points of the effective potential curves are only minima.

\subsubsection{Adiabatic invariance.} \label{subsubsec-adiabatic}

For $\bar x > \bar x_{\star}$, the turning points of the gyro-motion occur at the positions $x_+$ and $x_-$ where $U_{\perp} = \chi_R (x_{\pm}, \bar{x})$ on either side of the minimum of the effective potential: $x_- < x_{\text{m}} < x_+$.
A key result is that the particle motion conserves the adiabatic invariant
\begin{align}  \label{mugk-def}
\mu = \mu_{\alpha} (\bar{x}, U_{\perp}) \equiv \frac{1}{\pi} \int_{x_-}^{x_+} dx \sqrt{2\left( U_{\perp} - \chi_{R} \left( x, \bar{x} \right) \right) }
\end{align}
to lowest order in $\alpha \ll 1$ as it moves across the system \cite{Cohen-Ryutov-1998, Geraldini-2017}.
This quantity is conserved over the long timescale $1/(\alpha |\Omega_{s}|)$ in which the orbit parameters vary by order unity. 
The subscript in $\mu_{\alpha}$ serves to emphasise that this is a function of other particle variables ($\bar x$ and $U_{\perp}$) and thus distinguish it from $\mu$, which is the particle variable returned by $\mu_{\alpha}$.
Considering a particle that is sufficiently far from the target that the electric field can be neglected to lowest order, the adiabatic invariant reduces, from equation (\ref{mugk-def}) with constant $\phi_{R}$ and equation (\ref{chis}), to \cite{Geraldini-2019}
\begin{align} \label{mu-infty}
   \mu_{\infty} = \mu_{\alpha} (\bar{x}, U_{\perp}) \rvert_{\bar x \rightarrow \infty} = \frac{v_{x}^2 + v_{y}^2}{2|\Omega_s|} \rm .
\end{align}
Since, from equation (\ref{vxtil-def}), $v_{x}$ only differs from $v_{\tilde{x}}$ by corrections that are small in $\alpha \ll 1$, the adiabatic invariant $\mu$ and the magnetic moment $\tilde{\mu}$ of a particle far enough away from the target that its gyro-orbit is circular are equal to each other to within a small geometrical factor, $\mu = \tilde{\mu} (1 + O(\alpha))$, where $\tilde{\mu}$ is defined in (\ref{mu-tilde}).
Thus, for an ion at the magnetic presheath entrance, or an electron at the Debye sheath entrance, the value of $\mu$ given by (\ref{mugk-def}) can be equated, to lowest order in $\alpha \ll 1$, to the value of $\tilde{\mu}$ given by (\ref{mu-tilde}).
It can then be concluded, to a good approximation, that for such a particle the integral in (\ref{mugk-def}) evaluates to the same value of $\mu$ everywhere in its future trajectory (as long as the particle remains trapped in $\chi_R$).

\subsubsection{Open orbits.} \label{subsubsec-open}

Any trajectory that intersects the wall to lowest order in $\alpha$ is considered open.
The condition for a trajectory to be open is that its perpendicular energy $U_{\perp} $ be large enough that there is no bottom turning point $x_-$, that is, $U_{\perp} = \chi (x_-, \bar{x})$ does not have a solution for $0 \leqslant x_- < x_{\text{m}}$. 
Assuming $\bar x > \bar x_{\star}$, the maximum value of the effective potential between $x=0$ and the minimum $x= x_{\text{m}}$ is defined by
\begin{align}
\chi_{R, \text{M}} (\bar{x}) = \max_{x \in [0, x_{\text{m}}]} \chi_R (x, \bar{x}) \rm .
\end{align}
Therefore, a particle is considered to be in an open orbit if $U_{\perp} > \chi_{R, \text{M}}(\bar{x})$.
To lowest order in $\alpha$, open orbits are absent from the system. 
This is because the time $\sim 1/|\Omega_{s}|$ it takes for particles in open orbits to reach the wall is a factor of $\alpha$ smaller than the time $\sim 1/ (\alpha \Omega_{s})$ that it takes for the orbit parameters to change such that $U_{\perp}$ becomes significantly larger than $ \chi_{R, \text{M}}$.
Hence, particles in open orbits can be taken to satisfy $U_{\perp} \simeq \chi_{R, \text{M}}$ with a small spread in energy related to the change in perpendicular energy and orbit position over a single orbit.

The very small contribution to the density of the open ion orbits is crucial to solve the quasinetrality equation in the magnetic presheath.
Without it, the ion density at the entrance of the Debye sheath would be zero.
While this is consistent with the asymptotic ordering $\alpha \ll 1$, as the ion density effectively vanishes at the wall to lowest order in $\alpha$, it does not allow for a self-consistent solution of the electrostatic potential to lowest order in $\alpha$: the size of electron gyro-orbits is negligible in the magnetic presheath, and so the electron density can only be zero here if \emph{no} electrons reach $x=0$. 
Assuming that even a few electrons with arbitrarily high energies, $U_{\rm mp} \gg v_{\rm t, e}^2$, are present, equation (\ref{vztil-U-mu}) implies that an unphysical infinite potential drop in the magnetic presheath would be required to reflect all electrons. 
Including the small but finite number of ions in open orbits solves this problem: quasineutrality requires that \emph{most}, not all, electrons be reflected before reaching $x/\rho_{\rm S} = 0$. 

Electrons in open orbits also exist in the Debye sheath, but their contribution to the density 
is expected to be only a small correction to the charge density due to charge separation.
Accounting for electrons in open orbits is, however, necessary to calculate the distribution function of electrons reaching the target.

In order to calculate the density of particles in open orbits, it is useful to first calculate the small range of values of $v_{x}$ that a particle in an open orbit can have at a given position $x$ for $U_{\perp} > \chi_{R, \rm M}$ \cite{Geraldini-2018, Geraldini-2019}.
Ignoring some additional corrections which do not affect the density, the result of this calculation is
\begin{align} \label{vx-range}
- \sqrt{2\left( \chi_{R,\rm M}(\bar{x}) - \chi_{R}(x, \bar{x}) + \Delta_{\text{M}} (\bar{x}, U_{R}) \right)}  \leqslant v_{x} < - \sqrt{2\left( \chi_{R, \rm M}(\bar{x}) - \chi_R(x, \bar{x}) \right)} \rm ,
\end{align}
where
\begin{align} \label{DeltaM-def}
\Delta_{\rm M}(\bar{x}, U_R) = 2\pi \alpha \sqrt{2\left( U_{R} - \chi_{R,\text{M}}(\bar{x})  \right)}  \frac{d\mu_{\text{op}}}{d\bar x}(\bar{x}) \text{,}
\end{align}
and $ \mu_{\text{op}}(\bar{x})$ (where the subscript ``op'' stands for ``open'') is obtained by substituting $U_{\perp} = \chi_{R,\text{M}}(\bar{x})$ in (\ref{mugk-def}),
\begin{align} \label{muop-def}
\mu_{\text{op}}(\bar{x})  = \mu_{\alpha} (\bar{x}, \chi_{R,\text{M}}(\bar{x}) )\text{.}
\end{align}

\subsubsection{Electron reflection.} \label{subsubsec-reflection}

This section presents a new aspect of the analysis of particle motion in distorted gyro-orbits near a wall in a strongly tilted magnetic field, $\alpha \ll 1$.
Previous analyses were carried out only for ions undergoing net acceleration towards an absorbing wall, a case in which no ions with $\sigma_{z} = -1$ are present.
Since we consider the standard case in which electrons are repelled away from the wall, the values of $U_{\rm ds}$ and $\mu$ for which electrons can have $\sigma_{z} = -1$ must be calculated in order to account for reflected electrons. 
Concretely, the analysis of this section generalizes the realistic logical-sheath or conducting-wall boundary conditions typically used in kinetic plasma codes.

In an unmagnetised plasma, the logical-sheath boundary condition \cite{Parker-1993} only reflects electrons from the sheath back into the main plasma if their normal velocity component does not exceed a cutoff.
The physical principle underlying this boundary condition is the presence of the sheath potential barrier next to the wall at a scale (the Debye length) that is unresolved by the spatial grid of the simulation.
The size of the cutoff is determined locally by allowing through only as many electrons as are necessary to balance the ion current and achieve no net flux of charge (i.e., current density) out of the plasma at each point of the boundary, thus imposing local ambipolarity everywhere at the boundary.
In a magnetised plasma, the boundary conditions can be imposed in a similar way by invoking an unresolved potential drop on both the magnetic presheath and Debye sheath scales.
As well as the logical condition, a modification to the logical condition known as the conducting-wall boundary condition has been proposed \cite{Shi-2017}.
Although the global current into the wall (integrated over all plasma-facing material surfaces) must remain zero for a quasineutral plasma with a steady-state sheath, the conducting condition allows the current density reaching the wall to be non-zero \emph{locally}.
This allows local current fluctuations which have been shown to be essential to stabilise electromagnetic modes near the target \cite{Mandell-2022}.
When applied to a magnetised plasma, the logical-sheath and the conducting-wall models assume that the electrostatic potential of the magnetic presheath and of the Debye sheath only affect the electron \emph{parallel} velocity. 
While this assumption is accurate in the magnetic presheath (see section~\ref{subsec-electronsMP}), it also implicitly assumes that the electron gyroradius is negligible in the Debye sheath, $\gamma = \rho_{\rm e} / \lambda_{\rm D} = 0$.
Here, we present the analysis underlying the numerical calculation of the generalised logical-sheath and conducting-wall boundary conditions for a magnetised plasma where $\gamma \neq 0$, assuming $\alpha \ll 1$.
We derive the cutoff in parallel velocity at the Debye sheath entrance (from which the cutoff anywhere in the magnetic presheath can be easily recovered) as a function, to be calculated numerically in general, of the magnetic moment $\mu$.
This $\mu$-dependent cutoff was also recently observed in PIC simulations \cite{Castillo-2024}.
In the next subsection, we use the cutoff to obtain the self-consistent reflected electron distribution function, which can replace the one calculated in gyrokinetic codes from logical-sheath or conducting-wall boundary conditions.

The perpendicular energy $U_{\perp} = U_{\rm ds} - \frac{1}{2} v_z^2$ increases as electrons approach the wall.
This can be understood by noting that the electric field normal to the target has a small component pushing electrons away from the wall in the direction along the magnetic field, increasing the perpendicular energy $U_{\perp}$ at the expense of the kinetic energy related to the velocity component $v_{z}$ which is approximately parallel to the magnetic field.
For some electrons, the increase in $U_{\perp}$ occurs until $U_{\perp} = U_{\rm ds}$ is reached and the electron reflects, changing sign of $v_{z}$.
At that point the electric field component pushing along the magnetic field line makes $v_{z}$ become more negative, increasing $|v_{z}|$ and correspondingly decreasing $U_{\perp}$, until the electron eventually comes back out of the sheath.
For the rest of the electrons, the increase in $U_{\perp}$ occurs until the electron reaches the target, $x = 0$, and is absorbed, so that the sign of $v_{z}$ never changes.

\begin{figure} 
\centering
\includegraphics[width=0.8\textwidth]{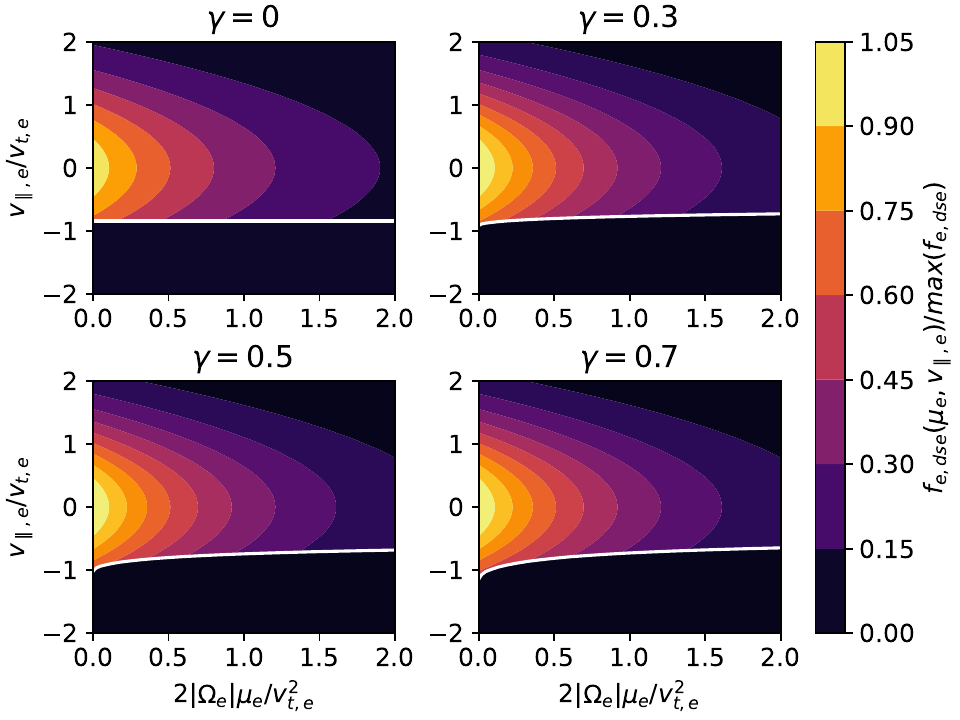} 
\caption{Contour plot of electron distribution function at the Debye sheath entrance, $f_{\rm e, dse}(\mu, v_{\parallel})$, resulting from a numerical solution of the ambipolar potential profile in the magnetised sheath for $M = 3600$ ($\approx $ deuterium), $\alpha = 2.5^{\circ}$ ($\approx $ ITER), and different values of $\gamma$, using the boundary conditions in section~\ref{subsec-num-bc} for the distribution functions. The effect of finite values of $\gamma$ is to make the cutoff parallel velocity $v_{\rm cut, dse}(\mu)$ (white solid line) dependent on $\mu$. 
The solution corresponding to $\gamma = 0.7$ is critical (see section~\ref{subsec-num-sol}), so monotonic electron-repelling sheath solutions are absent for higher values of $\gamma$ at this angle.}
\label{fig-vcut}
\end{figure}

The boundary between reflected electrons and absorbed ones for a given electrostatic potential profile $\phi_{\rm ds}$ can be obtained by identifying the electrons that only just touch the wall when $U_{\perp} = U_{\rm ds}$.
Electrons touching the wall must have $U_{\perp} = \chi_{\rm ds, M}(\bar{x}) =  \chi_{\rm ds}(0, \bar{x})$, where the second equality follows because the wall is electron-repelling. 
This gives the parametric equations
\begin{align} \label{mu-cut}
\mu = \mu_{\rm op} (\bar{x}) = \frac{1}{\pi} \int_{0}^{x_+}  \sqrt{2\left( \chi_{\rm ds} \left(0, \bar{x} \right) - \chi_{\rm ds} \left( x, \bar{x} \right) \right) } dx \rm ,
\end{align}
\begin{align} \label{U-cut}
U_{\rm ds} = \chi_{\rm ds} \left(0, \bar{x} \right) \rm ,
\end{align}
for the boundary, or cutoff, between reflected and absorbed electrons.
Recall that at the Debye sheath entrance, $\mu$ is, within geometric corrections that are small in $\alpha$, equivalent to the magnetic moment $\tilde{\mu}$, as concluded in section~\ref{subsubsec-adiabatic}, and that $U_{\rm mp} = U_{\rm ds} + \Omega_{\rm e} \phi_{\rm mp}(0) / B$.
Hence, the cutoff function $U_{\rm cut, ds} ( \tilde \mu ) $ introduced in (\ref{Fe-mp}) is, from (\ref{mu-cut})-(\ref{U-cut}),
\begin{align} \label{Ucutds}
U_{\rm cut, ds} ( \mu ) \equiv \chi_{\rm ds} \left(0, \mu_{\rm op}^{-1} (\mu)  \right) = \frac{1}{2} \Omega_{\rm e}^2 \left[ \mu_{\rm op}^{-1}(\mu) \right]^2 + \frac{\Omega_{\rm e} \phi_{\rm ds}(0)}{B}\rm ,
\end{align}
where $\mu_{\rm op}^{-1}$ is the inverse of the function $\mu_{\rm op}$ given in (\ref{mu-cut}).
For $U_{\rm ds} > U_{\rm cut, ds} (\mu)$ electrons reach the wall and are absorbed, while for $U_{\rm ds} < U_{\rm cut, ds} (\mu)$ they are reflected.
The cutoff parallel velocity at the Debye sheath entrance is, from (\ref{v-cut-mp}) evaluated at $x=0$,
\begin{align} \label{v-cut-dse}
v_{\rm cut, dse}(\mu) \equiv v_{\rm cut, mp}(0, \mu) = \sqrt{2\left(  U_{\rm cut, ds} (\mu) - |\Omega_e| \mu \right)} \rm .
\end{align}
This is plotted in figure~\ref{fig-vcut} for different values of $\gamma$.

When the electron gyroradius is neglected compared to the Debye length, $\gamma =  \rho_{\rm e} / \lambda_{\rm D}  = 0 $, the electrons can be considered to be tied to the magnetic field line.
Just as in the magnetic presheath, the variation of the electrostatic potential over an electron orbit is negligible.
Calculating $\mu_{\rm op}(\bar x)$ by taking $\phi_{\rm ds}(x) - \phi_{\rm ds}(0) = O( \rho_{\rm e} \phi_{\rm ds}'(0) ) = O( \gamma \phi_{\rm ds}(0) )$ with $\gamma = 0$ in (\ref{mu-cut}) gives, upon recalling the definition of $\chi_{R}$ in (\ref{chis}), $\mu_{\rm op}(\bar x) = \frac{1}{2} |\Omega_{\rm e}| \bar x^2$.
Hence, $\mu_{\rm op}^{-1}(\bar x) = \sqrt{2\mu / |\Omega_{\rm e}| }$ and, from (\ref{Ucutds}),
\begin{align} \label{Ucut-smallgamma}
U_{\text{cut,ds,}0}(\mu) = | \Omega_{\rm e} | \mu + \frac{ \Omega_{\rm e}  \phi_{\rm ds}(0) }{B} \rm .
\end{align}
Inserting (\ref{Ucut-smallgamma}) into (\ref{v-cut-dse}) recovers the constant parallel velocity cutoff presently assumed in logical-sheath or conducting-wall boundary conditions, 
\begin{align} \label{vcut-dse-smallgamma}
v_{\rm cut, dse,0}(\mu) = \sqrt{2\left(U_{\rm cut, ds,0}(\mu) - |\Omega_{\rm e}|\mu \right)} = \sqrt{\frac{2 e | \phi_{\rm ds}(0)|}{m_{\rm e}}} \rm .
\end{align}
Note that in $U_{\text{cut,ds,}0}$ and $v_{\rm cut, dse,0}$ the subscript $0$ refers to $\gamma = 0$.
Although $v_{\rm cut, dse}(\mu) = v_{\rm cut, dse,0}(\mu)$ holds only for $\gamma = 0$, the ordering $v_{\rm cut, dse}(\mu) \sim v_{\rm cut, dse,0}(\mu)$ can be considered to hold for all values of $\gamma \lesssim 1$, as seen in figure~\ref{fig-vcut}.

\subsubsection{Density and distribution function.} \label{subsubsec-dens}

The conservation of $\mu$ and $U_{R}$ implies that the particle velocity distribution for species $s$, when expressed as a function of $\mu$ and $U_{R}$, is constant for ions across the magnetic presheath and for electrons across the Debye sheath.
The particle distribution function can therefore be calculated at any position $x$ from the distribution of magnetic moment and energy of particles \emph{entering} ($\sigma_{\parallel} = +1$) the magnetised sheath from the bulk plasma, $F_{s\infty} (\tilde{\mu}, U_{\rm mp})$. 
Since there are no reflected ions, the ion velocity distribution in the magnetic presheath is 
\begin{align} \label{Fi-mp}
\bar{F}_{\rm i, mp} (\mu,  U_{\rm mp}, \sigma_{\parallel}) \simeq \begin{cases} F_{\rm i \infty} \left( \mu,  U_{\rm mp} \right)   & \text{ for } \sigma_{\parallel}  = 1 \rm , \\
0 & \text{ for } \sigma_{\parallel} = -1 \rm .  
\end{cases}
\end{align}
For electrons in the Debye sheath, the total distribution function, including reflected electrons, is (using (\ref{Fe-mp}) and $ U_{\rm mp} = U_{\rm ds} + \Omega_{\rm e} \phi_{\rm mp}(0)/B$)
\begin{align} \label{Fe-ds}
\bar{F}_{\rm e, ds} (\mu,  U_{\rm ds}, \sigma_{\parallel}) \simeq \begin{cases} F_{\text{e}\infty} (\tilde{\mu},  U_{\rm ds} + \frac{\Omega_{\rm e} \phi_{\rm mp}(0)}{B}) & \text{ for } \sigma_{\parallel}  = 1 \rm , \\
 F_{\text{e}\infty} (\tilde{\mu},  U_{\rm ds} + \frac{\Omega_{\rm e} \phi_{\rm mp}(0)}{B}) \Theta \left( U_{\rm cut, ds} (\tilde{\mu}) - U_{\rm ds} \right) & \text{ for } \sigma_{\parallel} = -1 \rm .
\end{cases} 
\end{align}
The electron distribution function expressed in terms of the parallel velocity (\ref{vztil-U-mu}) and the adiabatic invariant $\mu$ at the Debye sheath entrance is, from (\ref{fe-mp}) evaluated at $x=0$ and (\ref{v-cut-dse}),
\begin{align}
f_{\rm e, dse} (\mu, v_{\parallel}) & = f_{\rm e, mp}(0, \mu, v_{\parallel}) \nonumber \\
& = F_{\text{e}\infty} \left( \mu, |\Omega_{\rm e}| \mu + \frac{1}{2} v_{\parallel}^2 + \frac{\Omega_{\rm e} \phi_{\rm mp}(0)}{B} \right)   \Theta \left( v_{\parallel} + v_{\rm cut, dse}(\mu)  \right) \rm .
\end{align}

The density of ions in the magnetic presheath ($s=$i, $R = $mp) and of electrons in the Debye sheath ($s=$e, $R = $ds)  are both given by
\begin{align} \label{n-finorb}
& n_{s,R}(x)  = \nonumber  \\
&  \sum_{\sigma_{\parallel} = \pm 1} \int_{\bar{x}_{\rm m} (x)}^{\infty} |\Omega_s| d\bar{x} \int_{\chi_R (x, \bar x)}^{\chi_{R,\rm M} (\bar{x})} \frac{2 dU_{\perp} }{\sqrt{2\left( U_{\perp} - \chi_R (x, \bar{x} ) \right)}} \int_{U_{\perp}}^{\infty} \frac{\bar F_{s, R} (\mu_{\alpha}(\bar{x}, U_{\perp}), U_R, \sigma_{\parallel}) dU_R }{\sqrt{2\left( U_R - U_{\perp} \right)}} \nonumber \\
& +  \int_{\bar{x}_{\rm m,o}(x)}^{\infty} | \Omega_s| d\bar{x} \int_{\chi_{R,\rm M}(\bar{x})}^{\infty} \frac{ \bar F_{s, R}(\mu_{\rm op}(\bar{x}), U_R, 1) \Delta v_{x} (x, \bar{x}, U_R) }{\sqrt{2\left( U_R - \chi_{R, \rm M} (\bar{x}) \right)}} dU_R \rm ,
\end{align}
where 
\begin{align}
& \Delta v_{x} (x, \bar{x}, U) = \nonumber  \\
&  \sqrt{2\left( \chi_{R, \rm M}(\bar{x}) - \chi_R(x, \bar{x}) + \Delta_{\rm M} (\bar{x}, U) \right)} -  \sqrt{2\left( \chi_{R, \rm M}(\bar{x}) - \chi_R(x, \bar{x}) \right)} \rm .
\end{align}
The first term is the contribution from \textit{closed} orbits, while the second term is the much smaller contribution of particles in the open part of their trajectory just before intersecting the wall.
The added generalisation with respect to equations (70) and (86) in reference \cite{Geraldini-2018} is the sum, in the closed-orbit contribution to the density, over particles with $v_z >0$ and $v_z < 0$.
Since $\bar F_{\rm i, mp} = 0$ for $\sigma_{\parallel} = -1$, this sum does not affect the ion density in the magnetic presheath, but it includes the density contribution of electrons reflected within the Debye sheath.\footnote{In the second term in (\ref{n-finorb}), which is the contribution to the density of particles in open orbits, this sum is absent, because the fraction of electrons whose parallel velocity is reflected during their last gyration before reaching the target is small in $\alpha$.
We emphasise that, while the ion density in the magnetic presheath, $n_{\rm i, mp}(x)$, requires the second term in (\ref{n-finorb}) to solve for the electrostatic potential satisfying quasineutrality throughout the region, the electron density in the Debye sheath, $n_{\rm e, ds}(x)$, could be calculated without the second term, as this only provides a correction small in $\alpha$ to the charge density that develops on the Debye scale.
Even though it is not necessary, we include the second term in the electron density because it can be calculated at no additional cost.}
The limits of integration in equation (\ref{n-finorb}) depend also on the minimum allowed value of $\bar{x}$ for a particle passing through $x$ \cite{Geraldini-2018}: in a closed orbit, 
\begin{align} \label{xbarm-general}
\bar{x}_{\text{m}} \left( x \right) = \min_{x' \in [0,x)} \left\lbrace \frac{1}{2} \left( x + x' \right) + \frac{\phi(x) - \phi(x')}{\Omega_s B \left( x-x' \right)}  \right\rbrace \text{;}  
\end{align}
in an open orbit,
\begin{align} \label{xbarm-open}
\bar{x}_{\text{m,o}} (x) = \begin{cases} 
\bar{x}_{\star}  & \text{ for } x \leqslant x_{\star} \text{,} \\
\bar{x}_{\text{m}} (x) &  \text{ for } x >  x_{\star} \text{,}  
\end{cases}
\end{align}
where $\bar{x}_{\star}$ and $x_{\star}$ were defined in and below (\ref{xbarstar}).

By using the fact that only ions in open orbits reach the Debye sheath entrance in the limit $\lambda_{\rm D} / \rho_{\rm S} \rightarrow 0$, and recalling the range of allowed values of $v_x$ in (\ref{vx-range}), the outgoing distribution function of ions reaching the Debye sheath entrance and electrons reaching the target can be obtained,
\begin{align} \label{fopen}
& f_{s, R} (x=0; v_{x}, v_{y}, v_{z}) \simeq F_{\text{s} \infty} \left( \mu_{\text{op}} \left( \bar{x} \right), U_R \right) \Theta \left( \bar{x} - \bar{x}_{\star} \right)   \nonumber  \\
 & \times \hat{\Pi} \left(  v_{x} ,  - \sqrt{2\left( \chi_{R,\text{M}} (\bar{x}) - \chi_R (0, \bar{x}) + \Delta_{\rm M} (\bar{x}, U) \right)}, - \sqrt{2 \left( \chi_{R, \text{M}} (\bar{x}) - \chi_{R} (0, \bar{x}) \right) }  \right) \text{,}
\end{align}
where we have defined the top-hat function $\hat{\Pi}\left(r, l_1, l_2\right)$ as
\begin{align}
\hat{\Pi} \left( r, l_1, l_2 \right) = 
\begin{cases}
1 & \text{ if } l_1 \leqslant r < l_2 \text{,} \\
0 & \text{ otherwise.} 
\end{cases}
\end{align}
To evaluate (\ref{fopen}), we recall from (\ref{xbar-def}) and (\ref{U-def}) that $\bar x = v_y / |\Omega_{\rm  s}|$ and $U_R = (v_x^2 + v_y^2 + v_z^2)/2 + \Omega_{\rm s} \phi_R / B$, and that $\bar x_{\star}$ is given by (\ref{xbarstar}) and $\Delta_{\rm M}$ is given by (\ref{DeltaM-def}).

At $x/\lambda_{\rm D} \rightarrow \infty$, the electron density in the Debye sheath, given by (\ref{n-finorb}), is dominated by the first term, which becomes, upon using (\ref{mu-infty}) and integrating over $\bar{x}$,
\begin{align} \label{ne-mpinfty}
& n_{\rm e,ds}(\infty)  = 2\pi \sum_{\sigma_{\parallel} = \pm 1} \int_{0}^{\infty} |\Omega_{\rm e}| d\mu \int_{|\Omega_{\rm e}|\mu}^{\infty} \frac{\bar F_{\rm e, ds} (\mu, U_{\rm ds}, \sigma_{\parallel}) dU_{\rm ds} }{\sqrt{2\left( U_{\rm ds} - |\Omega_{\rm e}|\mu \right)}}  \rm .
\end{align}
This expression recovers $n_{\rm e, mp}(0)$ in (\ref{ne-mp}) upon using (\ref{Fe-mp}), (\ref{Fe-ds}) and $U_{\rm ds} = U_{\rm mp} - \Omega_{\rm e} \phi_{\rm mp}(0)/ B$.
For ions at $x/\rho_{\rm S} \rightarrow \infty$ in the magnetic presheath, the second term in (\ref{n-finorb}) also vanishes and we obtain, upon substituting (\ref{Fi-infty}),
\begin{align} \label{ni-mpinfty}
& n_{\rm i,mp}(\infty)  = 2\pi \int_{0}^{\infty} \Omega_{\rm i} d\mu \int_{\Omega_{\rm i} \mu}^{\infty} \frac{ F_{\rm i \infty} (\mu, U_{\rm mp}) dU_{\rm mp} }{\sqrt{2\left( U_{\rm mp} - \Omega_{\rm i}\mu \right)}}  \rm .
\end{align}
The ion flux in the $z$ direction is equivalent to the flux in the parallel direction to lowest order in $\alpha$, $J_{\parallel, \rm i} \simeq J_{z, \rm i}$, and is constant across the magnetised sheath.
We calculate it by multiplying the integrand in (\ref{ni-mpinfty}) by $v_{z} = \sqrt{2\left(U_{\rm mp} - \Omega_{\rm i} \mu \right)}$,
\begin{align} \label{Phixi-mp}
\Phi_{\parallel, \rm i} = 2\pi \int_0^{\infty} \Omega_{\rm i} d\mu \int_{\Omega_{\rm i} \mu }^{\infty} F_{\rm i \infty} (\mu, U_{\rm mp}) dU_{\rm mp}  \rm .
\end{align}
Since the ions are strongly tied to the magnetic field line at the entrance of the magnetic presheath, the flux perpendicular to the magnetic field is zero, $\Phi_{\tilde{x}, \rm i} \simeq 0$, and the ion current density in the direction normal to the wall is $J_{x, \rm i} = Ze\left( \Phi_{\tilde{x}, \rm i} \cos \alpha - \Phi_{\parallel, \rm i} \sin \alpha \right) \simeq -Ze \alpha \Phi_{\parallel, \rm i}  $.
Hence, we obtain
\begin{align} \label{Ji-mp}
J_{x, \rm i} = - 2\pi Ze \alpha \int_0^{\infty} \Omega_{\rm i} d\mu \int_{\Omega_{\rm i} \mu }^{\infty} F_{\rm i \infty} (\mu, U_{\rm mp}) dU_{\rm mp}  \rm .
\end{align}

\subsection{Ions in the Debye sheath} \label{subsec-ionsDS}

This section exploits the ordering $\lambda_{\rm D} \ll \rho_{\rm S}$ when the distance of an ion from the target is of the order of a Debye length, $x \sim \lambda_{\rm D}$.
The electric field is so strong and varies on such a short length scale that the electric force on an ion in the $x$ direction is much larger than any component of the magnetic force: from equations (\ref{vx-EOM-exact})-(\ref{vz-EOM-exact}) and $\phi_{\rm ds}' \sim T_{\rm e} / (e\lambda_{\rm D})$, we obtain $|\dot{v}_{x}| \simeq |\Omega_{\rm i}  \phi_{\rm ds}'/ B| \sim  \Omega_{\rm i} c_{\rm S} \rho_{\rm S} / \lambda_{\rm D} \gg  \Omega_{\rm i} c_{\rm S} \sim |\dot{v}_{y}| \sim |\dot{v}_{z}|$.
Thus, the ion motion can be assumed to be one-dimensional in the $x$ direction

The ion distribution function at the Debye sheath entrance is equal to the ion distribution function (\ref{fopen}) coming from the magnetic presheath,
\begin{align} \label{Fi-ds}
f_{\rm i, dse}(\vec{v}) = f_{\rm i, mp}(x=0; v_x, v_y, v_z) \rm .
\end{align}
In the Debye sheath, $v_y$ and $v_z$ are constant. 
Thus, using conservation of $U_{\rm ds} - \frac{1}{2} v_{y}^2 - \frac{1}{2} v_{z}^2 = \frac{1}{2} v_{x}^2 + \Omega_{\rm i} \phi_{\rm ds}(x)/B$, the density is
\begin{align} \label{ni-DS}
n_{\rm i, ds}(x) = \int  \frac{ |v_{x}| f_{\rm i, dse}\left(v_{x}, v_{y}, v_{z} \right) }{  \sqrt{v_{x}^2 - \frac{2\Omega_{\rm i} \phi_{\rm ds}(x)}{B}} } d^3 v \rm ,
\end{align}
where we have used $\int d^3 v = \int_{-\infty}^{\infty} dv_z \int_{-\infty}^{\infty} dv_y \int_{-\infty}^{\infty} dv_x$.
The ion distribution function at the target is, from (\ref{fopen}) and from conservation of $\frac{1}{2} v_{x}^2 + \Omega_{\rm i} \phi_{\rm ds}(x)/B$ for ions in the Debye sheath,
\begin{align} \label{f-target}
f_{\rm i, tar} (v_{x}, v_{y}, v_{z}) = & F_{\text{\rm i} \infty} \left( \mu_{\text{op}} \left( \bar{x} \right), U_{\rm mp} \right) \Theta \left( \bar{x} - \bar{x}_{\star} \right)   \nonumber  \\
& \times \hat{\Pi} \left(  v_{x} ,  - \sqrt{2\left( \chi_{R,\text{M}} (\bar{x}) - \chi_R (0, \bar{x}) + \Delta_{\rm M} (\bar{x}, U_{\rm mp}) - \frac{2\Omega_{\rm i} \phi_{\rm ds}(0)}{B} \right)}, \right. \nonumber \\
& \left. - \sqrt{2 \left( \chi_{R, \text{M}} (\bar{x}) - \chi_{R} (0, \bar{x}) - \frac{2\Omega_{\rm i} \phi_{\rm ds}(0)}{B} \right) }  \right) \rm .
\end{align}
Note that $\Omega_{\rm i} \phi_{\rm ds}(0) < 0$, consistent with the ions having accelerated towards the target.

\section{Analytical predictions}  \label{sec-analytical} 

In this work, the magnetic presheath and Debye sheath are solved as separate regions in the asymptotic limit $\epsilon_{\rm ms} = \lambda_{\rm D} / \rho_{\rm i} \rightarrow 0$.
The fundamental equation determining the electrostatic potential in both regions is Poisson's equation.
However, it is only necessary to use Poisson's equation to solve for the electrostatic potential in the Debye sheath,
\begin{align} \label{Poisson}
 \varepsilon_0  \phi_{\rm ds}'' (x)  = e\left( n_{\rm e,ds} (x) - Zn_{\rm i,ds} (x) \right) \rm .
\end{align}
In the magnetic presheath, the charge separation is negligible since $ \varepsilon_0  \phi_{\rm mp}'' (x) \sim  \varepsilon_0  T_{\rm e} / (e\rho_{\rm S}^2) \sim (\lambda_{\rm D}^2/ \rho_{\rm S}^2) e n_{\rm e,mp} (x) $ and $\lambda_{\rm D} / \rho_{\rm i} \rightarrow 0$, so the electrostatic potential is constrained to satisfy the quasineutrality relation,
\begin{align} \label{quasi}
n_{\rm e, mp} (x) = Zn_{\rm i, mp} (x) \rm .
\end{align}

Poisson's equation (\ref{Poisson}) is solved with $\phi_{\rm ds}(\infty) = 0$ and with the following boundary condition at the wall,
\begin{align} \label{phi(0)}
\phi_{\rm ds} (0) = \phi_{\rm w} - \phi_{\rm mp}(0) \rm .
\end{align}
The wall potential relative to the magnetic presheath entrance, $\phi_{\rm w}$, can be fixed directly or can be found by imposing an ambipolar flow (or by fixing any value of current to the wall).
A numerical scheme is necessary to solve equations (\ref{Poisson}) and (\ref{quasi}).
It will nonetheless be useful to derive some analytical results based on the assumed monotonicity of the profile of the electron-repelling electrostatic potential in the magnetised sheath. 

This section is structured as follows. 
In section~\ref{subsec-analytical-Chodura} we analyse the quasineutrality equation far from the wall in the magnetic presheath, thus recovering the kinetic Chodura condition that must be satisfied at the magnetic presheath entrance and the analytical form of the electrostatic potential decay at large $x$ \cite{Geraldini-2018}.
Then, in section~\ref{subsec-analytical-mpnear} we summarise the results of the analysis of the quasineutrality equation close to the wall in the magnetic presheath \cite{Geraldini-2018, Ewart-2021}, which are that the kinetic Bohm condition will be marginally satisfied and that the electric field must correspondingly diverge at $x=0$.
We additionally characterise the weakening of the electric field divergence as the potential drop $|\phi_{\rm ds}(0)|$ across the Debye sheath becomes small, which typically corresponds to the magnetic field angle $\alpha$ being made small.
In section~\ref{subsec-analytical-dsfar} we analyse Poisson's equation far from the wall in the Debye sheath and derive the analytical form of the electrostatic potential decay.
Finally, in section~\ref{subsec-analytical-critical} we argue that there exists a critical angle below which no monotonic solution of the magnetised sheath is possible, and estimate the critical angle and its dependence on the characteristic size of finite electron Larmor radii relative to the Debye length.

\subsection{Electrostatic potential decay at large $x$ in the magnetic presheath} \label{subsec-analytical-Chodura}

The expansion of the quasineutrality equation (\ref{quasi}) at $x/\rho_{\rm S} \gg 1$, where $\hat \phi_{\rm mp} = e|\phi_{\rm mp}| / T_{\rm e} \ll 1$ and $\kappa_{\rm mp} = \rho_{\rm i} \left| \phi'_{\rm mp} /\phi_{\rm mp} \right| \ll 1$, with the electron density given by (\ref{ne-mp}), is identical to the one in reference \cite{Ewart-2021}.
A brief review is given here.
The ion density close to the magnetic presheath entrance is, from \ref{app-densfinorb},
\begin{align} \label{niclosed-far}
n_{\rm i,\text{mp}} \left(x \right) \simeq & ~  2\pi  \int_{ 0 }^{\infty} \Omega_{\rm i} d\mu \left\lbrace  \int_{\Omega_{\rm i} \mu}^{\infty} \frac{ dU_{\rm mp}  }{\sqrt{2\left( U_{\rm mp} - \Omega_{\rm i} \mu \right)}}    \left[  F_{\rm i \infty} \left( \mu, U_{\rm mp} \right)  \phantom{\left( \frac{1}{1}\right)^2} \right. \right. \nonumber \\
&  \left. - \frac{\phi_{\rm mp}''(x) }{\Omega_{\rm i} B} \mu \partial_{\mu}  F_{\rm i \infty} \left( \mu, U_{\rm mp} \right) + \frac{\Omega_{\rm i} \phi_{\rm mp} (x)}{B} \partial_{U_{\rm mp}}  F_{\rm i \infty} \left( \mu, U_{\rm mp} \right) \right.  \nonumber \\
&  \left.  + \frac{1}{2} \left( \frac{\Omega_{\rm i} \phi_{\rm mp} (x)}{B} \right)^2 \partial_{U_{\rm mp}}^2  F_{\rm i \infty} \left( \mu, U \right)  \right] \nonumber \\
 & -  F_{\rm i \infty} \left( \mu, \Omega_{\rm i} \mu \right)  \sqrt{-\frac{ 2\Omega_{\rm i} \phi_{\rm mp} (x)  }{B}} \left[ 1 + O \left( \kappa_{\rm mp}^2 \right) \right]  \nonumber  \\
& \left. + \frac{1}{3} \left( - \frac{ 2 \Omega_{\rm i} \phi_{\rm mp} (x)  }{B} \right)^{3/2} \partial_{U_{\rm mp}} F_{\rm i \infty} \left( \mu, \Omega_{\rm i} \mu \right) \right\rbrace  \rm .
\end{align}
The electron density near the magnetic presheath entrance is
\begin{align} \label{nemp-far}
n_{\rm e, mp}(x) \simeq n_{\rm e, mp}(\infty) + \left. \frac{dn_{\rm e, mp}}{d\phi_{\rm mp}} \right\rvert_{x\rightarrow\infty} \phi_{\rm mp}(x) +  \left. \frac{d^2n_{\rm e, mp}}{d\phi_{\rm mp}^2} \right\rvert_{x\rightarrow\infty} \left[ \phi_{\rm mp}(x) \right]^2 \rm .
\end{align}
Inserting (\ref{niclosed-far}) and (\ref{nemp-far}) into (\ref{quasi}) and invoking $n_{\rm e, mp}(\infty) = Zn_{\rm i, mp}(\infty)$ gives, to lowest order in $\hat \phi_{\rm mp} \ll 1$ and $\kappa_{\rm mp} \ll 1$,
\begin{align} \label{quasi-phi1/2}
 \sqrt{-\frac{ 2\Omega_{\rm i} \phi_{\rm mp} (x)  }{B}}  2\pi  \int_{ 0 }^{\infty} \Omega_{\rm i} d\mu  \int_{\Omega_{\rm i} \mu}^{\infty} \frac{ dU_{\rm mp}  }{\sqrt{2\left( U_{\rm mp} - \Omega_{\rm i} \mu \right)}} F_{\rm i \infty} \left( \mu, \Omega_{\rm i} \mu  \right) = 0 \rm .
\end{align}
From (\ref{quasi-phi1/2}) and positivity of the distribution function, it follows that the ion distribution function must satisfy $F_{\rm i \infty} \left( \mu, \Omega_{\rm i} \mu  \right) = 0$, that is, that there must be no ions with zero parallel velocity at the magnetic presheath entrance.
To next order, we obtain an equation of the form $\phi_{\rm mp}''(x)/\phi_{\rm mp}(x) = $ constant.
Imposing that this constant be positive or zero for a monotonic potential profile results in the kinetic Chodura condition
\begin{align} \label{kinetic-Chodura}
 2\pi Z v_{\rm B}^2  \int_{ 0 }^{\infty} \Omega_{\rm i} d\mu  \int_{\Omega_{\rm i}\mu_{\rm i}}^{\infty} dU_{\rm mp} \frac{   \partial_{U_{\rm mp}}  F_{\rm i \infty} \left( \mu, U_{\rm mp} \right)  }{\sqrt{2\left( U_{\rm mp} - \Omega_{\rm i} \mu \right)}} \leqslant \frac{ T_{\rm e} }{e} \left. \frac{dn_{\rm e, mp}}{d\phi_{\rm mp}} \right\rvert_{x\rightarrow\infty} \rm .
\end{align}
Assuming that (\ref{kinetic-Chodura}) is marginally satisfied, that is, satisfied with the equality sign, and further assuming that $\partial_{U_{\rm mp}}  F_{\rm i \infty} \left( \mu, \Omega_{\rm i} \mu \right) \neq 0$, the next-order term that must be retained in (\ref{niclosed-far}) is that containing $[-\phi_{\rm mp}(x)]^{3/2}$.
Quasineutrality can then be rearranged to the form $-\phi''_{\rm mp}(x) / [-\phi_{\rm mp}(x)]^{3/2} = $ positive constant.
Hence, the electrostatic potential $\phi_{\rm mp}(x)$ for $x/\rho_{\rm S} \gg 1$ takes the form
\begin{align} \label{phi-mpe}
\frac{e\phi_{\rm mp}(x)}{T_{\rm e}}  = - \frac{a_{\rm mp} \rho_{\rm i}^4}{(x+c_{\rm mp})^4} \rm ,
\end{align}
where $a_{\rm mp}$ and $c_{\rm mp}$ are constants.
The constant $a_{\rm mp}$ is determined by integrals over the ion and electron distribution functions at the entrance of the magnetic presheath \cite{Geraldini-2018}, 
while $c_{\rm mp}$ is determined by the overall solution for $\phi_{\rm mp}$.

\subsection{Electric field divergence at $x=0$ in the magnetic presheath} \label{subsec-analytical-mpnear}

The expansion of equation (\ref{quasi}) at $x/\rho_{\rm S} \ll 1$ with the electron density given by (\ref{ne-mp}) is analogous to the one in references \cite{Geraldini-2018, Ewart-2021, Geraldini-2021} .
The marginal form of the Bohm condition emerges,
\begin{align} \label{Bohm-marginal}
Z\int d^3 v f_{\rm i,dse}(\vec{v}) \frac{v_{\rm B}^2}{v_x^2} = \frac{T_{\rm e}}{e} \left. \frac{dn_{\rm e, mp}}{d\phi_{\rm mp}} \right\rvert_{x=0} \rm ,
\end{align}
and the self-consistent form of the electrostatic potential variation near $x=0$ is
\begin{align} \label{phi-dse-mp}
\left(\frac{ e\left( \phi_{\rm mp}(x) - \phi_{\rm mp}(0) \right) }{T_{\rm e}} \right)^2  = p \frac{x}{\rho_{\rm B}} \rm ,
\end{align}
where $\rho_{\rm B} = v_{\rm B} / \Omega_{\rm i}$ is the Bohm gyroradius, and
\begin{align} \label{p-full}
p = \frac{2Z\int d^3 v f_{\rm i,dse}(\vec{v}) \frac{v_y v_{\rm B}}{v_x^2} }{3Z\int d^3 v f_{\rm i,dse}(\vec{v})  \frac{v_{\rm B}^4}{v_x^4} - \frac{T_{\rm e}^2}{e^2} \left.  \frac{d^2n_{\rm e,mp}}{d\phi_{\rm mp}^2} \right\rvert_{x=0} } \rm .
\end{align}
For a direct comparison, equations (\ref{phi-dse-mp}) and (\ref{p-full}) are in the same form as (6.10) and (6.11) in \cite{Geraldini-2021} (although the factors of $Z$ are incorrectly missing in equation (6.11) of \cite{Geraldini-2021}, but are present in equation (142) of \cite{Geraldini-2018} and equation (82) of \cite{Ewart-2021}).

We proceed to prove that $p > 0$ if the electron distribution function is Maxwellian.
This proof was carried out for $\gamma = 0$ in Appendix~B of reference \cite{Ewart-2021}; here, we show explicitly that the proof holds also when $\gamma \neq 0$.
The numerator of $p$ is manifestly positive because $v_y = \Omega_{\rm i} \bar x$ (from (\ref{xbar-def}) with $x=0$) and $\bar x \geqslant \bar{x}_{\rm m, o}(0) = \bar x_{\star} > 0$ (see (\ref{xbarstar}) and (\ref{xbarm-open}), and recall that $\phi_{\rm mp}'(x) \geqslant 0$).
Taking the first and second derivatives of the electron density (\ref{ne-mpre}) in the magnetic presheath and evaluating them at the Debye sheath entrance gives
\begin{align} \label{dne-mpre}
& \left. \frac{T_{\rm e}}{e} \frac{dn_{\rm e, mp}}{d\phi_{\rm mp}} \right\rvert_{x=0} = 2\pi v_{\rm t, e}^2 \int_0^{\infty} |\Omega_{\rm e}| d\tilde{\mu}  \left[  \frac{ F_{\rm e \infty} \left( \tilde{\mu}, U_{\rm cut, ds} + \frac{\Omega_{\rm e} \phi_{\rm mp}(0)}{B} \right) }{v_{\rm cut, dse}(\tilde \mu) }  \right. \nonumber \\
& \left. 
- \int_{-v_{\rm cut, dse}(\tilde \mu)}^{\infty} \partial_{U_{\rm mp}}F_{\rm e \infty} \left( \tilde{\mu}, |\Omega_{\rm e}| \tilde{\mu} + \frac{1}{2} v_{\parallel}^2 + \frac{\Omega_{\rm e} \phi_{\rm mp}(0)}{B} \right)  dv_{\parallel} \right] \rm ,
\end{align}
\begin{align} \label{d2ne-mpre}
& \left. \frac{T_{\rm e}^2}{e^2} \frac{d^2n_{\rm e, mp}}{d\phi_{\rm mp}^2} \right\rvert_{x=0} = 2\pi v_{\rm t, e}^4 \int_0^{\infty} |\Omega_{\rm e}| d\tilde{\mu} \nonumber \\
& \times \left[  - \frac{ F_{\rm e \infty} \left( \tilde{\mu}, U_{\rm cut, ds} + \frac{\Omega_{\rm e} \phi_{\rm mp}(0)}{B} \right) }{\left(v_{\rm cut, dse}(\tilde \mu)\right)^3}  
- \frac{ \partial_{U_{\rm mp}} F_{\rm e \infty} \left( \tilde{\mu}, U_{\rm cut, ds} + \frac{\Omega_{\rm e} \phi_{\rm mp}(0)}{B} \right) }{v_{\rm cut, dse}(\tilde \mu)}  \right. \nonumber \\
& \left. + \int_{-v_{\rm cut, dse}(\tilde \mu)}^{\infty} \partial_{U_{\rm mp}}^2 F_{\rm e \infty} \left( \tilde{\mu}, |\Omega_{\rm e}| \tilde{\mu} + \frac{1}{2} v_{\parallel}^2 + \frac{\Omega_{\rm e} \phi_{\rm mp}(0)}{B} \right)  dv_{\parallel} \right]
 \rm ,
\end{align}
where we recall from (\ref{v-cut-dse}) that $v_{\rm cut, dse}(\tilde \mu) = \sqrt{2\left( U_{\rm cut, ds}(\tilde \mu) - |\Omega_{\rm e}| \tilde \mu  \right)}$.
Specialising to a Maxwellian electron distribution function, which satisfies $v_{\rm t, e}^4 \partial_{U_{\rm mp}}^2 F_{\rm e, \infty} = - v_{\rm t, e}^2 \partial_{U_{\rm mp}} F_{\rm e, \infty} = F_{\rm e, \infty}$, equations (\ref{dne-mpre}) and (\ref{d2ne-mpre}) become
\begin{align} \label{dneMax}
\left. \frac{T_{\rm e}}{e} \frac{dn_{\rm e, mp}}{d\phi_{\rm mp}} \right\rvert_{x=0} & = n_{\rm e, mp}(0)  + 2\pi v_{\rm t, e}^2 \int_0^{\infty} |\Omega_{\rm e}| d\tilde{\mu}  \frac{ F_{\rm e \infty} \left( \tilde{\mu}, U_{\rm cut, ds} + \frac{\Omega_{\rm e} \phi_{\rm mp}(0)}{B} \right) }{v_{\rm cut, dse}(\tilde \mu)} \nonumber \\
& >  n_{\rm e, mp}(0) \rm ,
\end{align}
\begin{align}  \label{d2neMax}
\left. \frac{T_{\rm e}^2}{e^2} \frac{d^2n_{\rm e, mp}}{d\phi_{\rm mp}^2} \right\rvert_{x=0} & = \left. \frac{T_{\rm e}}{e} \frac{dn_{\rm e, mp}}{d\phi_{\rm mp}} \right\rvert_{x=0}  - 2\pi v_{\rm t, e}^4 \int_0^{\infty} |\Omega_{\rm e}| d\tilde{\mu}  \frac{F_{\rm e \infty} \left( \tilde{\mu}, U_{\rm cut, ds} + \frac{\Omega_{\rm e} \phi_{\rm mp}(0)}{B} \right) }{\left(v_{\rm cut, dse}(\tilde \mu)\right)^3} \nonumber \\
& <  \left. \frac{T_{\rm e}}{e} \frac{dn_{\rm e, mp}}{d\phi_{\rm mp}} \right\rvert_{x=0} \rm .
\end{align}
By rearranging the relation
\begin{align}
\left( \int d^3 v f_{\rm i,dse}(\vec{v}) \right) \left( \int d^3 v f_{\rm i,dse}(\vec{v}) \frac{v_{\rm B}^4}{v_x^4} \right) \geqslant \left( \int d^3 v f_{\rm i,dse}(\vec{v}) \frac{v_{\rm B}^2}{v_x^2} \right)^2 \rm,
\end{align}
deduced from Schwarz's inequality, we obtain, upon invoking (\ref{Bohm-marginal}),
\begin{align} \label{Schwarz-re}
Z\int d^3 v f_{\rm i,dse}(\vec{v})  \frac{v_{\rm B}^4}{v_x^4} \geqslant & \frac{1}{n_{\rm e, mp}(0)} \left[ \frac{T_{\rm e}}{e} \left. \frac{dn_{\rm e, mp}}{d\phi_{\rm mp}} \right\rvert_{x=0} \right]^2 \rm .
\end{align}
By combining the inequalities in (\ref{dneMax}), (\ref{d2neMax}) and (\ref{Schwarz-re}), we constrain the denominator of $p$ in (\ref{p-full}) to be positive,
\begin{align}
3Z\int d^3 v f_{\rm i,dse}(\vec{v})  \frac{v_{\rm B}^4}{v_x^4} - \frac{T_{\rm e}^2}{e^2} \left.  \frac{d^2n_{\rm e,mp}}{d\phi_{\rm mp}^2} \right\rvert_{x=0} > 2 \frac{T_{\rm e}}{e} \left. \frac{dn_{\rm e, mp}}{d\phi_{\rm mp}} \right\rvert_{x=0}  > 2n_{\rm e, mp}(0) \rm .
\end{align}
We can thus conclude that $p>0$ for a Maxwellian electron distribution function, independently of the value of $\gamma$.
Hence, the marginal form of the kinetic Bohm condition (\ref{Bohm-marginal}) at the Debye sheath entrance is a self-consistent result of the magnetic presheath solution, provided that the Debye sheath potential satisfies $\phi_{\rm ds} (x) < 0$ (no ions are reflected).

The size of the numerator of (\ref{p-full}) can be calculated from the marginal kinetic Bohm condition (\ref{Bohm-marginal}) together with the scaling $v_y \sim c_s \sim v_{\rm t,i} \sim v_{\rm B}$ ($T_{\rm i} \sim T_{\rm e}$),
\begin{align}
2Z\int d^3 v f_{\rm i,dse}(\vec{v}) \frac{v_y v_{\rm B}}{v_x^2} \sim \frac{T_{\rm e}}{e} \left. \frac{dn_{\rm e,mp}}{d\phi_{\rm mp}} \right\rvert_{x=0} \rm .
\end{align}
Using that the size of the denominator scales with the size of its second term, the scaling for $p$ is
\begin{align} \label{p-app}
p \sim  \frac{e}{T_{\rm e}} \frac{ \left. dn_{\rm e,mp} / d\phi_{\rm mp} \right\rvert_{x=0} }{ \left. d^2n_{\rm e,mp} / d\phi_{\rm mp}^2 \right\rvert_{x=0} } \rm .
\end{align}
The derivatives in (\ref{p-app}) are order unity unless the Debye sheath potential is small, $e\phi_{\rm ds}(0)/T_{\rm e} \ll 1$, in which case the cutoff velocity 
is small via the ordering $v_{\rm cut, dse} (\tilde \mu)  \sim \sqrt{ 2\Omega_{\rm e} |\phi_{\rm ds}(0)|/ B}$ which was justified after equation (\ref{vcut-dse-smallgamma}).
Hence, for $v_{\rm cut, dse}/v_{\rm t,e} \sim \sqrt{e|\phi_{\rm ds}(0)|/T_{\rm e}} \gtrsim 1$ we have the orderings
\begin{align} \label{dnedphi-ord}
\left. \frac{T_{\rm e}^2}{e^2} \frac{d^2n_{\rm e, mp}}{d\phi_{\rm mp}^2} \right\rvert_{x=0}  \sim \left. \frac{T_{\rm e}}{e} \frac{dn_{\rm e, mp}}{d\phi_{\rm mp}} \right\rvert_{x=0} \sim  n_{\rm e, mp}(0) \rm ,
\end{align}
while for $v_{\rm cut, dse}/v_{\rm t,e} \sim \sqrt{e\phi_{\rm ds}(0)/T_{\rm e}} \ll 1$ we have the orderings
\begin{align} \label{dnedphi-ord-smallphi}
\left. \left( \frac{e|\phi_{\rm ds}(0)|}{T_{\rm e}} \right)^{3/2} \frac{T_{\rm e}^2}{e^2} \frac{d^2n_{\rm e, mp}}{d\phi_{\rm mp}^2} \right\rvert_{x=0}  \sim \left. \left( \frac{e|\phi_{\rm ds}(0)|}{T_{\rm e}} \right)^{1/2} \frac{T_{\rm e}}{e} \frac{dn_{\rm e, mp}}{d\phi_{\rm mp}} \right\rvert_{x=0} \sim  n_{\rm e, mp}(0) \rm .
\end{align}
These orderings can be inserted into (\ref{p-app}) to obtain
\begin{align} \label{p-scaling}
p \sim \begin{cases} 1 & \text{ for }  \frac{e|\phi_{\rm ds}(0)|}{T_{\rm e}} \gtrsim 1 \text{,} \\
 \frac{e|\phi_{\rm ds}(0)|}{T_{\rm e}} & \text{ for } \frac{e|\phi_{\rm ds}(0)|}{T_{\rm e}} \ll 1 \text{.}
\end{cases} 
\end{align}
The electric field divergence at $x=0$, expressed by (\ref{phi-dse-mp}), is thus weak ($p$ is small) for small values of the Debye sheath potential drop.
The orderings (\ref{dnedphi-ord}) and (\ref{dnedphi-ord-smallphi}) can also be applied to the right hand side of (\ref{Bohm-marginal}) to obtain an ordering for the Bohm integral at small values of potential drop across the Debye sheath,
\begin{align} \label{Bohmint-ordered}
Z\int d^3 v f_{\rm i,dse}(\vec{v}) \frac{v_{\rm B}^2}{v_x^2} \sim \begin{cases} n_{\rm e, ds}(\infty)  & \text{ for } \frac{e|\phi_{\rm ds}(0)|}{T_{\rm e}} \gtrsim 1 \rm , \\
n_{\rm e, ds}(\infty) \left( \frac{T_{\rm e}}{e|\phi_{\rm ds}(0)|} \right)^{1/2} & \text{ for } \frac{e|\phi_{\rm ds}(0)|}{T_{\rm e}} \ll 1 \rm .
\end{cases}
\end{align}
Hence, as $\phi_{\rm ds}(0)$ decreases to zero, the weakening of the electric field divergence (\ref{phi-dse-mp}) at the Debye sheath entrance, characterised by (\ref{p-scaling}), is interconnected with a corresponding weakening of the constraint posed by the kinetic Bohm criterion (\ref{Bohm-marginal}): from (\ref{Bohmint-ordered}), more slow ions with small values of $|v_x|$ can be present at the Debye sheath entrance.

Equations (\ref{phi-dse-mp}) and (\ref{p-scaling}) will serve to discretize the $x$-coordinate in section~\ref{sec-num}.

\subsection{Electrostatic potential decay at large $x$ in the Debye sheath entrance} \label{subsec-analytical-dsfar}

The ion density (\ref{ni-DS}) in the Debye sheath expanded far from the wall (for $e|\phi_{\rm ds}| /T_{\rm e} \ll 1$) gives
\begin{align} \label{nids-infty}
n_{\rm i, ds} (x) \simeq \int d^3 v f_{\rm i,dse}(\vec{v}) +  \frac{2\Omega_{\rm i} \phi_{\rm ds}(x) }{B}  \int d^3 v \frac{f_{\rm i,dse}(\vec{v})}{v_x^2 } 
+  3 \left(  \frac{\Omega_{\rm i} \phi_{\rm ds}(x) }{B} \right)^2  \int d^3 v \frac{f_{\rm i,dse}(\vec{v}) }{v_x^4 }  \rm .
\end{align}
The expansion can be carried out by Taylor expanding in $\phi_{\rm ds}$ because $f_{\rm i, dse} (\vec{v})$ tends to zero exponentially for $v_x \rightarrow 0$ if the electric field diverges at $x=0$ in the magnetic presheath \cite{Geraldini-2021}, which in section~\ref{subsec-analytical-mpnear} was shown to be the case provided that the potential drop across the Debye sheath remains finite, $\phi_{\rm ds}(0) \neq 0$.
The expansion of the electron density at $x/\lambda_{\rm D} \gg 1$ is carried out in \ref{app-densfinorb} and leads to 
\begin{align} \label{neclosed-dse-ds-1}
n_{\text{e,ds}} \left(x \right) =  & \sum_{\sigma_{\parallel} = \pm 1} 2\pi  \int_{ 0 }^{\infty} | \Omega_{\rm e}| d\mu \int_{|\Omega_{\rm e}| \mu}^{\infty} \frac{ dU_{\rm ds}  }{\sqrt{2\left( U_{\rm ds} - |\Omega_{s}| \mu\right)}}    \left[  \bar{F}_{\rm e, ds} \left( \mu, U_{\rm ds}, \sigma_{\parallel} \right) 
\right. \nonumber \\
&  \left. - \frac{\phi_{\rm ds}''(x) \mu }{\Omega_{\rm e} B} \partial_{\mu}  \bar{F}_{\rm e, ds} \left( \mu, U_{\rm ds}, \sigma_{\parallel}  \right)  
  + \frac{\Omega_{\rm e} \phi_{\rm ds} (x)}{B} \partial_{U_{\rm ds}}  \bar{F}_{\rm e, ds} \left( \mu, U_{\rm ds}, \sigma_{\parallel}  \right) \right. \nonumber \\
&  \left.  + \frac{1}{2} \left( \frac{\Omega_{\rm e} \phi_{\rm ds} (x)}{B} \right)^2 \partial_{U_{\rm ds}}^2  \bar{F}_{\rm e, ds} \left( \mu, U_{\rm ds}, \sigma_{\parallel}  \right)  \right]
 \text{.}
\end{align}
We can identify the third and fourth terms as being equivalent to the ones occurring in the expansion of $n_{\rm e,mp}(x)$ near the Debye sheath entrance,
\begin{align} \label{neclosed-dse-ds}
n_{\text{e,ds}} \left(x \right) =  n_{\rm e, ds}(\infty) - \sum_{\sigma_{\parallel} = \pm 1} 2\pi  \int_{ 0 }^{\infty} | \Omega_{\rm e}| d\mu  \int_{|\Omega_{\rm e}| \mu_{\rm e}}^{\infty} \frac{ \partial_{\mu} \bar{F}_{\rm e, ds} \left( \mu, U_{\rm ds}, \sigma_{\parallel} \right)   dU  }{\sqrt{2\left( U_{\rm ds} - |\Omega_{e}| \mu \right)}}   \frac{\phi_{\rm ds}''(x) \mu }{\Omega_{\rm e} B}   \nonumber \\
+  \phi_{\rm ds} (x) \left. \frac{dn_{\rm e,mp}}{d\phi_{\rm mp}} \right\rvert_{x=0} +  \frac{1}{2}  \phi_{\rm ds}^2(x) \left. \frac{d^2n_{\rm e,mp}}{d\phi_{\rm mp}^2} \right\rvert_{x=0} 
 \text{.}
\end{align}
The second term in (\ref{neclosed-dse-ds-1}) and (\ref{neclosed-dse-ds}) is different, and comes from the polarization drift of the electrons, which becomes significant as the electron gyro-orbits distort to non-circular in the Debye sheath.

At $x/\lambda_{\rm D} \rightarrow \infty$, where $\phi_{\rm ds}'' = 0$, Poisson's equation (\ref{Poisson}) simplifies to quasineutrality at the Debye sheath entrance, 
\begin{align}
n_{\rm e, \rm ds}(\infty) = Zn_{\rm i, ds}(\infty) \rm .
\end{align}
For $e \phi_{\rm ds}(x) / T_{\rm e} \ll 1$, inserting (\ref{nids-infty}) and (\ref{neclosed-dse-ds}) into (\ref{Poisson}) gives
\begin{align} \label{Poisson-1}
\lambda_{\rm D}^2 \frac{\phi_{\rm ds}''(x)}{ \phi_{\rm ds}(x) }  = 
\frac{ \frac{T_{\rm e}}{e} \left. \frac{dn_{\rm e,mp}}{d\phi_{\rm mp}} \right\rvert_{x=0} -   Z\int d^3 v \frac{v_{\rm B}^2 f_{\rm i,dse}(\vec{v})}{v_x^2 }  }{ n_{\rm e, ds}(\infty) - \gamma^2 \sum_{\sigma_{\parallel} = \pm 1} 2\pi  \int_{ 0 }^{\infty} | \Omega_{\rm e}| d\mu  \int_{|\Omega_{\rm e}| \mu}^{\infty} \frac{ \mu \partial_{\mu}  \bar{F}_{\rm e, ds} \left( \mu, U_{\rm ds}, \sigma_{\parallel}  \right)  }{\sqrt{2\left( U_{\rm ds} - |\Omega_{\rm e}| \mu \right)}} dU_{\rm ds}   } 
 \text{,}
\end{align}
where we have defined the Debye length calculated at the Debye sheath entrance,
\begin{align} \label{lambdaD}
\lambda_{\rm D} = \sqrt{\frac{\varepsilon_0 T_{\rm e}}{e^2 n_{\rm e, ds}(\infty)}} \rm ,
\end{align}
effectively setting $n_{\rm e, ref} = n_{\rm e, ds}(\infty)$ in the second paragraph of section~\ref{sec-intro} and in the definition of $\gamma$ in (\ref{gamma-def}).
The right hand side of (\ref{Poisson-1}) must be greater than or equal to zero in order to have a monotonic decay of the potential near the Debye sheath entrance.
The sign of the denominator of the right hand side of (\ref{Poisson-1}) depends on whether the second term can reverse the positive first term.
We therefore proceed to estimate the size of the second term in the denominator  assuming a Maxwellian electron distribution function, for which $\partial_{\mu} F_{\rm e \infty} = 0$.
From (\ref{Fe-ds}) we have 
\begin{align}
\partial_{\mu} \bar F_{\rm e, ds} = \begin{cases} 0 & \text{ for } \sigma_{\parallel} = +1 \text{,} \\
U_{\rm cut, ds}'(\mu) F_{\rm e\infty} \left( \mu, U_{\rm ds} +  \frac{\Omega_{\rm e} \phi_{\rm mp}(0) }{ B} \right) \delta_{\rm Dirac}(U_{\rm cut, ds}(\mu) - U_{\rm ds}) & \text{ for } \sigma_{\parallel} = -1 \rm ,
\end{cases}
\end{align}
which implies
\begin{align}
\sum_{\sigma_{\parallel} = \pm 1} \int_{|\Omega_{\rm e}| \mu}^{\infty} \frac{ \mu \partial_{\mu}  \bar{F}_{\rm e, ds} \left( \mu, U_{\rm ds}, \sigma_{\parallel}  \right)  }{\sqrt{2\left( U_{\rm ds} - |\Omega_{\rm e}| \mu \right)}} dU_{\rm ds} =  \frac{ \mu U_{\rm cut, ds}'(\mu)}{v_{\rm cut, dse}(\mu) } F_{\rm e\infty} \left( \mu, U_{\rm cut, ds}(\mu)  +  \frac{\Omega_{\rm e} \phi_{\rm mp}(0) }{ B} \right) \rm . 
\end{align}
Estimating the size of this term using $v_{\rm cut, dse}(\mu) \sim v_{\rm cut, dse, 0}(\mu) = \sqrt{ 2\Omega_{\rm e} \phi_{\rm ds}(0) / B }$, $U_{\rm cut, ds}(\mu) \sim U_{\rm cut, ds, 0}(\mu) = |\Omega_{\rm e}| \mu + \Omega_{\rm e} \phi_{\rm ds}(0) / B$ and 
\begin{align}
F_{\rm e \infty} \left( \mu, U_{\rm cut, ds}(\mu) + \frac{\Omega_{\rm e} \phi_{\rm mp}(0)}{B} \right) & \sim \frac{n_{\rm e, mp}(\infty)}{v_{\rm t, e}^3} \exp\left( - \frac{U_{\rm cut, ds}(\mu)}{v_{\rm t,e}^2} + \frac{e \phi_{\rm mp}(0) }{ T_{\rm e}} \right) \nonumber \\
& \sim \frac{n_{\rm e, ds}(\infty)}{v_{\rm t, e}^3} \exp\left( -\frac{U_{\rm cut, ds}(\mu)}{v_{\rm t, e}^2} \right)  \nonumber
\end{align}
gives
\begin{align}
\gamma^2 \pi \sum_{\sigma_{\parallel} = \pm 1}  \int_{ 0 }^{\infty} | \Omega_{\rm e}| d\mu  \int_{|\Omega_{\rm e}| \mu}^{\infty} \frac{ \mu \partial_{\mu}  \bar{F}_{\rm e, ds} \left( \mu, U_{\rm ds}, \sigma_{\parallel}  \right)  }{\sqrt{2\left( U_{\rm ds} - |\Omega_{\rm e}| \mu \right)}} dU_{\rm ds}  \sim \frac{\exp \left( e \phi_{\rm ds}(0)/ T_{\rm e} \right) }{\sqrt{e |\phi_{\rm ds}(0)| / T_{\rm e} }} \gamma^2 n_{\rm e, ds}(\infty) \rm .
\end{align}
Hence, there are two cases in which the denominator on the right hand side of (\ref{Poisson-1}) is guaranteed to be positive: for $\gamma \sim 1 \ll e|\phi_{\rm ds}(0)|/T_{\rm e}$ and for $\gamma \ll \left( e|\phi_{\rm ds}(0)|/T_{\rm e} \right)^{1/4} \lesssim 1$.
Assuming then that the denominator is positive, the kinetic Bohm condition is recovered by imposing that the numerator of the right hand side of (\ref{Poisson-1}) be positive,
\begin{align} \label{kinetic-Bohm}
Z \int d^3 v \frac{v_{\rm B}^2 f_{\rm i,dse}(\vec{v})}{v_x^2 }    \leqslant   \left.  \frac{T_{\rm e}}{e} \frac{dn_{\rm e,mp}}{d\phi_{\rm mp}} \right\rvert_{x=0} \rm . 
\end{align}

From the separate analysis of the potential near the wall at the magnetic presheath scale, we know that the Bohm condition is self-consistently marginally satisfied, as in (\ref{Bohm-marginal}).
To derive the electrostatic potential decay far away from the target on the Debye sheath scale, we must analyse Poisson's equation to higher order in $e\phi_{\rm ds}(x)/T_{\rm e} \ll 1$, to obtain
\begin{align} \label{Poisson-higher-order}
\lambda_{\rm D}^2 \frac{e \phi_{\rm ds}''(x)}{T_{\rm e}}  \left[ n_{\rm ds}(\infty) - \frac{1}{2} \gamma^2 \sum_{\sigma_{\parallel} = \pm 1} \pi  \int_{ 0 }^{\infty} | \Omega_{\rm e}| d\mu \int_{|\Omega_{\rm e}| \mu}^{\infty} \frac{ \mu \partial_{\mu}  \bar{F}_{\rm e, ds} \left( \mu, U_{\rm ds}, \sigma_{\parallel} \right)  }{\sqrt{2\left( U_{\rm ds} - |\Omega_{\rm e}| \mu \right)}} dU_{\rm ds}  \right]  = \nonumber \\
\left( \frac{e\phi_{\rm ds}(x)}{T_{\rm e}} \right)^2 \left[ \frac{T_{\rm e}}{e} \left. \frac{d^2n_{\rm e,mp}}{d\phi_{\rm mp}^2} \right\rvert_{x=0} - 3Z \int \frac{v_{\rm B}^4 f_{\rm i,dse}(\vec{v})}{v_x^4} d^3 \vec{v} \right] \rm ,
 \end{align}
leading to
\begin{align} \label{phidstwoprime-dse}
-\lambda_{\rm D}^2 \frac{e\phi_{\rm ds}''}{T_{\rm e}} = k_{2,\rm ds}  \left( \frac{e\phi_{\rm ds}}{T_{\rm e}} \right)^{2} \rm ,
\end{align}
with 
\begin{align} \label{k2ds}
k_{2,\rm ds} =  \frac{  3Z \int \frac{v_{\rm B}^4 f_{\rm i,dse}(\vec{v})}{v_x^4} d^3 \vec{v}  -   \frac{T_{\rm e}}{e} \left. \frac{d^2n_{\rm e,mp}}{d\phi_{\rm mp}^2} \right\rvert_{x=0}  }{  n_{\rm ds}(\infty) - \frac{1}{2} \gamma^2 \sum_{\sigma_{\parallel} = \pm 1} 2\pi \int_{ 0 }^{\infty} | \Omega_{\rm e}| d\mu  \int_{|\Omega_{\rm e}| \mu}^{\infty} \frac{\mu \partial_{\mu} \bar{F}_{\rm e, ds} \left( \mu, U_{\rm ds}, \sigma_{\parallel} \right)  dU_{\rm ds}  }{\sqrt{2\left( U_{\rm ds} - |\Omega_{\rm e}| \mu \right)}} }   \rm .
\end{align}
The potential profile far from the target is thus
\begin{align} \label{phi-dse-ds}
\frac{e\phi_{\rm ds}(x)}{T_{\rm e}}  = -\frac{a_{\rm ds} \lambda_{\rm D}^2}{(x+c_{\rm ds})^2} \rm ,
\end{align}
where $a_{\rm ds} = 6/k_{\rm ds, 2}$ and $c_{\rm ds}$ is a constant of integration.

For a Maxwellian incoming electron distribution function, we showed in the previous subsection that the numerator of (\ref{k2ds}) is always positive. 
We have shown in this section that the denominator of (\ref{k2ds}) is positive if $\gamma \sim 1 \ll e|\phi_{\rm ds}(0)|/T_{\rm e}$ and $\gamma \ll \left( e|\phi_{\rm ds}(0)|/T_{\rm e} \right)^{1/4} \lesssim 1$, thus guaranteeing that for these values (\ref{phidstwoprime-dse}) is consistent with a monotonic electron-repelling electrostatic potential variation near the Debye sheath entrance.
When the denominator on the right hand side of (\ref{Poisson-1}) and (\ref{k2ds}) becomes negative, the inequality of the kinetic Bohm criterion (\ref{kinetic-Bohm}) reverses, and, in the typical case of equality (see section~\ref{subsec-analytical-mpnear} and (\ref{Bohm-marginal})), (\ref{k2ds}) then predicts $\phi_{\rm ds}''(x) > 0$ near the Debye sheath entrance.
This is inconsistent with the assumption that $\phi_{\rm ds}(x) \rightarrow 0_-$ for $x/\lambda_{\rm D} \rightarrow \infty$.
Hence, the local monotonicity (and electron-repelling) constraint at the Debye sheath entrance imposes that the potential drop within the Debye sheath be larger than a critical value, such that $e|\phi_{\rm ds}(0)| /T_{\rm e} \gtrsim \gamma^4$ for $\gamma \lesssim 1$.
However, in the following subsection we show that an even stronger global monotonicity constraint exists in the Debye sheath for $\gamma \ll 1$, such that $e|\phi_{\rm ds}(0)| /T_{\rm e} \gtrsim \gamma^2$ for $\gamma \lesssim 1$.

\subsection{Conditions for a monotonic and electron-repelling magnetised sheath potential} \label{subsec-analytical-critical}

The electrostatic potential drop across the magnetic presheath, $|\phi_{\rm mp}(0)|$, increases as the magnetic field angle is reduced.
Recalling that $|\phi_{\rm mp}(0)| + |\phi_{\rm ds}(0)| = |\phi_{\rm w}|$, at fixed total potential drop $|\phi_{\rm w}|$, this situation corresponds to a decrease in the Debye sheath potential drop, $|\phi_{\rm ds}(0)|$.
In this section, we first analyse the dependence of $|\phi_{\rm ds}(0)|$ on $\alpha$ and $|\phi_{\rm w}|$, and we use it to obtain the scaling for critical angles at which the Debye sheath potential drop vanishes, $|\phi_{\rm ds}(0)| = 0$.
We then argue that a monotonic potential profile across the magnetised sheath does not allow $|\phi_{\rm ds}(0)| = 0$ when finite electron gyro-orbits are accounted for through finite values of $\gamma$, and we calculate an approximate minimum Debye sheath potential drop for which the potential profile in the Debye sheath can remain monotonic, as well as the corresponding critical angle.

\begin{figure}
\centering
\includegraphics[scale=0.65]{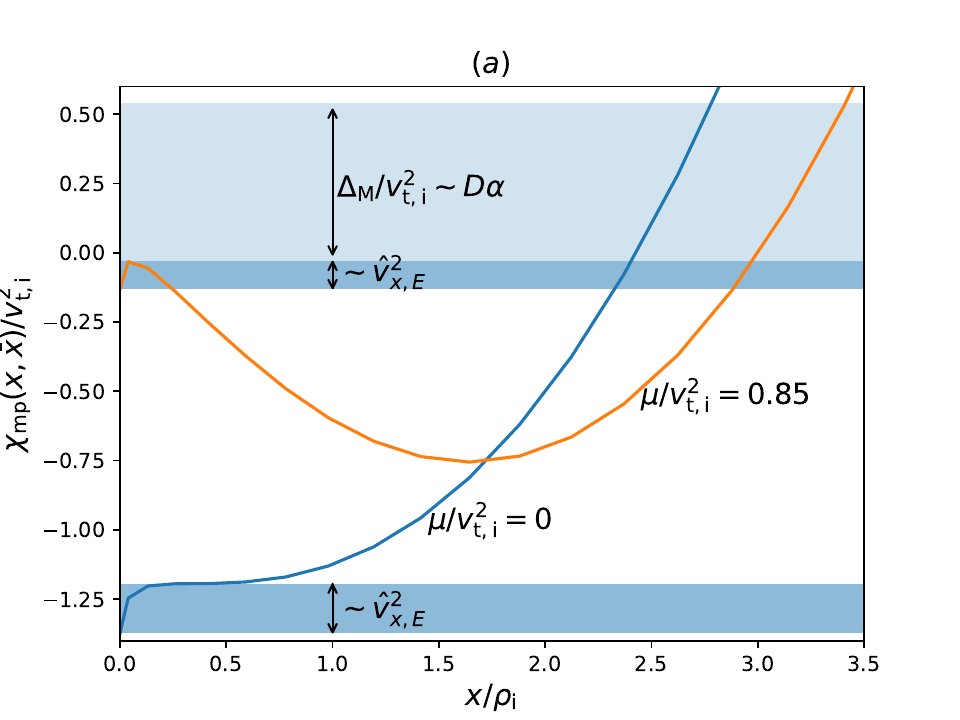}
\includegraphics[scale=0.65]{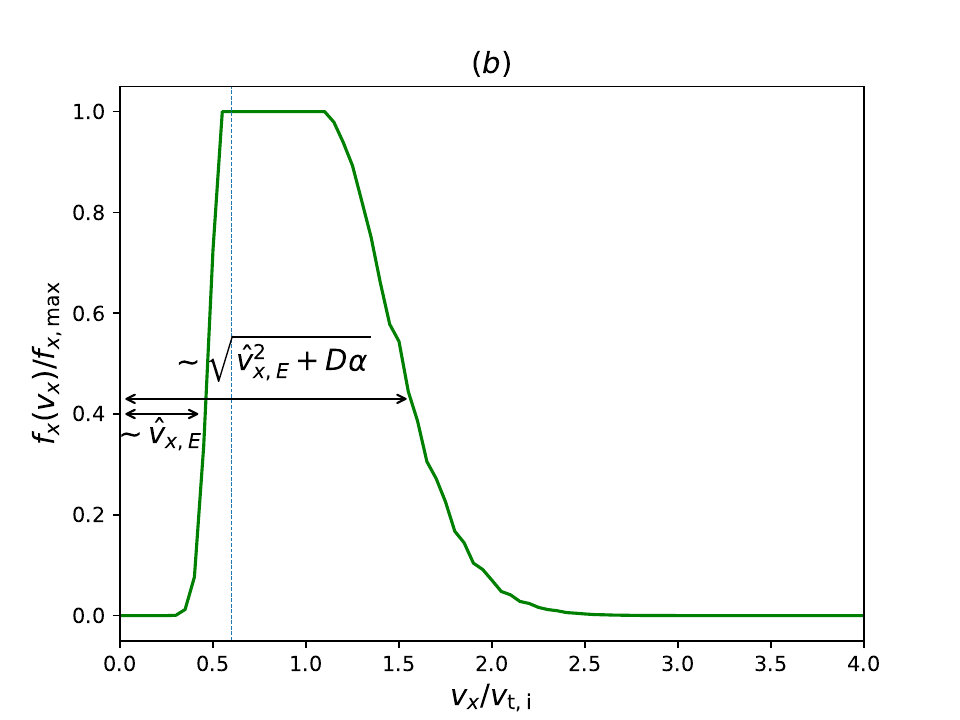}
\caption{(a) The effective potential $\chi_{\rm mp}(x,\bar{x})$ of ions in the magnetic presheath for the parameters $\alpha = 3^{\circ}$, $J_x = 0$, $\gamma = 0$, $\bar{x}/\rho_{\rm i} = 1.36$ ($\mu = 0$) and $\bar{x}/\rho_{\rm i} = 1.85$ ($\mu / v_{\rm t,i}^2 = 0.85$), as a function of $x$. The width of the darker and lighter shaded regions corresponds to the size of $\hat{v}_{x,E}^2$ and $\Delta_{\rm M} / v_{\rm t, i}^2 $, respectively. It can be seen that $\Delta_{\rm M}$ is zero for $\mu = 0$ and quickly increases with $\mu$, while $\hat{v}_{x,E}^2$ is finite at $\mu = 0$ and decreases with $\mu$, such that $\Delta_{\rm M}/v_{\rm t,i}^2 \gtrsim \hat v_{x,E}$ is satisfied for most ions. 
(b) For the same parameters, the marginalised velocity distribution of ions at the Debye sheath entrance, $f_x (v_x ) = \int_{-\infty}^{\infty} dv_y \int_{-\infty}^{\infty} dv_z f_{\rm i,dse}(\vec{v})$, normalised to its maximum value $f_{x,\rm max} = \max f_x (v_x)$. Its structure is related to $\hat{v}_{x,E}$ and $\Delta_{\rm M} /v_{\rm t,i}^2 \sim D \alpha$, as indicated. The dashed vertical line is the largest value of $\sqrt{2\left(\chi_{\rm mp, M}(\bar x) - \chi_{\rm mp}(0, \bar x) \right)} / v_{\rm t,i} \sim \hat v_{x, E}$, i.e., the one corresponding to $\mu = 0$.}
\label{fig-chi}
\end{figure}

To obtain the scaling for the potential drop $|\phi_{\rm ds}(0)|$ across the Debye sheath with the angle $\alpha$, we must first calculate the characteristic velocity component $v_x$ with which ions enter the Debye sheath. 
Recall that the range of possible values of $v_x$ of ions in open orbits at the Debye sheath entrance is given by (\ref{vx-range}), with $\Delta_{\rm M}$ given by (\ref{DeltaM-def}).
To estimate the size of $v_x$, we will consider cases in which the electron and ion temperatures are similar, $\tau \sim 1$, such that $v_{\rm B} \sim c_{\rm S} \sim v_{\rm t,i}$ and $\rho_{\rm B} \sim \rho_{\rm S} \sim \rho_{\rm i}$.
Using $\sqrt{2(U-\chi_{\rm M})} \sim v_{\rm t,i}$, and recalling the definition of $\mu_{\rm op}$ in (\ref{muop-def}), and that of $\mu_\alpha$ in (\ref{mugk-def}), we obtain
\begin{align}
\frac{d\mu_{\rm op}}{d\bar x} = \frac{1}{\pi} \int_{x_-}^{x_+} dx \frac{\Omega^2 (x-x_{\rm M})}{\sqrt{2\left(\chi_{\rm M}(\bar x) - \chi(x, \bar x)\right)}} \sim \frac{\Omega_{\rm i}^2 \rho_{\rm t,i}^2}{v_{\rm t,i}} \sim v_{\rm t,i} \rm ,
\end{align}
which gives $\Delta_{\rm M} \sim \alpha v_{\rm t,i}^2$ using (\ref{DeltaM-def}).
Hence, the characteristic size of the velocity component $v_x$ of ions entering the Debye sheath is given by
\begin{align} \label{vx-dse}
|v_x| \sim v_{\rm t, i} \sqrt{\hat{v}_{x,E}^2 + D\alpha}  \rm,
\end{align}
where $D$ is a number of order unity, and we have defined a velocity scale $\hat{v}_{x,E} v_{\rm t,i}$ such that 
\begin{align}
\sqrt{ \chi_{\rm mp, \rm M}(\bar x) - \chi_{\rm mp}(0, \bar x) } \sim \hat{v}_{x,E} v_{\rm t,i} \rm .
\end{align}
In figure~\ref{fig-chi}(a), the two different contributions to $v_x^2$ are shaded on the effective potential curve of an ion orbit for a typical set of parameters and magnetic field angle $\alpha = 3^{\circ}$.
The contribution $ \chi_{\rm mp, \rm M}(\bar x) - \chi_{\rm mp}(0, \bar x) \sim \hat{v}_{x,E}^2 v_{\rm t,i}^2$ is larger for smaller gyro-orbits and tends to zero for large gyro-orbits \cite{Geraldini-2021}, while the piece $\Delta_{\rm M} \sim D \alpha v_{\rm t,i}^2$ quickly increases from zero to become dominant, such that $D \alpha \gtrsim \hat{v}_{x,E}^2 $.
Figure~\ref{fig-chi}(b) shows how the size of the two contributions to $v_x^2$ affect the structure of the ion velocity distribution at the sheath entrance.
Despite its inadequacy for the parameters considered in this paper, we take the ordering $D \alpha \ll \hat{v}_{x,E}^2 $ to shed light on the link between $\hat{v}_{x,E}$ and $|\phi_{\rm ds}(0)|$ via the strength of the electric field divergence described by (\ref{phi-dse-mp}) and (\ref{p-scaling}).
For $D \alpha \ll \hat v_{x,E}^2$, the size of the Bohm integral on the left hand side of (\ref{Bohmint-ordered}) is controlled by the characteristic size of $\hat{v}_{x,E}$, 
\begin{align} \label{Bohmint-re}
Z\int d^3 v f_{\rm i,dse}(\vec{v}) \frac{v_{\rm B}^2}{v_x^2} \sim  \frac{n_{\rm e, ds}(\infty)}{\hat{v}_{x,E}^2 }  \rm .
\end{align}
Combining (\ref{Bohmint-re}) and (\ref{Bohmint-ordered}) (and using $v_{\rm B} \sim v_{\rm t,i}$) leads to
\begin{align} \label{ux-typeII}
\hat{v}_{x,E} \sim \begin{cases} 1 & \text{ for } \frac{e|\phi_{\rm ds}(0)|}{T_{\rm e}} \gtrsim 1 \rm , \\
\left( \frac{ e  |\phi_{\rm ds} (0)| }{ T_{\rm e} } \right)^{1/4}  & \text{ for } \frac{e|\phi_{\rm ds}(0)|}{T_{\rm e}} \ll 1 \rm .
\end{cases}  
\end{align}
The scaling (\ref{ux-typeII}) highlights that $\hat{v}_{x,E}$ depends on $|\phi_{\rm ds}(0)|$ and must satisfy $\hat v_{x, E} = 0$ for $\phi_{\rm ds}(0) = 0$, such that the left hand side of (\ref{Bohmint-ordered}) can diverge when the right hand side diverges;
that is, when the Debye sheath collapses ($|\phi_{\rm ds}(0)| = 0$), the Bohm criterion ceases to enforce the absence of ions with $v_x = 0$ at the Debye sheath entrance, and such ions will be present.

The ion current density at the Debye sheath entrance is given by $|J_{x, \rm i}| = e n_{\rm i, ds}(\infty) |v_x|$, with $|v_x|$ given by (\ref{vx-dse}).
For an electron distribution function that is exponentially decaying in energy (such as a Maxwellian), the electron density at the Debye sheath entrance is $n_{\rm e, ds}(\infty) \sim n_{\rm e, mp}(\infty) \exp \left( e\phi_{\rm mp}(0) / T_{\rm e} \right) $.
Since quasineutrality still holds at the Debye sheath entrance, $n_{\rm i, ds}(\infty) = n_{\rm e, ds}(\infty) / Z$ and so the ion density satisfies the same scaling (taking $Z \sim 1$).
Hence, and using $\phi_{\rm mp}(0) = \phi_{\rm w} - \phi_{\rm ds}(0)$, we obtain
\begin{align} \label{Ji-dse}
\frac{|J_{x, \rm i}| \exp \left( - e\phi_{\rm w}/T_{\rm e} \right) }{e n_{\rm e, mp}(\infty) v_{\rm t,i}} \sim  \exp \left( - \frac{e\phi_{\rm ds}(0)}{T_{\rm e}} \right)   \sqrt{\hat v_{x,E}^2 + D\alpha }  \rm .
\end{align}
Imposing ion continuity (we assumed that there are no ion sources in the magnetised sheath), such that the ion current (\ref{Ji-dse}) at the Debye sheath entrance is equal to the ion current (\ref{Ji-mpe}) at the magnetic presheath entrance, gives
\begin{align} \label{alpha-phids-full-notclean}
\alpha \exp\left(\frac{e|\phi_{\rm w}| }{ T_{\rm e}} \right) \sim \exp \left( \frac{e|\phi_{\rm ds}(0)| }{ T_{\rm e} } \right)   \sqrt{\hat v_{x,E}^2 + D\alpha }  \rm .
\end{align}
Using (\ref{alpha-phids-full-notclean}) and the fact that $\hat v_{x,E} = 0$ for $\phi_{\rm ds}(0) = 0$, the magnetic field angle for which the Debye sheath potential drop vanishes can be predicted to be $\alpha \sim D\exp \left( 2e\phi_{\rm w} / T_{\rm e} \right)$ (recall $\phi_{\rm w} <0$).
At smaller angles, the left hand side of equation (\ref{alpha-phids-full-notclean}) is too small to balance the right hand side, which suggests that a monotonic solution in the magnetised sheath can only exist for 
\begin{align} \label{alphacrit-gamma0}
\alpha \gtrsim \exp \left( \frac{2e\phi_{\rm w}}{T_{\rm e}} \right) \rm ,
\end{align}
where we have dropped the order-unity factor $D$.
This result was first derived by Ewart \emph{et al.} \cite{Ewart-2021}.
From (\ref{ordering-phiw-float}), inequality (\ref{alphacrit-gamma0}) becomes $\alpha \gtrsim M^{-1}$ in ambipolar conditions.

The physical picture of the constraint (\ref{alphacrit-gamma0}) is as follows.
Without an electric field across the magnetic presheath (and without an ion-repelling field in the Debye sheath), the ion density would naturally drop, from its value $n_{\rm i, mp}(\infty) = n_{\rm e, mp}(\infty)/Z$ at $x/\rho_{\rm i } \rightarrow \infty$, as $x$ decreases towards $x/\rho_{\rm i} = 0$.
At $x=0$, the ion density is composed only of the open orbit contribution, which is the second term in equation (\ref{n-finorb}), with the velocity spread of ions in open orbits at the Debye sheath entrance given by (\ref{vx-range}) and $\Delta_{\rm M}$ given by (\ref{DeltaM-def}).
Using $\Delta_{\rm M} \sim \alpha v_{\rm t,i}^2$ and setting $\chi_{\text{M}, R}(\bar x) \geqslant \chi_R (0, \bar x)$, the second term in (\ref{n-finorb}) satisfies $n_{\rm i, mp}(0) \lesssim \alpha^{1/2} n_{\rm i, mp}(\infty)$.
This is the maximum value that $n_{\rm i, mp}(0)$ can acquire, as adding an electron-repelling potential drop will accelerate the ions towards the wall, decreasing the ion density at $x/\rho_{\rm i} \rightarrow 0$ even further.
Such a potential drop must exist because the electron density can only be equal to $n_{\rm i, mp}(0) \sim \alpha^{1/2} n_{\rm i, mp}(\infty)$ at the Debye sheath entrance insofar as enough electrons are reflected across the magnetic presheath, such that $n_{\rm e, mp}(0) \sim n_{\rm e, mp}(\infty) \exp \left( e\phi_{\rm mp}(0)/T_{\rm e} \right) \sim n_{\rm i, mp}(0)$.
Quasineutrality therefore constrains $\exp \left( e\phi_{\rm mp}(0)/T_{\rm e} \right) \lesssim \alpha^{1/2} $.
If $\phi_{\rm mp}(0) < \phi_{\rm w}$, such that $\exp \left( e\phi_{\rm mp}(0)/T_{\rm e} \right) < \exp \left( e\phi_{\rm w}/T_{\rm e} \right)$, the Debye sheath potential profile must reverse, if possible\footnote{The question of whether such an ion-repelling Debye sheath could exist in a steady state next to an electron-repelling magnetic presheath is not addressed here.}, because $\phi_{\rm ds}(0) = \phi_{\rm w} - \phi_{\rm mp}(0) > 0$.
Excluding this reversal, and thus imposing $\exp \left( e\phi_{\rm w}/T_{\rm e} \right) < \exp \left( e\phi_{\rm mp}(0)/T_{\rm e} \right)$, leads to (\ref{alphacrit-gamma0}).

\begin{figure}
\centering
\includegraphics[scale=0.8]{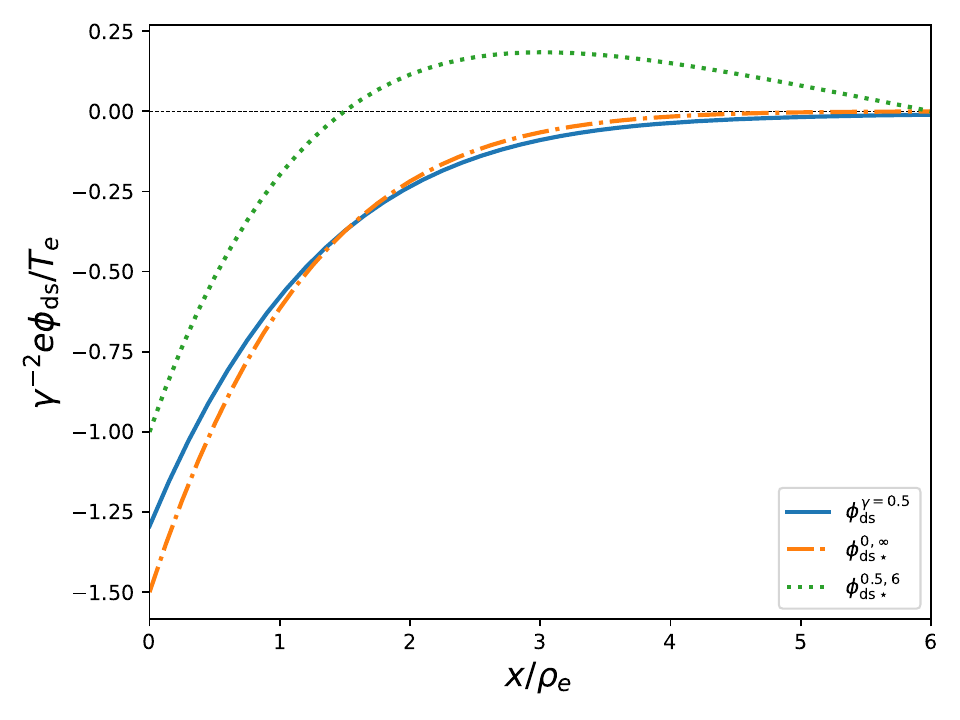}
\caption{The functions $\phi_{\rm ds}^{\gamma = 0.5}$ (solid line), $\phi_{\rm ds\star}^{0,\infty}$ (dash-dotted line) and $\phi_{\rm ds\star}^{0.5,6}$ (dotted line), normalised to $\gamma^{2} T_{\rm e} / e$, are plotted as a function of $x/\rho_{\rm e}$. Here, $\phi_{\rm ds}^{\gamma = 0.5}$ denotes the numerical solution (see section~\ref{sec-num}) of the potential profile in the Debye sheath at $\alpha = 1.8^{\circ}$, corresponding to the smallest Debye sheath potential drop for which a monotonic solution could be found for the parameters $J_x = 0$ (ambipolarity) and $M = 3600$. The function $\phi_{\rm ds\star}^{0,\infty}(x)$ is given by (\ref{phids-flat-gammacorr}), while $\phi_{\rm ds\star}^{0.5,6}(x)$ is obtained by solving (\ref{Poisson-flat}) with the boundary conditions (\ref{phiflat-modbound}), using $g = 0.5$ and $a = 6$.}
\label{fig-phiflat}
\end{figure}

The monotonicity requirement (\ref{alphacrit-gamma0}) has been calculated by assuming that the Debye sheath potential drop $|\phi_{\rm ds}(0)|$ could be made arbitrarily small, and that the Debye sheath potential profile could correspondingly remain monotonic.
However, it is strictly impossible to achieve exact sheath collapse, understood as a flat potential profile $\phi_{\rm ds}(x)=0$ on the Debye sheath scale, at finite values of $\gamma = \rho_{\rm e} / \lambda_{\rm D}$.
To see this, suppose first that a collapsed Debye sheath exists.
With a Maxwellian electron distribution, the electron density profile without electric field is given, to lowest order in $\alpha \ll 1$, by $n_{\rm e, ds}(\infty) \eta_{\rm flat}(x/\rho_{\rm e})$, with \cite{Geraldini-2019, Cohen-Ryutov-1998, Krasheninnikova-2010}\footnote{Equation (\ref{n-closed-cold}) is derived in section~4.2 of reference \cite{Geraldini-2019} (note the difference coming from the definition of $\rho_{\rm i}$ in \cite{Geraldini-2019} containing a factor of $\sqrt{2}$) by considering ions in the magnetic presheath with cold electrons. Equation (\ref{n-closed-cold}) had been previously derived in reference \cite{Krasheninnikova-2010} by considering a wall-parallel magnetic field.
Limiting cases of this expression for small and large arguments also appear in reference \cite{Cohen-Ryutov-1998}.}
\begin{align} \label{n-closed-cold}
\eta_{\text{flat}}(\hat x) = \frac{1}{\sqrt{2\pi}}  \int_{-\frac{\hat x}{2}}^{\infty}  \exp \left(- \frac{w^2}{2} \right) \text{erf}  \left( \sqrt{  \hat x \left( w + \frac{\hat x}{2} \right) } \right) dw \text{,}
\end{align}
and with the error function defined by
\begin{align} \label{erf}
\text{erf}(\xi) = \sqrt{\frac{2}{\pi}} \int_0^{\sqrt{2}\xi} \exp \left(-\frac{\xi'^2}{2} \right) d\xi' \text{.}
\end{align}
The ion density is uniform due to the absence of an electric field.
This situation would result in a potential profile across the Debye sheath, denoted $\phi_{\rm ds \star}(x)$, which is obtained by solving Poisson's equation
\begin{align} \label{Poisson-flat}
\varepsilon_0 \phi_{\rm ds \star}''(x) = en_{\rm e, ds}(\infty) \left( \eta_{\rm flat}\left( \frac{x}{\gamma \lambda_{\rm D} } \right) - 1 \right) \rm .
\end{align}
Proceeding to integrate twice (\ref{Poisson-flat}) using the boundary conditions $\phi_{\rm ds \star}'(\infty) = \phi_{\rm ds \star}(\infty) = 0$ gives
\begin{align} \label{phids-flat-gammacorr}
\frac{ e \phi_{\rm ds \star}^{0, \infty} (x) }{T_{\rm e}}  = -\gamma^2  \int_{\frac{x}{\rho_{\rm e}}}^{\infty} d\hat{x} \int_{\hat{x}}^{\infty}   \left[ 1 - \eta_{\rm flat}\left( \hat{x}'  \right) \right] d\hat{x}'  \rm ,
\end{align}
which is shown by the dash-dotted line in figure~\ref{fig-phiflat}.
It will soon be explained why we use the superscripts $0, \infty$ for the potential profile (\ref{phids-flat-gammacorr}).
The potential drop associated with (\ref{phids-flat-gammacorr}) is given by
\begin{align} \label{phids-min-gammasmall-exact}
\frac{ e \phi_{\rm ds \star}^{0,\infty} (0) }{T_{\rm e}} ~  = - \gamma^2 \int_{0}^{\infty} d\hat{x} \int_{\hat{x}}^{\infty}   \left[ 1 - \eta_{\rm flat}\left( \hat{x}'  \right) \right] d\hat{x}' = - \frac{3}{2} \gamma^2 \rm ,
\end{align}
where the last equality in (\ref{phids-min-gammasmall-exact}) follows from an analytical evaluation of the integral carried out in \ref{app-integral}. 
Since the potential profile (\ref{phids-flat-gammacorr}) is non-flat for $\gamma \neq 0$, but is generated by the spatial charge distribution (right hand side of (\ref{Poisson-flat})) associated with a flat potential profile, it follows that a flat potential profile is not self-consistent in the Debye sheath when $\gamma \neq 0$.

Since a non-flat electrostatic potential profile changes both the electron and the ion density profiles compared to the ones appearing in (\ref{Poisson-flat}), the non-flat potential profile (\ref{phids-flat-gammacorr}) is also not exactly self-consistent in the Debye sheath.
For $\gamma \ll 1$, the density profiles are modified very little by the small electrostatic potential variation resulting from (\ref{phids-flat-gammacorr}). 
This causes an additional small correction to the charge density on the right hand side of (\ref{Poisson-flat}), which is spatially restricted to the small region where $x \sim \rho_{\rm e} \sim \gamma \lambda_{\rm D} \ll \lambda_{\rm D}$, and therefore generates an additional smaller (in $\gamma$) correction to the potential profile (\ref{phids-flat-gammacorr}).
Therefore, the flat potential profile $\phi(x)= 0$ and the profile $\phi_{\rm ds\star}^{0,\infty}(x)$ in (\ref{phids-flat-gammacorr}) are 
increasingly accurate (in $\gamma \ll 1$) asymptotic approximations to a self-consistent monotonic potential profile $\phi_{\rm ds}(x)$ which varies weakly only on the electron Larmor radius scale, $x \sim \rho_{\rm e}$, and is flat on the Debye scale, $x \sim \lambda_{\rm D} \gg \rho_{\rm e}$.

One might wonder why we found a fixed potential drop across the sheath when it should be possible to achieve any potential drop by modifying the current into the sheath.
It is in fact possible to modify the small value of the potential at $x=0$ by solving (\ref{Poisson-flat}), since the potential remains approximately flat when $\phi_{\rm ds \star}(0) \sim \gamma^2 T_{\rm e} / e$ and the right hand side of (\ref{Poisson-flat}) thus remains accurate in the region $x \sim \rho_{\rm e}$. 
The simplest way to tweak the potential drop by $\gamma^2 T_{\rm e} / e$ is to regard the potential profile as the solution of (\ref{Poisson-flat}) with the Dirichlet boundary conditions 
\begin{align} \label{phiflat-modbound}
 \phi_{\rm ds \star}(0)  = \left(- \frac{3}{2} + g \right) \gamma^2 \frac{T_{\rm e}}{e} \text{ and } \phi_{\rm ds \star}(a \rho_{\rm e})  = 0 \rm .
\end{align} 
Equation (\ref{phids-flat-gammacorr}) thus results by taking $g = 0$ and $a \rightarrow  \infty$, which is why we denote this solution with superscripts $0$ and $\infty$.
Imposing on (\ref{Poisson-flat}) a smaller potential drop via the Dirichlet boundary conditions (\ref{phiflat-modbound}), with $g>0$ and the constant $a$ large enough that $\phi_{\rm ds \star}''(a\rho_{\rm e})/\gamma^2$ is negligible, the electric field is forced to change sign, as exemplified by the dotted line in figure~\ref{fig-phiflat} for $a = 6$ and $g=0.5$.
Hence, the steady-state solution for the electrostatic potential in the Debye sheath for potential drops smaller than (\ref{phids-min-gammasmall-exact}) is forced to be non-monotonic, and must additionally be non-flat in the region $x \sim \lambda_{\rm D} \gg \rho_{\rm e}$ to ensure that the potential asymptotes to $\phi(\infty) = 0$.

For $\gamma \lesssim 1$, equation (\ref{phids-flat-gammacorr}) gives an accurate estimate of the potential profile in the Debye sheath corresponding to the smallest potential drop $|\phi_{\rm ds}(0)|$ for which this profile is monotonic.
This can be seen in figure~\ref{fig-phiflat} by comparing the profile obtained from (\ref{phids-flat-gammacorr}) (dash-dotted line) to one obtained from a self-consistent numerical solution of the Debye sheath (solid line) with $\gamma = 0.5$ for a case in which the Debye sheath potential drop is (approximately) the smallest one allowing a monotonic numerical solution to be found.
Considering $e|\phi_{\rm ds}(0)|/T_{\rm e} \lesssim 1$ and taking once again the limit $\hat v_{x, E}^2 \gg D\alpha$, we insert equation (\ref{ux-typeII}) for $\hat v_{x, E}$ into the estimate (\ref{alpha-phids-full-notclean}) to obtain a relation between magnetic field angle and Debye sheath potential drop,
\begin{align}  \label{alpha-phids-full-notclean-largevxE}
\alpha \exp\left(\frac{e|\phi_{\rm w}| }{ T_{\rm e}} \right) \sim \hat v_{x,E} \sim \left( \frac{e|\phi_{\rm ds}(0)|}{T_{\rm e}} \right)^{1/4} \rm .
\end{align}
Applying the constraint $|\phi_{\rm ds}(0)| \gtrsim |\phi_{\rm ds \star}(0)| \sim \gamma^2 T_{\rm e}/ e$ to (\ref{alpha-phids-full-notclean-largevxE}) returns an  estimate for the critical magnetic field angle below which a monotonic solution cannot exist,
\begin{align} \label{alphacrit-new}
\alpha \gtrsim \gamma^{1/2} \exp \left( \frac{e\phi_{\rm w}}{T_{\rm e}} \right) \rm .
\end{align}
Equation (\ref{alphacrit-new}) extends the requirement (\ref{alphacrit-gamma0}) for a monotonic electron-repelling magnetised sheath to finite values of $\gamma \lesssim 1$.
By comparing (\ref{alphacrit-gamma0}) and (\ref{alphacrit-new}), the enhancement of the critical angle is significant for $\gamma \gtrsim \exp \left( 2e\phi_{\rm w} / T_{\rm e} \right) $.
From (\ref{ordering-phiw-float}), inequality (\ref{alphacrit-new}) becomes $\alpha \gtrsim \gamma^{1/2} M^{-1/2}$ in ambipolar conditions, giving an enhanced critical angle for $ \gamma \gtrsim M^{-1}$.
Even though the limit $\hat v_{x,E}^2 \gg D\alpha$ used to derive (\ref{alphacrit-new}) is not well satisfied for the angles $\alpha$ we consider in this paper (see figure~\ref{fig-chi}), this asymptotic result serves as an illustration of how the critical angle is enhanced due to the presence of finite electron gyro-orbits in the Debye sheath.

\section{Numerical solutions} \label{sec-num}

To numerically solve for the steady state of the magnetised sheath, we employ a sequence of iterative steps to improve the accuracy of an initial guess of the electrostatic potential solution in each of the magnetic presheath and Debye sheath regions.
From a guess of the potential profile in the magnetic presheath and in the Debye sheath, we calculate the electron and ion densities.
Using the calculated densities, we calculate a new guess for the potential in each region, and we repeat the density calculation using the corresponding new guess.
The process is stopped when the calculated densities are found to numerically satisfy Poisson's equation (\ref{Poisson}) in the Debye sheath and quasineutrality (\ref{quasi}) in the magnetic presheath.
The numerical scheme presented here is a significant extension of the one presented in \cite{Geraldini-2018} and \cite{Ewart-2021}, which only solved for the magnetic presheath and took $\gamma = 0$ by default. 

The code, which has been made available at \url{https://github.com/alessandrogeraldini/GYRAZE}, is called \url{GYRAZE} as a reminder of the fundamental underlying analytical basis of the scheme: a \textbf{gyrokinetic} treatment valid at \textbf{grazing} magnetic field incidence ($\alpha \ll 1$) to compute the ion density in the magnetic presheath and the electron density in the Debye sheath.
The physical inputs of \url{GYRAZE} are: the magnetic field angle $\alpha$, the particle distribution functions entering the magnetised sheath (section~\ref{subsec-num-bc}), the mass ratio $M$, the ratio $T_{\rm i}/T_{\rm e}$ of the ion temperature to the electron temperature, 
the ratio $\gamma$ of the electron gyroradius to the Debye length \textit{at the Debye sheath entrance} (in this work, we consider $\gamma = 0$ and $0.3 \leqslant \gamma \leqslant 1$), a flag whose value indicates which one between the current $J_x$ and the wall potential $\phi_{\rm w}$ is given as an input, and the corresponding value of the normalised current $J_x/c_{\rm S}$ or the magnitude of the total potential drop $e|\phi_{\rm w}|/T_{\rm e}$.
The code can run with multiple ion species and with different incoming distribution functions, although in this work we focus on a case with a single ion species and with a prescribed distribution function for each species (electrons and ions), with $T_{\rm i} = T_{\rm e}$.

This section is organised as follows.
In section~\ref{subsec-num-bc}, we introduce the assumed distribution functions of electrons and ions entering the magnetised sheath.
In section \ref{subsec-num-disc}, we introduce the numerical discretization of the potential and density profiles, thus describing the non-uniform grid used in the magnetic presheath.
In section~\ref{subsec-num-iter} we describe the iterative scheme used to obtain the numerical solution for the electrostatic potential. 
We then present the key numerical results in section~\ref{subsec-num-sol};
in particular, we numerically demonstrate the existence of a critical value $\alpha_{\star}$ of the magnetic field angle for a monotonic electron-repelling sheath solution, and its enhancement due to the finite size of electron gyro-orbits.

\subsection{Distribution functions of electrons and ions entering the magnetised sheath} \label{subsec-num-bc}

To solve for the ion and electron densities in the magnetised sheath using the expressions in section~\ref{sec-particle},
we must specify the distribution functions of incoming particles at the system boundaries.
In this study we have assumed the wall to be perfectly absorbing with respect to charged particles, so that no electrons and ions enter the magnetised sheath from the wall boundary.
Hence, we only need to impose the form of the distribution functions of ions and electrons entering the magnetised sheath at the magnetic presheath entrance.

For the ions, we use the following \textit{ad hoc} distribution function as appearing in (\ref{Fi-mp}),
\begin{align} \label{Fi-infty}
& F_{\rm i\infty} \left( \mu, U_{\rm mp} \right)  = n_{\rm i, mp}(\infty) \frac{2(U_{\rm mp} - \Omega_{\rm i} \mu )}{\pi^{3/2}\left( 2\bar T_{\rm i} / m_{\rm i} \right)^{5/2}}  \exp \left( - \frac{m_{\rm i}U_{\rm mp}}{\bar T_{\rm i}} \right)   \text{,}
\end{align} 
and pick $\bar T_{\rm i} = Z T_{\rm e}$.
Since this distribution function is non-Maxwellian, $\bar T_{\rm i}$ is a parameter that determines the width of the velocity distribution but is strictly \textit{not} the ion temperature as conventionally defined from a Maxwellian velocity distribution.
However, since this parameter is related to the characteristic energy spread of the ion population, we choose to define $\bar T_{\rm i} = T_{\rm i}$ and thus $v_{\rm t,i} = \sqrt{\bar T_{\rm i}/m_{\rm i}}$.
It can be verified that (\ref{Fi-infty}) is appropriately normalised, satisfying
\begin{align} \label{findNi}
n_{\rm i, mp}(\infty) = 2\pi \int_0^{\infty} \Omega_{\rm i} d\mu \int_{\Omega_{\rm i} \mu}^{\infty} \frac{F_{\rm i \infty} (\mu, U_{\rm mp})}{\sqrt{2\left( U_{\rm mp} - \Omega_{\rm i} \mu  \right) }} dU  \rm .
\end{align}

For the electrons, we assume a Maxwellian distribution function
\begin{align} \label{Fe-infty}
F_{\rm e\infty} (\mu, U_{\rm mp}) =  \frac{ \bar n_{\rm e \infty}  }{2v_{\rm t,e}^{3} \pi^{3/2}} \exp \left[ - \frac{U_{\rm mp}}{v_{\text{t,e}}^2}   \right] \rm .
\end{align}
The constant $\bar n_{\rm e\infty}$ is related to the electron density at the magnetic presheath entrance, $n_{\rm e, mp}(\infty)$,
\begin{align} \label{findNe}
n_{\rm e, \rm mp}(\infty) = 2\pi \int_0^{\infty} |\Omega_{\rm e}| d\tilde{\mu} \int_{-v_{\rm cut, mp}(\infty, \tilde{\mu})}^{\infty}  F_{\rm e \infty} \left( \tilde{\mu}, \frac{1}{2} v_{\parallel}^2 + |\Omega_{\rm e}| \tilde \mu \right)  dv_{\parallel}  \rm .
\end{align}
Since electrons are repelled away from the wall by the magnetised sheath, and most of the incoming electrons are reflected, the precise value of the electron density at the magnetic presheath entrance depends on the parallel velocity cutoff $v_{\rm cut, mp}(\infty, \tilde{\mu})$ in (\ref{v-cut-mp}) which results from the Debye sheath and magnetic presheath potential profiles.
As mentioned in section~\ref{subsec-electronsMP}, since we solve the magnetised sheath as an isolated system, we choose to adapt the normalisation $\bar n_{\rm e \infty}$ of the electron distribution function to ensure that quasineutrality is satisfied at the magnetised sheath entrance for any given magnetised sheath potential,
\begin{align}
n_{\rm e, mp}(\infty) = Z n_{\rm i, mp}(\infty) \rm .
\end{align}
Note that inserting (\ref{Fe-infty}) and expression (\ref{Ucut-smallgamma}) for the electron parallel velocity cutoff for $\gamma = \rho_{\rm e} / \lambda_{\rm D} = 0$ into expression (\ref{Je-mp}) for the electron current density through the magnetised sheath results in $J_{x, \rm e} = J_{x, \rm e,Max,0}$, with $J_{x, \rm e, Max,0}$ defined in (\ref{Je-mpe}).
The electron current generally takes a different value, while still satisfying $J_{x, \rm e} \sim J_{x, \rm e,Max,0}$, because the distribution function may be different from the one assumed in (\ref{Fe-infty}) and because in general $\gamma \neq 0$ changes (\ref{Ucut-smallgamma}).

The chosen distribution functions satisfy the kinetic Chodura condition (\ref{kinetic-Chodura}), slightly oversatisfying it by an amount small in $\exp\left( e\phi_{\rm w} / T_{\rm e} \right)$.
Recalling (\ref{ordering-phiw-float}), the floating wall potential satisfies $\exp\left( e\phi_{\rm w, fl} / T_{\rm e} \right) \sim M^{-1/2} \ll 1$ for $\tau \sim 1$.
Since we consider values of the wall potential not too far from $\phi_{\rm w, fl}$, we take the analytical form (\ref{phi-mpe}) of the electrostatic potential far from the wall, derived assuming a marginally satisfied Chodura condition, to be valid to a very good approximation. 
This differs from what was done in reference \cite{Ewart-2021}, where the ion distribution function was slightly corrected to ensure that (\ref{kinetic-Chodura}) was exactly marginally satisfied.

\subsection{Discretization of the electrostatic potential} \label{subsec-num-disc}

The electrostatic potential function is discretized by introducing a grid of position values indexed $l$ in each region $R \in \left\lbrace \text{mp,ds} \right\rbrace$, $x_{R,l}$, and a corresponding grid of values of the electrostatic potential $\phi_{R,l}$.
The aim of the numerical scheme is to make the difference between the discretized potential $\phi_{R,l}$ and the true electrostatic potential solution evaluated at each position on the grid, $\phi_{R} (x_{l}) $, i.e. $\phi_{R,l} - \phi_{R} (x_{l})$, as small as possible for all values of $l$ in both regions $R$ (to be quantified in subsection~\ref{subsec-num-iter}).
The distance from the wall is discretised into a number $\nu_R$ of grid points indexed from $0$ to $\nu_R - 1$,
\begin{align} \label{xi-disc}
x_{R,l} = \left( \sqrt{g_{R} + l h_{R}} -\sqrt{g_R} \right)^2  \text{ for } l \in [0, \nu_R -1] \rm .
\end{align}
In (\ref{xi-disc}), a non-zero value of the parameter $g_{R}$ ensures that the grid is finer near $x=0$, while becoming approximately equally spaced at large $x_{l}$, where the spacing tends to $h_{R}$.
For Debye sheath simulations, an evenly spaced grid in $x$ was chosen, with $g_{\rm ds} = 0$ and $h_{\rm ds} = 0.125 \rho_{\rm e}$, such that $x_{\text{ds},l} = l h_{\rm ds}$.
For magnetic presheath simulations, we exploit scaling (\ref{p-scaling}) to adapt the grid to the value of $\phi_{\rm ds, 0} $ at a given iteration.
Equation (\ref{xi-disc}) can be rearranged to
\begin{align} \label{f-disc-def}
f_{l} \equiv l h_{\rm mp} = g_{\rm mp} \left[ \left( 1 + \sqrt{\frac{x_{\text{mp},l}}{g_{\rm mp}} } \right)^2 - 1 \right] \simeq \begin{cases} 2\sqrt{g_{\rm mp}x_{\text{mp},l}} & \text{ for } x_{\text{mp},l} \ll g_{\rm mp} \text{,} \\
x_{\text{mp},l} & \text{ for } x_{\text{mp},l} \gg  g_{\rm mp} \text{,} \end{cases}
\end{align}
for the newly-defined discrete variable $f_{l}$.
Comparing the expansion in (\ref{f-disc-def}) for $x_{\text{mp},l} \ll g_{\rm mp}$ to (\ref{phi-dse-mp}) in section \ref{sec-analytical}, we notice a correspondence between $g_{\rm mp}$ and $p$.
To make $\phi_{\text{mp},l}$ roughly equally spaced in $f_l$ near $x=0$, we choose
\begin{align} \label{gridparam-def}
g_{\rm mp} = \begin{cases}
\rho_{\rm B} & \text{ for } 2 \frac{e|\phi_{\rm ds,0}|}{T_{\rm e}}  \geqslant 1  \text{,} \\
2 \frac{e|\phi_{\rm ds,0}|}{T_{\rm e}} \rho_{\rm B} & \text{ for } 2 \frac{e|\phi_{\rm ds,0}|}{T_{\rm e}} < 1 \text{,}
\end{cases}
\end{align}
which is justified from the scaling (\ref{p-scaling}).
The interval between grid points in magnetic presheath simulations was chosen to be $h_{\rm mp} = 0.4 \rho_{\rm i}$. 
This is coarser than in the Debye sheath because the magnetic presheath iteration does not require calculating the derivative of the potential.
Leaving the iterable electrostatic potential grid relatively coarse helps stabilise the iteration scheme (described in section~\ref{subsec-num-iter}), which is less stable at finer grid sizes due to numerical error in the second derivative of the potential \cite{Geraldini-2018}.
The accuracy of the integrals needed for the ion density calculation in the magnetic presheath (see reference \cite{Geraldini-2018} for details of the numerical integration) is increased by introducing a refined grid $f_{l'} = l' h_{\rm mp} / 3$ for $l' \in [ 0, 3\nu_{\rm mp} - 3]$, with roughly three times more grid points than the coarse grid, where the relation $x_{\text{mp},l'} = \left( \sqrt{g_{\rm mp} + f_{l'}} -\sqrt{g_{\rm mp}} \right)^2$ (from (\ref{f-disc-def}) and (\ref{xi-disc})) is used to obtain the refined position grid $x_{\text{mp},l'}$, and a spline interpolation of the mapping $f_{l} \mapsto \phi_{\text{mp},l}$ is used to obtain the corresponding potential values $\phi_{l'}$.
The choice of grid in (\ref{gridparam-def}) and the grid refinement used to calculate velocity space integrals makes the marginal Bohm condition (\ref{Bohm-marginal}) more closely and more robustly satisfied at $x=0$ in the magnetic presheath, indicating a more accurately resolved numerical solution close to $x=0$ (recall that this is where the electric field diverges on the magnetic presheath scale).

The tentative system size is chosen to be
\begin{align}
\bar L_{\rm mp} = 25 \rho_{\rm B} 
\end{align}
for the magnetic presheath and 
\begin{align}
\bar L_{\rm ds} = \begin{cases}  25 \lambda_{\rm D} \text{ for } \rho_{\rm e} \leqslant \lambda_{\rm D} \text{,} \\
 25 \rho_{\rm e} \text{ for } \rho_{\rm e} > \lambda_{\rm D} \text{} \end{cases}
\end{align}
for the Debye sheath.
The actual system size $L_R$ is set by $h_R$, $g_R$ and 
\begin{align}
\nu_R = \text{floor}  \left( \frac{ \left(\sqrt{g_R} + \sqrt{\bar L_R} \right)^2 - g_R }{h_R} \right)  \rm ,
\end{align}
such that the system size is $L_R = \left( \sqrt{g_{R} + \nu_R h_{R}} -\sqrt{g_R} \right)^2$ according to (\ref{xi-disc}).
For the Debye sheath, $g_{\rm ds} = 0$ and so $\nu_{\rm ds} = \text{floor} \left( \bar L_{\rm ds} / h_{\rm ds} \right)$ gives $L_{\rm ds} = h_{\rm ds} \text{floor} \left( \bar L_{\rm ds} / h_{\rm ds} \right)$.
For the magnetic presheath, (\ref{gridparam-def}) is applied at every iteration to adapt the grid, implying that the actual system size changes depending on the value of $g_{\rm mp}$ at each iteration.
Hence, $L_{\rm mp}$ can actually end up differing from $\bar L_{\rm mp}$ by more than the grid size $h_{\rm mp}$.
The size of the domain nonetheless remains large enough to capture almost all of the electrostatic potential variation occurring across the magnetic presheath.

\subsection{Iterative method of solution} \label{subsec-num-iter}

In this section, we present the three different iterative schemes to solve for:
the electrostatic potential profile in the magnetic presheath; the wall potential corresponding to a fixed value of current through the magnetised sheath (if imposed); the electrostatic potential profile in the Debye sheath. 

To significantly speed up the convergence to a full solution in the magnetic presheath and Debye sheath, the solver is run in two phases.
First the magnetic presheath is solved on its own, fixing either $\phi_{\rm w}$ or $J_x$ and using an analytical model for the electron parallel velocity reflection cutoff.
If $J_x$ is fixed (e.g. $J_x = 0$), the wall potential value is iterated together with the magnetic presheath potential profile.
When the magnetic presheath potential profile is such that, to within a numerical tolerance, the electron and ion densities satify quasineutrality, and the current, if imposed, takes the desired value, the Debye sheath potential profile is iterated as well in a second phase of the iteration. 
The iterative scheme is summarised below.

\noindent \textbf{Phase 1}
\begin{enumerate}
\item Set the iteration index $N=0$.
\item Introduce the initial magnetic presheath potential guess $\phi_{\text{mp},l}^0$ (see section~\ref{subsubsec-num-mp}).
\item If the current $J_x$ is imposed, introduce the initial wall potential $\phi_{\rm w}^0$ (see section~\ref{subsubsec-num-j}); if the wall potential $\phi_{\rm w}$ is imposed instead, it always takes the imposed value and need not be iterated in all subsequent steps.
\item Calculate:
\begin{enumerate}
\item the particle densities in the magnetic presheath, $n_{\text{e,mp},l}^N$ and $n_{\text{i,mp},l}^N$, with the electron density calculated using an analytical model for the electron reflection cutoff in the Debye sheath (see section~\ref{subsubsec-num-mp});
\item the electron flux and thus the current through the system. 
\end{enumerate}
\item Verify whether:
\begin{enumerate}
\item quasineutrality is satisfied in the magnetic presheath up to the desired tolerance (see section~\ref{subsubsec-num-mp}); 
\item the current is equal to the prescribed value (if imposed) up to the desired tolerance (see section~\ref{subsubsec-num-j}).
\end{enumerate} 
\item If all conditions in (v) are satisfied, proceed to (x).
\item If condition (v)(a) is satisfied, set $\phi_{\text{mp},l}^{N+1} = \phi_{\text{mp},l}^N$; if not, guess $\phi_{\text{mp},l}^{N+1}$ (see section \ref{subsubsec-num-mp}).
\item If condition (v)(b) is satisfied, set $\phi_{\text{w}}^{N+1} = \phi_{\text{w}}^{N}$; if not, guess $\phi_{\text{w}}^{N+1}$ (see section \ref{subsubsec-num-j}).
\item Add one to the iteration index $N$, and return to (iv).
\item If $\gamma \geqslant 0.3$, proceed to \textbf{Phase 2}. Otherwise, the iteration is terminated.
\end{enumerate}
\noindent \textbf{Phase 2}
\begin{enumerate}
\item Introduce a Debye sheath potential guess $\phi_{\text{ds},l}^N$.
\item Calculate:
\begin{enumerate}
\item the particle densities in the magnetic presheath, $n_{\text{e,mp},l}^N$ and $n_{\text{i,mp},l}^N$, with the electron density calculated using the electron reflection cutoff from the previous iteration;
\item the particle densities in the Debye sheath, $n_{\text{e,ds},l}^N$ and $n_{\text{i,ds},l}^N$;
\item the electron flux and thus the current through the system.
\end{enumerate}
\item Verify whether:
\begin{enumerate}
\item quasineutrality is satisfied in the magnetic presheath up to the desired tolerance (see section~\ref{subsubsec-num-mp}); 
\item the current is equal to the prescribed value (if imposed) up to the desired tolerance (see section~\ref{subsubsec-num-j});
\item Poisson's equation is satisfied in the Debye sheath up to the desired tolerance (see section~\ref{subsubsec-num-ds}).
\end{enumerate} 
\item If all conditions in (iii) are simultaneously satisfied for the second time in a row, 
proceed to (ix). 
(The second time is needed to allow the electron density in the magnetic presheath to be calculated using the electron reflection cutoff from the same Debye sheath potential profile for which the electron density in the Debye sheath is calculated.)
\item if condition (iii)(a) is satisfied, set $\phi_{\text{mp},l}^{N+1} = \phi_{\text{mp},l}^N$; if not, guess $\phi_{\text{mp},l}^{N+1}$ as described in section \ref{subsubsec-num-mp}.
\item if condition (iii)(b) is satisfied, set $\phi_{\text{w}}^{N+1} = \phi_{\text{w}}^{N}$; if not, guess $\phi_{\text{w}}^{N+1}$ as described in section \ref{subsubsec-num-j}.
\item if all conditions (iii)(a)-(c) are simultaneously satisfied, set $\phi_{\text{ds},l}^{N+1} = \phi_{\text{ds},l}^N$; if not, guess $\phi_{\text{ds},l}^{N+1}$ as described in section \ref{subsubsec-num-ds}.
\item Add one to the iteration index $N$ and return to (ii).
\item The iteration is terminated.
\end{enumerate}

In the rest of this section, we describe in detail some of the crucial steps in the iterative scheme summarised above. 
We focus first on the magnetic presheath iteration (section~\ref{subsubsec-num-mp}), then on the iteration for the wall current (section~\ref{subsubsec-num-j}), and finally on the Debye sheath iteration (section~\ref{subsubsec-num-ds}).

\subsubsection{Potential profile in the magnetic presheath.} \label{subsubsec-num-mp}

A flat potential is generally chosen as the initial guess in the magnetic presheath
\begin{align}
\phi^0_{\text{mp},l} = 0 \text{ for } l \in [0 , \nu_{\rm mp} - 1] \rm .
\end{align}

The electron density in the magnetic presheath at the $N$th iteration, $n^N_{\text{e,mp},l}$, should depend on the numerically computed cutoff function $U_{\rm ds, cut}^N(\tilde \mu)$, which in turn depends on the electron reflection occurring in the Debye sheath.
During the first phase of the iteration, which iterates only over the magnetic presheath potential profile, a model is adopted for the electron reflection cutoff.
For $\gamma = 0$, we use the analytical cutoff $U_{\rm ds, cut, 0}(\tilde \mu)$ described in section~\ref{subsec-electronsMP}. 
Since this model is exact, a second phase of the iteration will not be necessary.
For $\gamma \geqslant 0.3$, we use the cutoff $U_{\rm ds, cut, \infty}(\tilde \mu)$ described in \ref{app-gammainf}, which is only appropriate in the limit $\gamma \rightarrow \infty$.
The reason $U_{\rm ds, cut, \infty}(\tilde \mu)$ is chosen over $U_{\rm ds, cut, 0}(\tilde \mu)$ is that it ensures that the guessed wall potential $\phi_{\rm w}^{N+1}$ is overestimated, rather than underestimated, when the current $J_x$ is imposed (see figure~\ref{fig-phiamb}).
This ensures that the imposed potential drop across the Debye sheath is overestimated during phase 1 of the iteration, which is important because the numerical scheme cannot handle this potential drop crossing zero. 
Underestimating the Debye sheath potential drop could lead to an incorrect reversal of the Debye sheath during phase 1 of the iteration, even if this would not happen with the more accurate electron cutoff calculated during phase 2.  
We plan to develop a more accurate first-principles model for the electron reflection cutoff for finite values of $\gamma \lesssim 1$. 
In phase 2 of the iteration scheme, at each iteration we choose to calculate the ion and electron densities first in the magnetic presheath, and subsequently in the Debye sheath.
Hence, at the stage of the iteration where the densities in the magnetic presheath are calculated, $U_{\rm ds, cut}^N(\tilde \mu)$ is not known.
Therefore, the function $U_{\rm ds, cut}^{N-1}(\tilde \mu)$, calculated at the previous iteration using (\ref{mu-cut})-(\ref{U-cut}), or using the model in phase 1 if $N$ corresponds to the first density evaluation in phase 2, is used to calculate $n^N_{\text{e,mp},l}$.

The iterative scheme to solve for the electrostatic potential in the magnetic presheath is a generalization of the one presented in reference \cite{Geraldini-2018} (for adiabatic electrons) and modified in reference \cite{Ewart-2021} to include the distribution of parallel electron velocities and electron reflection.
At each iteration $N$, we first calculate a tentative guess for the next iteration, $\bar \phi_{\text{mp,l}}^{N+1}$.
Note that we use an overline because this guess will not be our final guess $\phi_{\text{mp,l}}^{N+1}$, which is instead obtained via
\begin{align}
\phi_{\rm mp}^{N+1} = w_{\rm mp} \bar \phi_{\rm mp}^{N+1} + (1-w_{\rm mp}) \phi_{\rm mp}^{N} \rm .
\end{align}
Here the weight $w_{\rm mp}$ is set to $0.3$, but is reduced to $0.1$ if the potential drop across the Debye sheath becomes very small, such that $e|\phi_{\rm ds, 0}| / T_{\rm e} < 0.02$.
This reduction is necessary because the iteration becomes more sensitive as the potential drop across the Debye sheath becomes smaller.
The problem is that our iteration scheme does not allow for an intermediate iterative step where the Debye sheath potential profile reverses.

To calculate the tentative potential guess $\phi_{\text{mp}, l}^{N+1}$, instead of numerically inverting the expression for the electron density as a function of potential, as was done in reference \cite{Ewart-2021}, here we add a Boltzmann factor on both sides of the quasineutrality equation (\ref{quasi}) to re-express it as
\begin{align} \label{quasi-rearr}
n_{\text{e}, \rm mp}(\infty) \exp \left( \frac{e\phi_{\rm mp}(x) }{T_e}  \right) = Zn_{\text{i,mp}}(x) - n_{\text{e},{\rm mp}} (x)  + n_{\text{e}, \rm mp}(\infty) \exp \left( \frac{e\phi_{\rm mp} (x) }{T_e} \right)  \rm ,
\end{align}
and then we prescribe that the left hand side evaluated using a tentative potential guess at the \textit{next} iteration $N+1$, $\bar \phi_{\rm mp}^{N+1}$, be equal to the right hand side evaluated using the potential at the present iteration $N$, $\phi_{\rm mp}^{N}$.
This leads to
\begin{align} \label{quasi-rearr-it}
n_{\text{e}, \rm mp}(\infty) \exp \left( \frac{e \bar \phi_{\text{mp},l}^{N+1}}{T_e}  \right) = Zn_{\text{i,mp},l}^N - n_{\text{e,mp},l}^N (x)  + n_{\text{e,mp}}(\infty) \exp \left( \frac{e\phi_{\text{mp},l}^N (x) }{T_e} \right)  \rm ,
\end{align}
which can be inverted to obtain $\bar \phi_{\text{mp},l}^{N+1}$. 
Although this iterative scheme works well, we prefer another version which is linearised about $\phi_{\text{mp},l}^{N+1} - \phi_{\text{mp},l}^{N}$,
\begin{align} \label{tentative-mp}
 \frac{e\bar{\phi}^{N+1}_{l}  }{T_e} = \frac{Zen^N_{\text{i,mp},l} - en^N_{\text{e,mp},l} }{n_{\rm e, mp}(\infty) }  \exp \left( - \frac{e\phi^N_{\text{mp},l}  }{T_e}  \right)  + \frac{e\phi^N_{\text{mp},l}  }{T_e}   \text{ for } l \in [ 0 , q_{\rm mp} - 1] \rm .
\end{align}
In these equations the index $l$ runs from $0$ to $q_{\rm mp}-1$, where $l = q_{\rm mp}-1$ is the largest value of the index for which $ x_l < L_{\rm mp} - 5\rho_{\rm i} $.
It is not possible to apply equation (\ref{tentative-mp}) for all values of $l$ on the grid because the density evaluation at a given position $x$ requires the profile of the effective potential up to a distance of a few thermal gyroradii in both directions.
This is because the value of the adiabatic invariant $\mu$ of any ion orbit crossing the position $x$ depends on an integral of the effective potential curve about the entire gyro-orbit, whose typical size is $\rho_{\rm i}$. 
Gyro-orbits whose size is larger than $5\rho_{\rm i}$ have an exponentially small contribution to the density, which can be neglected.

The tentative new potential guess for $x > x_{q_{\rm mp}}$ can be obtained from the analytical form in (\ref{phi-mpe}).
The derivative of (\ref{phi-mpe}) is 
\begin{align} \label{phi'-mpe}
\frac{e\phi_{\rm mp}'(x)}{T_{\rm e}}  = \frac{4 a_{\rm mp} \rho_{\rm i}^4}{(x+c_{\rm mp})^5} \rm .
\end{align}
By combining (\ref{phi-mpe}) and (\ref{phi'-mpe}), we obtain
\begin{align}
c_{\rm mp} = - \frac{4 \phi_{\text{mp}}(x)}{\phi_{\text{mp}}'(x)} - x \rm .
\end{align}
The expressions used to numerically evaluate the constants $c_{\rm mp}$ and $a_{\rm mp}$ at the $N$th iteration are
\begin{align} \label{cmp-num}
c_{\rm mp}^N = - \frac{4 \phi_{\text{mp},q_{\rm mp}-1 }^N ( x_{q_{\rm mp}-1} - x_{q_{\rm mp}-2} )}{\phi_{\text{mp},q_{\rm mp}-1} - \phi_{\text{mp},q_{\rm mp}-2}} - x_{q_{\rm mp}-1} \rm ,
\end{align}
based on the approximation $\phi_{\text{mp}}^N(x_{q_{\rm mp}-1}) \approx (\phi_{\text{mp},q_{\rm mp}-1}^N - \phi_{\text{mp},q_{\rm mp}-2}^N)/ ( x_{q_{\rm mp}-1} - x_{q_{\rm mp}-2} )$, and
\begin{align} \label{amp-num}
a_{\rm mp}^N = - \frac{e\phi^N_{\text{mp},q_{\rm mp} -1}}{T_{\rm e}} \left( \frac{x_{q_{\rm mp}-1} + c_{\rm mp}^N }{\rho_{\rm i}} \right)^4  \rm .
\end{align}
The numerical electrostatic potential guess at large values of $x$ is
\begin{align}
\frac{e \bar{\phi}^{N+1}_{\text{mp},l} }{T_{\rm e}}  = - \frac{ a_{\rm mp}^N \rho_{\rm i}^4 }{\left( x_l + c_{\rm mp}^N \right)^4}   & \text{ for }  q_{\rm mp} \leqslant l < \nu_{\rm mp} \rm ,
\end{align}
with $c_{\rm mp}^N$ and $a_{\rm mp}^N$ given by equations (\ref{cmp-num}) and (\ref{amp-num}), respectively.

The electrostatic potential at the $N$th iteration is deemed an acceptable numerical solution if 
\begin{align}
\frac{1}{q_{\rm mp}} \sum_{l=0}^{q_{\rm mp}-1} \frac{ \left| n_{\text{e,mp},l}^N - Zn_{\text{i,mp},l}^N \right|}{Zn_{\text{i,mp},l}^N} < 5 \times 10^{-3} \rm .
\end{align}
As explained in reference \cite{Geraldini-2018}, the numerical scheme exits with an error if more than one minimum or maximum of the effective potential of $\chi(x, \bar x)$ are found for any $\bar x$.
This occurrence is very rare owing to the fact that the potential is iterated on a coarser grid, as described in section~\ref{subsec-num-disc}, from which it is then interpolated during the density calculation.
Contrary to references \cite{Geraldini-2018, Ewart-2021}, no smoothing procedure is necessary in the initial stages of the iteration, thanks to the coarser grid.

\subsubsection{Wall potential value.} \label{subsubsec-num-j}

If the wall potential $\phi_{\rm w}$ is prescribed as an input, we impose this value at every iteration step, such that $\phi_{\rm w}^{N} = \phi_{\rm w}$ by default. 
If the current density $J_x$ reaching the target is instead prescribed as an input, including the special case $J_x = 0$ that corresponds to a locally ambipolar flow, an iterative scheme is necessary to specify $\phi_{\rm w}^{N+1}$ such that the converged solution has the desired value of $J_x$.
We normally start from a very large potential drop as an initial guess, 
\begin{align}
\frac{e |\phi_{\rm w}^0| }{T_{\rm e}} = 5 \rm .
\end{align}

The total current in the $x$ direction, normal to the wall, is, from (\ref{Je-mp}) and (\ref{Ji-mp}), given by
\begin{align}
J_{x} = J_{x, \rm i} + J_{x, \rm e} = 2\pi e \alpha \left[ \int_0^{\infty} |\Omega_{\rm e}| d\tilde{\mu} \int_{\frac{\Omega_{\rm e} \phi_{\rm mp}(0)}{B} + U_{\rm cut, ds} (\tilde{\mu})}^{\infty} F_{\rm e \infty} (\tilde{\mu}, U_{\rm mp}) dU_{\rm mp} \right. \nonumber \\
\left. - Z\int_0^{\infty} \Omega_{\rm i} d\mu \int_{\Omega_{\rm i} \mu }^{\infty} F_{\rm i \infty} (\mu, U_{\rm mp}) dU_{\rm mp}  \right] \rm .
\end{align}
Solving for the wall potential $\phi_{\text{w}}$ requires an iteration because the relationship between $\phi_{\text{w}}$ and the electron current density $J_{x,\text{e}}$ depends on the spatial profile of the magnetised sheath potential via the numerically computed cutoff function $U_{\rm cut, ds}(\mu)$.
The iteration of the wall potential makes use of the explicit formula in (\ref{Je-mpe}) for the electron current as a function of the wall potential for an incoming Maxwellian electron distribution functions and for $\gamma = 0$, $J_{x, \rm e, Max,0}$.

The deviation of the electron current from that evaluated from a Maxwellian for $\gamma = 0$ can be defined as
\begin{align} \label{devcur}
\tilde{J}_{x,\rm e} = J_{x, \rm e} - J_{x, \text{e,Max},0} \rm .
\end{align}
The equation $J_x = J_{x,\rm i} + J_{x, \rm e}$ for the total current $J_x$ through the sheath can thus be re-arranged as
\begin{align} \label{newj-prelim}
J_{x,\text{e,Max},0} =  J_{x} - \tilde{J}_{x,\text{e}} - J_{x,\rm i} \rm .
\end{align}
An iteration scheme emerges by prescribing that the left hand side correspond to the potential at the iteration $N+1$ while the right hand side correspond to the potential calculated at the iteration $N$,
\begin{align} \label{newj-1}
J^{N+1}_{x,\text{e, Max},0} =  J_{x} - \tilde{J}^N_{x,\text{e}} - J_{x,\rm i}  \rm .
\end{align}
The ion current $J_{x,\rm i}$ only depends on the incoming ion distribution function and not on the iteration number $N$, since we only consider monotonic and electron-repelling magnetised sheath potential profiles which cannot reflect any ions.
From equation (\ref{Je-mpe}), we have
\begin{align} \label{Je-mpe-disc}
J_{x, \rm e, Max,0}^{N+1} =  \frac{1}{\sqrt{2\pi}} \alpha e n_{\rm e, mp}(\infty) v_{\rm t, e} \exp \left( \frac{e\phi_{\rm w}^{N+1} }{ T_{\rm e} } \right) \rm ;
\end{align}
hence, (\ref{newj-1}) can be re-expressed as
\begin{align} \label{newphiwall}
 \frac{e\phi^{N+1}_{\rm w}}{T_e} =  \ln \left[ \frac{ \sqrt{2\pi} ( J_{x} - \tilde J^N_{x,\text{e}} - J_{x,\rm i} ) }{ \alpha e n_{\rm e, mp}(\infty) v_{\rm t,e} } \right] \rm ,
\end{align}
with
\begin{align}
\tilde J^N_{x,\text{e}} = J^N_{x,\text{e}} -  \frac{1}{\sqrt{2\pi}} \alpha e n_{\rm e, mp}(\infty) v_{\rm t, e} \exp \left( \frac{e\phi_{\rm w}^N }{ T_{\rm e} } \right) \rm .
\end{align}
When $\gamma = 0$ and the electron distribution function is a Maxwellian, a single iterative step (\ref{newphiwall}) would always return the exact wall potential, since $\tilde{J}_{x,\text{e}}^N = 0$. 

The iteration scheme must be changed to avoid cases in which the argument of the logarithm in (\ref{newphiwall}) becomes negative when $\gamma \neq 0$.
Linearising $J^{N+1}_{x,\text{e,Max,}0}$ in (\ref{Je-mpe-disc}) about $J^N_{x,\text{e,Max,}0}$ for small $\phi^{N+1}_{\rm w} - \phi^N_{\rm w}$ gives
\begin{align} \label{newj-lin}
J^{N+1}_{x,\text{e,Max},0}  = J^N_{x,\text{e,Max,}0} \left( 1+\frac{e(\phi^{N+1}_{\rm w} - \phi^{N}_{\rm w})}{T_{\rm e}} \right) \rm .
\end{align} 
The tentative new guess for the wall potential is obtained by combining (\ref{newj-1}) and (\ref{newj-lin}), and replacing $\phi^{N+1}_{\text{w}}$ with $\bar{\phi}^{N+1}_{\text{w}}$,
\begin{align} \label{phiW-impose}
\frac{e\bar \phi^{N+1}_{\text{w}}}{T_e}  =  \frac{e\phi^{N}_{\text{w}}}{T_e}   +  \frac{ J_{x} - J_{x}^N }{J_{x,\text{e,Max},0}^N} \rm.
\end{align}
To obtain the value of the wall potential at the next iteration, we use the equation
\begin{align} \label{phiW-tentativetopermanent}
\frac{e\phi^{N+1}_{\text{w}}}{T_e} = w_J \frac{e\bar{\phi}^{N+1}_{\text{w}}}{T_e} + (1-w_J) \frac{e\phi^{N}_{\text{w}}}{T_e} \rm ,
\end{align}
where $w_J$ is a weight.
The weight $w_{J}$ is set to $0.2$ initially, but, just as for $w_{\rm mp}$, it is reduced by a factor of 3, to $0.066$, if $e|\phi_{\rm ds,0}|/T_{\rm e} < 0.02$.

\subsubsection{Potential profile in the Debye sheath.} \label{subsubsec-num-ds}

The first potential guess on the Debye sheath scale, which happens at some iteration number $N$ corresponding to a converged magnetic presheath solution with a model electron reflection cutoff, is chosen arbitrarily to be a function with the same form as (\ref{phi-dse-ds}) and with an initial potential drop $\phi^N_{\rm ds,0} = \phi^N_{\rm w} - \phi_{\rm mp, 0}^N$,
\begin{align}
\phi^N_{\text{ds},l} = \frac{\phi^N_{\text{ds},0}}{\left( \frac{x_l}{\lambda_{\rm D}} + 1 \right)^2} \text{ for } l \in [0, n_{\rm ds})\rm .
\end{align}

The iterative scheme developed here to solve for the electrostatic potential in the Debye sheath is a generalization of the magnetic presheath iteration described in section~\ref{subsubsec-num-mp}.
As done before with quasineutrality, we subtract a Boltzmann factor on both sides of Poisson's equation (\ref{Poisson}), thus re-expressing it as
\begin{align} \label{Poisson-rearr}
& \epsilon_0 \phi_{\rm ds}''(x) - en_{\text{e,ds}}(\infty) \exp \left( \frac{e\phi_{\rm ds}(x) }{T_e}  \right) \nonumber \\
& = en_{\text{e,ds}} (x)  - en_{\text{e,ds}}(\infty) \exp \left( \frac{e\phi_{\rm ds} (x) }{T_e} \right)- Zen_{\text{i,ds}}(x) \rm .
\end{align}
To turn (\ref{Poisson-rearr}) into an iterative scheme, we prescribe that the left hand side be calculated using a tentative guess to the potential at the $N+1$ iteration, $\bar{\phi}_{\rm ds}^{N+1}$, and the right hand using using the potential at the $N$ iteration, $\phi_{\rm ds}^{N}$, such that
\begin{align} \label{iterative-ideal}
\epsilon_0 \frac{d^2\bar{\phi}_{\rm ds}^{N+1}(x)}{dx^2} - en_{\text{e}, \rm ds}(\infty) \exp \left( \frac{e\bar{\phi}_{\rm ds}^{N+1} (x) }{T_{\rm e}}  \right) = en_{\text{e,ds}}^{N} (x) - Zen_{\text{i,ds}}^{N}(x) \nonumber \\
- en_{\text{e,ds}}(\infty) \exp \left( \frac{e\phi_{\rm ds}^{N} (x) }{T_{\rm e}} \right)  \rm .
\end{align} 
To make the operator acting on $\bar{\phi}^{N+1}_{\rm ds}(x)$ on the left hand side readily invertible, we modify (\ref{iterative-ideal}) by linearising for small  $e(\bar \phi_{\rm ds}^{N+1}(x) - \phi_{\rm ds}^N(x))/T_{\rm e} $,
\begin{align} \label{iterative-general}
\epsilon_0 \frac{d^2\bar{\phi}_{\rm ds}^{N+1}(x)}{dx^2} -  \frac{e\bar{\phi}_{\rm ds}^{N+1} (x)}{T_e} en_{\text{e,ds}}(\infty) \exp \left( \frac{e\phi_{\rm ds}^{N} (x) }{T_e}  \right) = en_{\text{e,ds}}^{N}(x) - Zen_{\text{i,ds}}^{N}(x) \nonumber \\  -  \frac{e\phi_{\rm ds}^N (x) }{T_e}  en_{\text{e,ds}}(\infty) \exp \left( \frac{e\phi_{\rm ds}^{N} (x) }{T_e}  \right) \rm .
\end{align}
Equation (\ref{iterative-general}) is the iterative scheme used in this work.
After obtaining $\bar{\phi}^{N+1}_{\rm ds}$ from $\phi_{\rm ds}^N$, a constant weight $w_{\rm ds}$ is used to compute the (N+1)th iteration of the electrostatic potential,
\begin{align} \label{tentativetopermanent-general}
 \phi_{\rm ds}^{N+1} = w_{\rm ds} \bar{\phi}_{\rm ds}^{N+1} + (1-w_{\rm ds}) \phi_{\rm ds}^{N}  \rm .
\end{align}
In the simulations, the weight $w_{\rm ds}$ is set to $1$, such that the tentative potential guess coincides with the potential at the next iteration.

A second-order accurate discretization of the second derivative $d^2/dx^2$ is used, such that, for $l \in [1, \nu_{\rm ds} - 2]$,
\begin{align} \label{Laplacian-discrete}
\epsilon_0 \frac{ \bar{\phi}^{N+1}_{\text{ds},l+1} - 2\bar{\phi}^{N+1}_{\text{ds},l} + \bar{\phi}^{N+1}_{\text{ds},l-1} }{ h_{\rm ds}^2}  - \frac{e^2\bar{\phi}^{N+1}_{\text{ds}, l} }{T_{\rm e}}  n_{\rm e, mp, 0} \exp \left( \frac{e\phi^N_{\text{ds},l} }{T_{\rm e}}  \right)  \nonumber \\  = e n^N_{\text{e,ds}, l} - Zen^N_{\text{i,ds}, l} - \frac{e^2\phi^N_{\text{ds},l}  }{T_{\rm e}}  n_{\rm e, mp, 0} \exp \left( \frac{e\phi^N_{\text{ds},l} }{T_{\rm e}}  \right)  \rm .
\end{align}
Note that we have replaced $n_{\rm e,ds}(\infty)$, which is the electron density in the Debye sheath infinitely far from the wall (far beyond any grid point present in our numerical discretisation), by the equivalent electron density calculated at $x=0$ in the magnetic presheath scale, $n_{\rm e, mp, 0}$.
Dirichlet boundary conditions are imposed at the first and last grid points, $l=0$ and $l=n_{\rm ds} -1$,  such that
\begin{align} \label{bcPoissonwall}
\bar{\phi}_{\text{ds},0}^{N+1} = w_{\rm ds}^{-1} \left( \phi_{\rm w}^{N+1} - \phi_{\text{mp},0}^{N+1} - (1-w_{\rm ds})\phi_{\text{ds},0}^N \right) \rm ,
\end{align}
\begin{align} \label{bcPoissonfar}
\bar{\phi}_{\text{ds}, \nu_{\rm ds}-1}^{N+1} = 0 \rm .
\end{align}
Note that condition (\ref{bcPoissonwall}) is on the \textit{tentative} potential profile $\bar \phi_{\text{ds},l}^{N+1}$ and not on the next iteration of the potential, $\phi_{\text{ds},l}^{N+1} = w_{\rm ds} \bar \phi_{\text{ds},l}^{N+1} + (1-w_{\rm ds}) \phi_{\text{ds},l}^{N}$.
From (\ref{tentativetopermanent-general}), condition (\ref{bcPoissonwall}) is equivalent to imposing the wall potential at the next iteration to be $\phi_{\text{ds},0}^{N+1} = \phi_{\rm w}^{N+1} - \phi_{\text{mp},0}^{N+1}$. 
Using (\ref{bcPoissonwall}) and (\ref{bcPoissonfar}), equation (\ref{Laplacian-discrete}) can be re-expressed as 
\begin{align} \label{iterative-Poisson}
\sum_{j=1}^{\nu_{\rm ds}-2} \frac{e}{T_{\rm e}} \mathsf{L}^N_{lj} \bar{\phi}^{N+1}_{\text{ds},j} =  \frac{n^N_{\text{e,ds},l} - Z n^N_{\text{i,ds}, l} }{n_{\rm e, mp, 0}} - \lambda_{\rm D}^2 h_{\rm ds}^{-2} \frac{e\bar{\phi}^{N+1}_{\text{ds}, 0}}{T_{\rm e}} \delta_{l,1} -  \frac{e\phi^N_{\text{ds},l}}{T_{\rm e}} \exp \left( \frac{e\phi^N_{\text{ds},l}}{T_{\rm e}}  \right)  \nonumber \\
\text{ for } 1 \leqslant l \leqslant \nu_{\rm ds}-2 \rm ,
\end{align}
with
\begin{align} \label{L-operator}
\mathsf{L}^N_{lj} = 
\begin{cases}
 \lambda_{\rm D}^2 h_{\rm ds}^{-2} \left( \delta_{l,j+1} - 2 \delta_{l,j} \right)  - \delta_{l,j} \exp \left( e\phi^N_{\text{ds},l} / T_e \right)    & \text{ for } l = \nu_{\rm ds} - 2 \text{,} \\
 \lambda_{\rm D}^2 h_{\rm ds}^{-2} \left( \delta_{l,j-1} - 2 \delta_{l,j} \right)  - \delta_{l,j} \exp \left( e\phi^N_{\text{ds},l} / T_e \right)  &  \text{ for } l = 1 \text{,} \\
 \lambda_{\rm D}^2 h_{\rm ds}^{-2} \left( \delta_{l,j+1} + \delta_{l,j-1} - 2 \delta_{l,j} \right)  - \delta_{l,j} \exp \left( e\phi^N_{\text{ds},l} / T_e \right)  & \text{ otherwise,}
\end{cases}
\end{align} 
where $\delta_{l,j}$ is the Kronecker delta, and the right hand side of (\ref{iterative-Poisson}) depends on the wall boundary condition (\ref{bcPoissonwall}).

Analogously to the ion density in the magnetic presheath (recall section~\ref{subsubsec-num-mp}), the electron density in the Debye sheath can only be evaluated for $i \leqslant q_{\rm ds} -1$, with $q_{\rm ds} < \nu_{\rm ds}$, due to the finite size of electron gyro-orbits.
In order to impose the analytically derived functional form (\ref{phi-dse-ds}) for the potential decay far from the wall in the Debye sheath, for $i \geqslant q_{\rm ds}$ we make the following replacement on the right hand side of (\ref{iterative-Poisson}),
\begin{align} \label{iterative-Poisson-replace}
&\frac{n^N_{\text{e,ds},l} - Z n^N_{\text{i,ds}, l} }{n_{\rm e, mp, 0}} =  - k_{2, \rm ds}^N \left( \frac{e\phi^N_{\text{ds}, l}}{T_{\rm e}} \right)^2  \text{ for }  l \in [q_{\rm ds}, \nu_{\rm ds}-1] \rm ,
\end{align}
where $k_{2, \rm ds}^N$ is the numerical value at the $N$th iteration of the constant $k_{2, \rm ds}$ appearing in equation (\ref{phidstwoprime-dse}). 
Equation (\ref{iterative-Poisson-replace}) results from equating the right hand sides of Poisson's equation (\ref{Poisson}) and of equation (\ref{phidstwoprime-dse}), and then discretizing. 
The constant $k_{2, \rm ds}^N$ can be determined by applying the equality (\ref{iterative-Poisson-replace}) at the grid point furthest away from the wall where the density can be evaluated, $l = q_{\rm ds} - 1$,
\begin{align}
k_{2, \rm ds}^N = \frac{ Zn^N_{\text{i,ds},q_{\rm ds}-1} - n^N_{\text{e,ds},q_{\rm ds}-1} }{n_{\rm e, mp, 0}} \left(   \frac{e\phi^N_{\text{ds}, q_{\rm ds}-1}}{T_{\rm e}}\right)^{-2}  \rm .
\end{align}

The tentative potential profile resulting from inverting (\ref{iterative-Poisson}) has the disadvantage that it is equal to zero at the largest grid point, and therefore cannot capture the functional form of the potential decay very far from the wall, beyond $x_{q_{\rm mp}}$.
To circumvent this problem, after $\bar \phi_{\text{ds},l}^{N+1}$ is calculated using (\ref{iterative-Poisson}), we \emph{replace} the potential guess for $i \geqslant q_{\rm ds}$ with a new one at large values of $x$, which is based on the analytically predicted potential decay in (\ref{phi-dse-ds}).
The derivative of (\ref{phi-dse-ds}) is
\begin{align} \label{phi'-dse-ds}
\frac{e\phi_{\rm ds}'(x)}{T_{\rm e}}  = \frac{2 a_{\rm ds} \lambda_{\rm D}^2}{(x+c_{\rm ds})^3} \rm .
\end{align}
By combining (\ref{phi-dse-ds}) and (\ref{phi'-dse-ds}) we obtain
\begin{align}
c_{\rm ds} = - \frac{2 \phi_{\text{ds}}(x)}{\phi_{\text{ds}}'(x)} - x \rm .
\end{align}
The numerical evaluation of the constants $a_{\rm ds}$ and $c_{\rm ds}$ at iteration $N+1$ are given by
\begin{align}
c_{\rm ds}^{N+1} = - \frac{ 2 \bar \phi_{\text{ds},q_{\rm ds}-1}^{N+1} h_{\rm ds} }{ \bar \phi_{\text{ds},q_{\rm ds}-1}^{N+1} -\bar \phi_{\text{ds},q_{\rm ds}-2}^{N+1} } - x_{q_{\rm ds}-1} \rm ,
\end{align}
and
\begin{align}
a_{\rm ds}^{N+1} = \phi_{\text{ds},q_{\rm ds} -1}^{N+1} \left( x_{q_{\rm ds}-1} + c_{\rm ds}^{N+1} \right)^2  \rm .
\end{align}
Hence, the tentative potential guess for the next iteration is modified according to
\begin{align}
\frac{e \bar{\phi}^{N+1}_{\text{ds},l} }{T_{\rm e}}  = - \frac{ a_{\rm ds}^{N+1} \lambda_{\rm D}^2 }{\left( x_l + c_{\rm ds}^{N+1} \right)^2}   & \text{ for }  q_{\rm ds} \leqslant l < \nu_{\rm ds} \rm .
\end{align}
The electrostatic potential at the $N$th iteration is deemed an acceptable numerical solution if
\begin{align}
\sum_{l=1}^{q_{\rm ds}-1} \frac{1}{Zn_{\text{i,ds},l}^N}  \left| n_{\text{e,ds},l}^N - Zn_{\text{i,ds},l}^N - n_{\text{e,mp},0}^N \frac{1}{\gamma^2} \frac{\rho_{\rm e}^2}{ h_{\rm ds}^{2}} \frac{e}{T_{\rm e}} \mathsf{L}_{lj}^N \phi_{\text{i,ds},l}^N \right| < 5 \times 10^{-3} \rm .
\end{align}

One important point regards how $\lambda_{\rm D}$ and $\gamma$ are defined.
From equation (\ref{lambdaD}), the size of the Debye sheath scales with the Debye length calculated using the density at the Debye sheath entrance, where the electron density is much smaller than at the magnetic presheath entrance: $n_{\rm e, mp}(0) / n_{\rm e, mp}(\infty) \approx 0.07-0.12 $ in the range of angles $\alpha = 2.5^{\circ}-5^{\circ}$ in ambipolar conditions.
Hence, the parameter $\gamma$, which is fixed in the simulations, is given by (\ref{gamma-def}) with $n_{\rm e, ref} = n_{\text{e,mp},0}^N$.
One could have fixed a different parameter $\gamma_{\rm mpe} = \rho_{\rm e} / \lambda_{\text{D,mpe}} $, with $\lambda_{\rm D, mpe} = \sqrt{\epsilon_0 T_{\rm e} / e^2 n_{\rm e, mp}(\infty)}$, defined using the density at the magnetic presheath entrance, and calculated the values of $\lambda_{\rm D}$ and $\gamma$ corresponding to the density at the Debye sheath entrance at every iteration, which would typically satisfy $\lambda_{\rm D} / \lambda_{\rm D, mpe} \approx 3-4$ and $\gamma / \gamma_{\rm mpe} \approx 1/4-1/3$.
In order to study the direct effect of the size of electron Larmor orbits on the magnetised sheath solution, as done here, $\gamma$ is a more appropriate parameter.
Using data from \cite{Militello-Fundamenski-2011}, we estimate $\gamma_{\rm mpe} \approx 0.1-0.4$ for JET and $\gamma_{\rm mpe} = 0.15-0.25$ for ITER, so that $\gamma \lesssim 0.1$ for both machines\footnote{The quoted range of values was obtained by considering data for the Near SOL and the Far SOL in \cite{Militello-Fundamenski-2011}. For JET, the range includes data for both H-mode and L-mode.}. 
In detached conditions, we instead estimate the slightly larger value $\gamma_{\rm mpe} \approx 0.5$, which implies $\gamma \approx 0.1-0.2$. 
Hence, the effect of $\gamma$ is expected to be small in most cases in current and future fusion devices.
It is possible that the effect of a small value of $\gamma$ on the sheath characteristics can be adequately included via an expansion in $\gamma \ll 1$.


\subsection{Numerical solutions of monotonic electron-repelling electrostatic potential} \label{subsec-num-sol}

\begin{figure}
\centering
\includegraphics[width=0.8\textwidth]{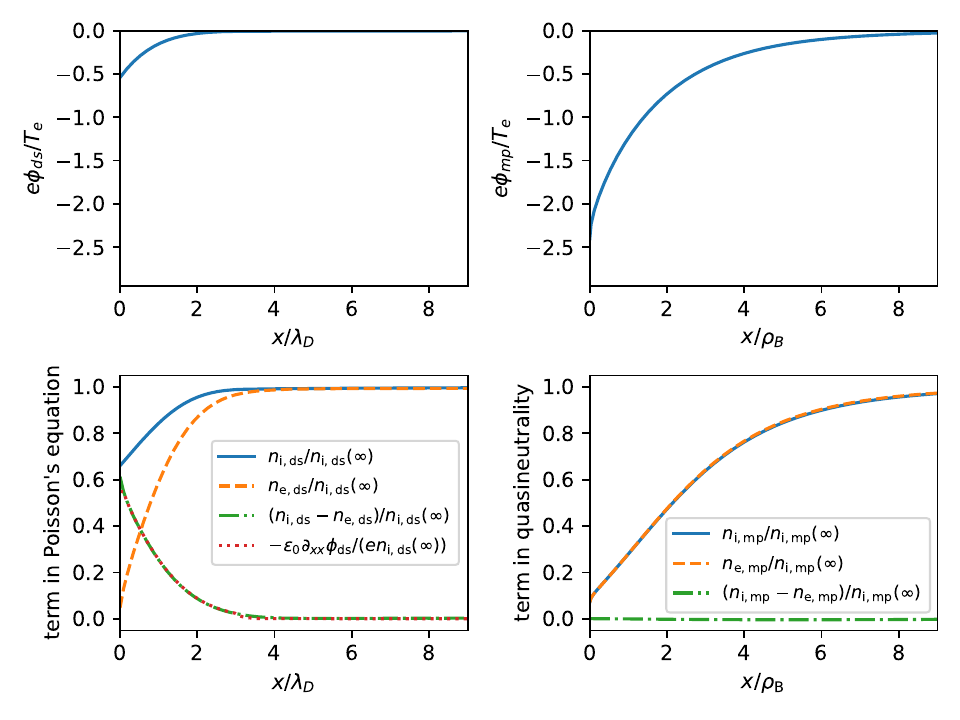}
\caption{Electrostatic potential (top) and density profiles (bottom) in the Debye sheath (left) and in the magnetic presheath (right) 
for $M = 3600$, $Z=1$, $\gamma = 0.7$, $J = 0$ (ambipolar) and $\alpha = \alpha_{\star} = 2.5^{\circ}$ (critical).}
\label{fig-phisol1}
\end{figure}
\begin{figure} 
\centering
\includegraphics[width=0.8\textwidth]{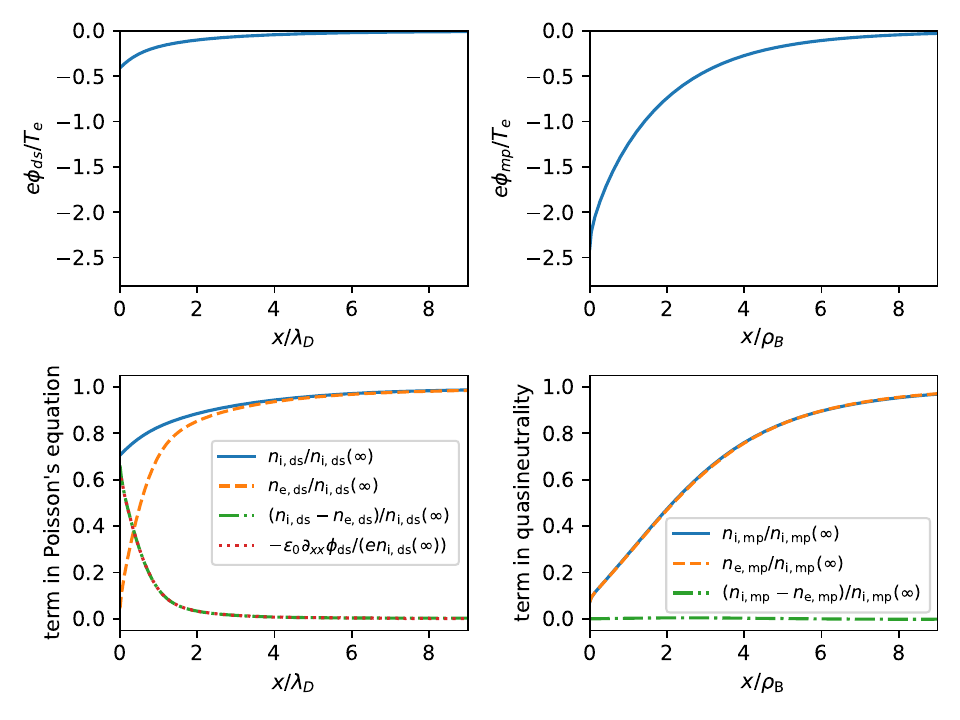}
\caption{Electrostatic potential (top) and density profiles (bottom) in the Debye sheath (left) and in the magnetic presheath (right) 
for $M = 3600$, $Z=1$, $\gamma = 0.3$, $J = 0$ (ambipolar) and $\alpha = 2.5^{\circ}$.}
\label{fig-phisol2}
\end{figure}
\begin{figure} 
\centering
\includegraphics[width=0.8\textwidth]{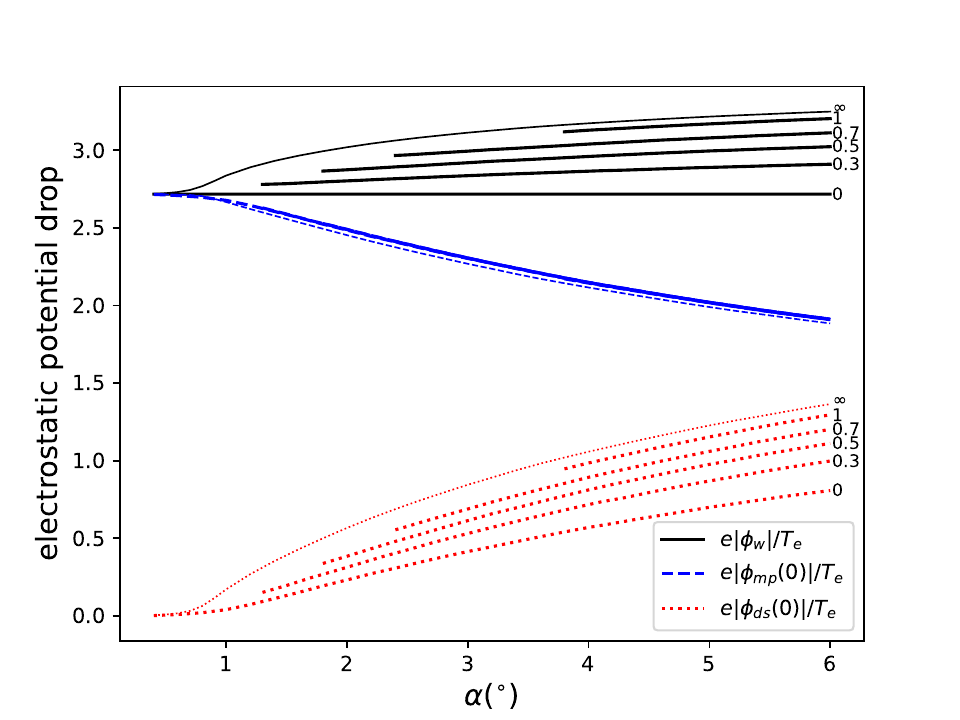} 
\caption{The magnitude of the floating potential drop across the magnetised sheath ($|\phi_{\rm w, fl}|$, corresponding to $J_x = 0$), and the corresponding potential drops across the Debye sheath ($|\phi_{\rm ds}(0)|$) and magnetic presheath ($|\phi_{\rm mp}(0)|$) for $M=3600$, are shown as a function of $\alpha$ for different values of $\gamma$ (labelled). The lines are truncated on the left hand side at the critical angle $\alpha_{\star}$. The case $\gamma \rightarrow \infty$ corresponds to the model cutoff in \ref{app-gammainf}.}
\label{fig-phiamb}
\end{figure}
\begin{figure} 
\centering
\includegraphics[width=0.8\textwidth]{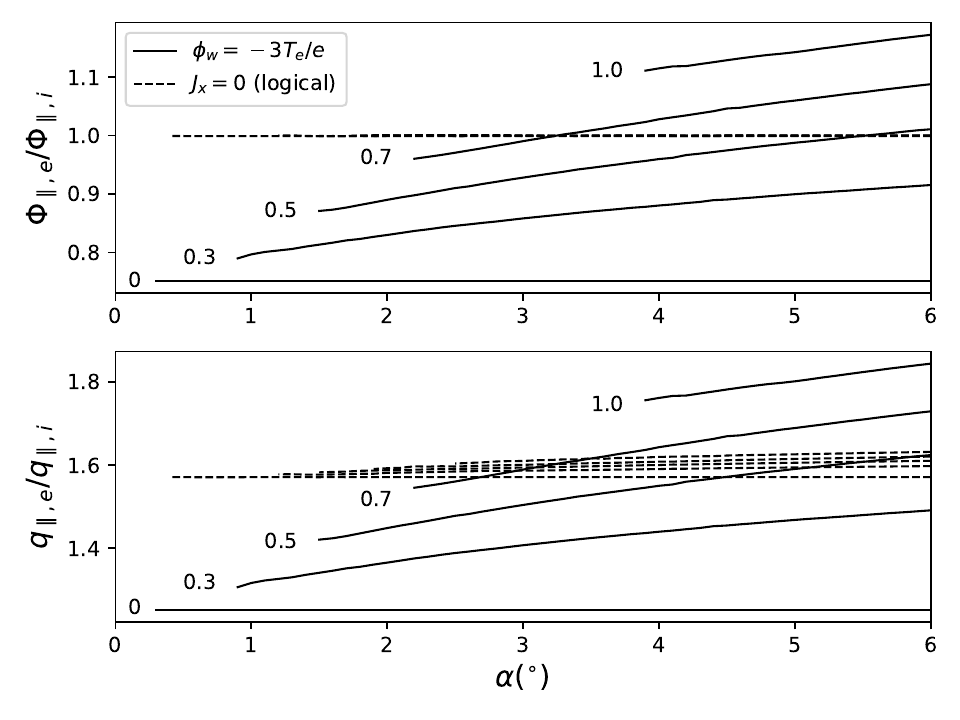} 
\caption{Parallel electron flux $\Phi_{\parallel, \rm e}$ (defined in (\ref{Phipare-mp})) through the magnetised sheath (top) and parallel electron heat flux $q_{\parallel, \rm e}$ at the magnetised sheath entrance (bottom) as a function of magnetic field angle $\alpha$ and electron gyro-orbit size $\gamma$ (labelled values), obtained with two different boundary conditions: $e\phi_{\rm w}/T_{\rm e} = -3$ (solid line), and $J_{x} = 0$ (dashed lines). The electron fluxes are normalised to the respective ion flux: $\Phi_{\parallel, \rm i}$ (defined in (\ref{Phixi-mp})) and $q_{\parallel, \rm i}$. The parallel heat flux of species s is defined by $q_{\parallel, s} = 2\pi \int_0^{\infty} |\Omega_{s}| d\mu \int_{-\infty}^{\infty}  \bar F_{s, \rm mp} (\mu, U_{\rm mp}, \sigma_{\parallel}) m_{\rm s} U_{\rm mp} v_{\parallel}  dv_{\parallel}$ with $U_{\rm mp} = |\Omega_{s} | \mu + \frac{1}{2}v_{\parallel}^2$ and $\sigma_{\parallel}=v_{\parallel}/|v_{\parallel}|$. The dashed lines in the top plot overlap (to within a small tolerance) due to the equivalence of $J_{x, \rm e} = -J_{x, \rm i}$ and $\Phi_{\parallel, \rm e} = \Phi_{\parallel, \rm i}$ ($Z=1$). In the bottom plot, the dashed lines are too clustered to be labelled individually, but the values of $\gamma$ are distributed in the same order.}
\label{fig-sehtc}
\end{figure}
\begin{figure} 
\centering
\includegraphics[width=0.5\textwidth]{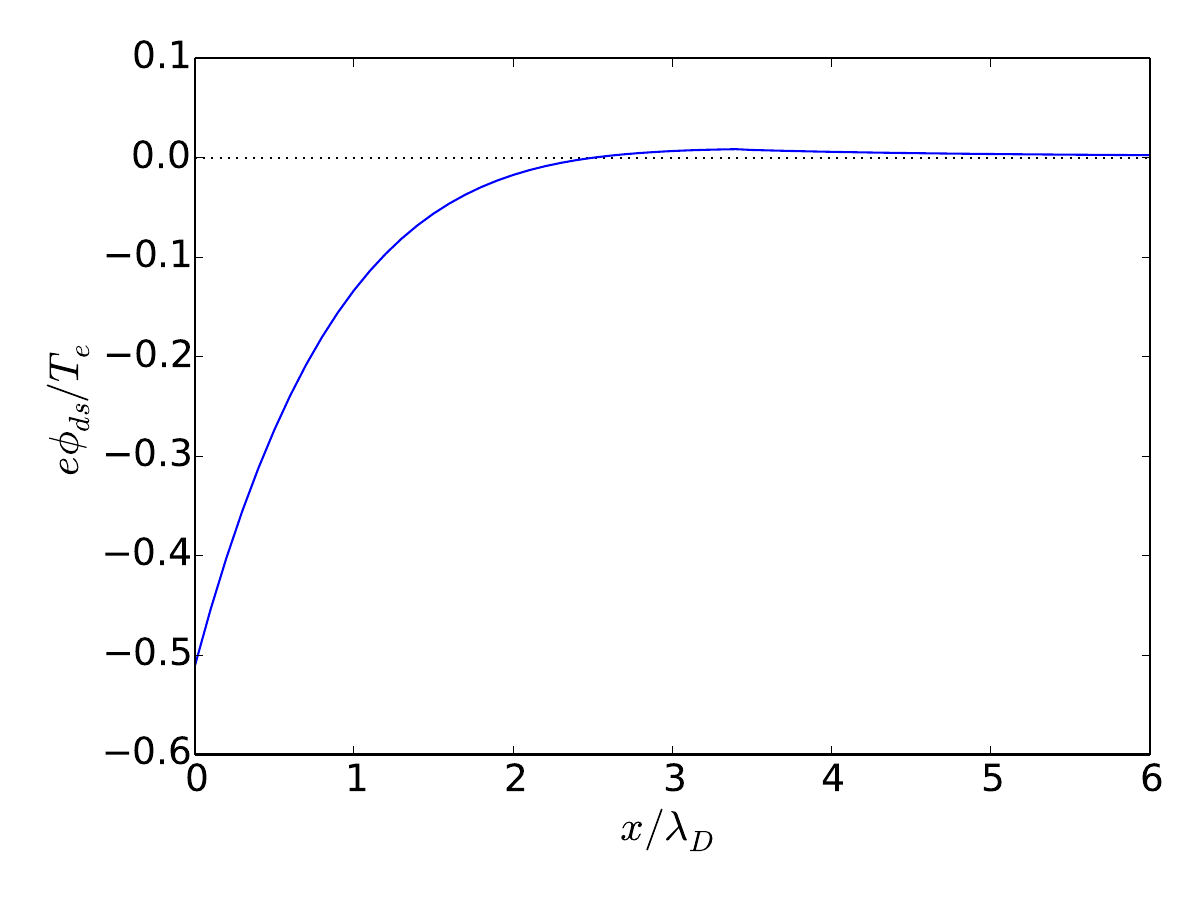} 
\caption{Non-converged profile of the electrostatic potential in the Debye sheath for $\alpha = 2.2^{\circ}$, $\gamma = 1$ and $M = 3600$. The angle is below critical ($\alpha_{\star} = 2.4^{\circ}$), and the iteration has reached a non-monotonic guess to the solution of the electrostatic potential, with $\phi_{\rm ds} >0$ at $x/\lambda_{\rm D} \approx 3$. The iteration is halted because the calculation of the ion density assumes a monotonic potential profile.}
\label{fig-nonmono}
\end{figure}
\begin{figure} 
\centering
\includegraphics[width=0.8\textwidth]{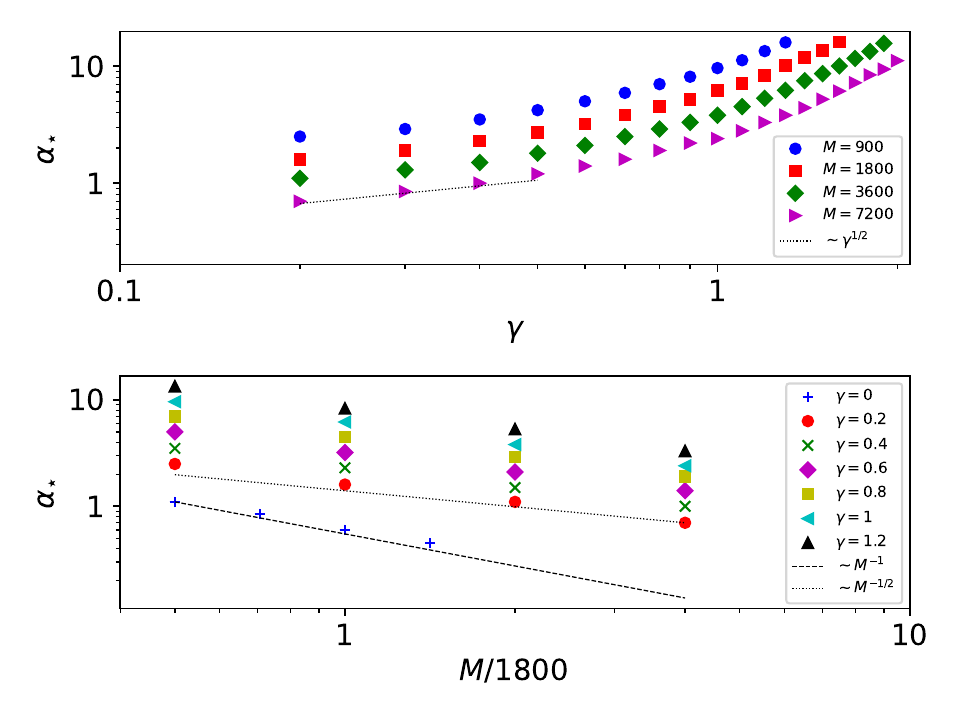} 
\caption{The critical angle $\alpha_{\star}$ as a function of $\gamma = \rho_{\rm e}/\lambda_{\rm D}$ and $M = m_{\rm i} / m_{\rm e}$ for $\tau = T_{\rm i} / (ZT_{\rm e}) = 1$. No results are shown for angles below $0.45^{\circ}$, where the numerical scheme does not work reliably.}
\label{fig-alphacrit}
\end{figure}

Numerical profiles of electrostatic potential and densities in the magnetic presheath, Debye sheath, and magnetised sheath obtained for $M = 3600$, $Z=1$ and $\alpha = 2.5^{\circ}$ and two different values of $\gamma$ are shown in figure~\ref{fig-phisol1} ($\gamma = 0.7$) and figure~\ref{fig-phisol2} ($\gamma = 0.3$).
The bottom figures show, for the magnetic presheath, the ion density and the electron density, and, for the Debye sheath, the ion density, the electron density, the charge density and the normalised Laplacian of the electrostatic potential.
It can be seen that the magnetic presheath is quasineutral, and that in the Debye sheath the normalised Laplacian of the electrostatic potential is equal to the charge density, such that the potential profiles correspond to a good numerical solution.
In figure~\ref{fig-phiamb} we plot the values of the potential drop across the magnetised sheath, magnetic presheath and Debye sheath as a function of magnetic field angle $\alpha$ for $M = 3600$ and for different values of $\gamma$.
For $\alpha \approx 5^{\circ}$, the ambipolar potential drop is larger by $15-20\%$ for $\gamma = 1$ compared to $\gamma = 0$.

By reflecting most of the electrons, the magnetised sheath controls the electron heat flux reaching the wall.
The presence of finite electron orbits in the Debye sheath, often neglected in logical-sheath or conducting-wall boundary conditions, can impact the electron heat and particle fluxes reaching the sheath entrance, and thus ultimately impact the kinetic and fluid boundary conditions.
Indeed, these effects mean that the parallel velocity cutoff is different for different magnetic moments, as was plotted in figure~\ref{fig-vcut} (section~\ref{subsubsec-reflection}).
To illustrate this, in figure~\ref{fig-sehtc} we plot the electron particle and heat flux at the sheath entrance as a function of magnetic field angle and magnetic field strength, i.e. $\gamma$, for two cases: a conducting wall in which $\phi_{\rm w} = -3T_{\rm e} / e$, and a logical sheath in which $J_x = 0$.
Since the logical sheath ensures that the electron flux always equals the ion flux, the heat flux exhibits a smaller variation.
For the conducting sheath, however, local particle and especially heat fluxes can vary significantly from the value obtained by neglecting the electron Larmor orbits and setting $\gamma = 0$.

\subsection{Critical parameters with finite electron gyro-orbits}

The numerical scheme is, in its present form, not designed to account for the possibility of a non-monotonic or a purely ion-repelling monotonic magnetised sheath potential profile.
However, there are parameters for which the numerical iteration reaches a non-converged solution guess for the electrostatic potential in the Debye sheath which is non-monotonic, having a local maximum that satisfies $\phi_{\text{ds}, j}^N > 0$ that typically occurs at large values of $x_j/\lambda_{\rm D}$, as exemplified by figure~\ref{fig-nonmono}.
Such an electrostatic potential ``bump'' in the Debye sheath is a potential barrier for the ions and would reflect some of them within the Debye sheath back into the magnetic presheath.
This possibility, which would significantly alter the nature of the magnetised sheath and thus the analytical predictions of section~\ref{sec-analytical}, is not presently considered in our numerical scheme.
Hence, the numerical code is forced to exit when encountering such non-monotonic potential profiles.
These situations occur when the numerical scheme searches for a solution in which $\phi_{\rm w}$, either imposed directly or via ambipolarity, takes a value that is not negative enough to sustain a monotonic solution throughout the magnetised sheath for a given magnetic field angle (see discussion in section~\ref{subsec-analytical-critical}). 

In section~\ref{subsec-analytical-critical} we saw that a monotonic potential profile across a magnetised sheath with a given value of $\phi_{\rm w}$ or $J_x$ can only exist for magnetic field angles larger than a critical angle $\alpha_{\star}$.
We present results for the ambipolar critical angles obtained for $J_x = 0$, for the set of mass ratios $M \in \{ 900, 1800, 3600, 7200 \}$ and for normalized electron gyroradii (or normalized inverse magnetic field strengths) $\gamma \in \{0.0, [0.3, 1.2] \}$.
These results were obtained by using the velocity distributions fixed in section~\ref{subsec-num-bc}, although the critical angles would naturally depend on the specific velocity distributions of particles entering the magnetised sheath.
Note that some of the critical angles exceed the range of applicability of the expansion in $\alpha \ll 1$, exploited in section~\ref{subsec-IMPEDS}, which was argued to be valid for $\alpha \lesssim 5^{\circ}$ \cite{Cohen-Ryutov-1998, Geraldini-2019}, and so the values larger than $5^{\circ}$ could be inaccurate.
We chose to run the numerical solver only for magnetic field angles which are multiples of $0.1^{\circ}$ for $\alpha > 1^{\circ}$ and multiples of $0.05^{\circ}$ for $\alpha \leqslant 1^{\circ}$, and for values of $\gamma$ which are multiples of $0.1$.
A critical angle of $\alpha_{\star}$ implies that the iteration converges to a numerical solution at that angle, but does not converge at the angle $\alpha_{\star} - 0.1^{\circ}$ or $\alpha_{\star} - 0.05^{\circ}$.
The critical angles in an ambipolar magnetised sheath are shown in figure~\ref{fig-alphacrit}, and are seen to be significantly enhanced by the inclusion of finite electron gyro-orbit effects (finite $\gamma$) in the Debye sheath, consistent with the analytical predictions of section~\ref{subsec-analytical-critical}.
This is a further argument in support of the viewpoint that the behaviour of magnetised sheaths at grazing magnetic field incidence can only be accurately predicted within a kinetic framework. 

The ambipolar scaling $\alpha_{\star} \sim \gamma^{1/2} M^{-1/2}$ for the critical angles at small values of $\gamma$, derived in section~\ref{subsec-analytical-critical}, is supported by the numerical results shown in figure~\ref{fig-alphacrit}.
Our numerical results indicate that for $\gamma \gtrsim 1$ the critical angle $\alpha_{\star}$ continues increasing even faster than $\gamma^{1/2}$ with $\gamma$.
In fusion devices, where $\gamma \approx 0.1$ near the plasma-wall boundary (see the end of section~\ref{subsubsec-num-ds}) and $M = 3600$, the numerically predicted critical angle is very close to being exactly $1^{\circ}$, a value which coincidentally confirms the basic estimate made by Stangeby \cite{Stangeby-2012}. 
The angle between the magnetic field and divertor targets is usually greater than the angle $2.5^{\circ}$ expected for ITER, and is thus larger than the critical angle we predict.
However, the limited configurations adopted during the start-up phase of the operation of fusion devices, as well as the island divertors in stellarators, have locations where the incidence angle of the magnetic field on the wall crosses zero and is very small over a finite region.
Therefore, the investigation of what happens below the critical angles remains nonetheless important.

\section{Conclusions} \label{sec-conc}

This work significantly advances the physical understanding and numerical simulation of magnetised plasma sheaths.
We highlight two key novel contributions:
\begin{itemize}
\item a numerical scheme obtaining directly the kinetic steady-state solution, and associated characteristics, of the magnetised plasma sheath (combined magnetic presheath and Debye sheath) in $\sim 1$ minute on a laptop;
\item numerical characterisation, supported by analytical predictions, of the magnetic field angles below which a monotonic potential profile across the magnetised sheath cannot exist.  
\end{itemize}

We have presented fully kinetic simulations obtaining numerical solutions of a steady-state magnetised sheath in the asymptotic limit where its thickness, several ion sound gyroradii $\rho_{\rm S}$, is infinitely smaller than the bulk plasma length scale $L$, $\rho_{\rm S} / L \rightarrow \infty$, thus exploiting the ordering $\rho_{\rm S} / L \ll 1$.
The method of solution also exploits the smallness of the ratio of Debye length to ion sound gyroradius, $\lambda_{\rm D} / \rho_{\rm S} \ll 1$, to solve the magnetised sheath in the limit $\lambda_{\rm D} / \rho_{\rm S} \rightarrow 0$.
The magnetised sheath is thus split into two distinct regions: the Debye sheath where $x \sim \lambda_{\rm D}$, and the magnetic presheath, where $x \sim \rho_{\rm S}$.
The ordering $\alpha \ll 1$ is used to obtain expressions for the ion density in the magnetic presheath and the electron density in the Debye sheath for a given potential profile.
The non-circular finite electron gyro-orbits present in the Debye sheath when $\gamma = \rho_{\rm e} / \lambda_{\rm D} \sim 1$ are retained.
The parallel velocity cutoff below which an electron is reflected, and above which it reaches the wall, is not constant as is usually assumed in conducting boundary conditions \cite{Shi-2017} (which were adapted from the logical boundary conditions of \cite{Parker-1993}), as shown in figure~\ref{fig-vcut}.
The dependence of the parallel velocity cutoff on magnetic moment, also seen in recent PIC simulations \cite{Castillo-2024}, can be computed from properties of the electron gyro-orbits for any given potential profile in the Debye sheath, and then used in the magnetic presheath to calculate the correct number of reflected electrons. 

To solve for the self-consistent potential profiles both in the magnetic presheath and in the Debye sheath, we have generalised the iterative scheme which solves quasineutrality in the magnetic presheath of references \cite{Geraldini-2018, Ewart-2021} to also solve Poisson's equation in the Debye sheath via equations (\ref{iterative-general})-(\ref{tentativetopermanent-general}).
An additional (third) iteration (\ref{phiW-impose})-(\ref{phiW-tentativetopermanent}) is carried out to calculate the total potential drop in the Debye sheath for a given current density at the wall (e.g. the ambipolar potential drop $\phi_{\rm w, fl}$ corresponding to zero current density, $J_x = 0$).
We obtained numerical solutions of the magnetised sheath by using the velocity distributions of particles entering the magnetised sheath specified in section~\ref{subsec-num-bc}.
In our numerical scheme, where the magnetic presheath and Debye sheath are treated as distinct regions, the particles leaving the magnetic presheath in the direction of the wall enter the Debye sheath, and vice versa.
This is expressed by imposing that the ion and electron distribution functions at the wall on the magnetic presheath scale be equal to those infinitely far from the wall on the Debye sheath scale. 
However, we remark that the iterative scheme for the magnetic presheath (\ref{tentative-mp}) is just the limit $\lambda_{\rm D} \rightarrow 0$ of the one in the Debye sheath (\ref{iterative-Poisson}), which opens the possibility of obtaining steady state solutions resolving both scales simultaneously (instead of separately).
Examples of converged potential profiles in the Debye sheath and magnetic presheath, obtained using the iterative scheme, are shown in figures~\ref{fig-phisol1} and \ref{fig-phisol2}. 
The ambipolar ($J_x = 0$) potential drop $\phi_{\rm w, fl}$ across the magnetised sheath (combined potential drop of Debye sheath and magnetic presheath) is shown in figure~\ref{fig-phiamb} as a function of different values of $\gamma$ and $\alpha$.
More generally, a useful result of the iterative solution of the magnetised sheath is the relation---nontrivial for finite $\gamma$---between the magnetised sheath potential drop and the outgoing electron fluxes, which depend on the electron distribution function and the self-consistent parallel velocity cutoff function (figure~\ref{fig-vcut}). 
To illustrate this, the electron particle and energy flux are plotted in figure~\ref{fig-sehtc} for different magnetic field angles and strengths for two different boundary conditions: an imposed floating-wall potential (fixing zero current) and an imposed fixed potential drop of $3T_{\rm e} / e$.

One of the main results of this work is the calculation of a small critical magnetic field angle below which a self-consistent electron-repelling monotonic potential profile across the magnetised sheath cannot be found.
For $\gamma = \rho_{\rm e} / \lambda_{\rm D} = 0$ this critical angle is connected to the disappearance of the Debye sheath.
The fact that the potential drop across the Debye sheath tends to vanish at small angles was realised in the earliest studies of the magnetic presheath \cite{Chodura-1982}. 
It was a natural consequence of the result that the combined potential drop across the magnetic presheath and Debye sheath is not strongly dependent on the magnetic field angle, while the potential drop across the magnetic presheath increases with angle.
Later, many authors employed the ordering $\alpha \gg M^{-1/2}$ to explicitly rule out this regime. 
Stangeby \cite{Stangeby-2012} proposed that there is a first critical magnetic field angle, denoted here $\alpha_{\star 1}$ and estimated to be $ \approx 3^{\circ}$ for $M=3600$, below which the Debye sheath disappears. 
The Bohm condition also disappears, such that the magnetic presheath potential drop can be equal to the total required ambipolar potential drop.
At an even smaller angle, $\alpha_{\star 2} \approx M^{-1/2} \approx 1^{\circ} < \alpha_{\star 1} $, Stangeby claimed that the Debye sheath should become ion repelling, as the ion gyration tends to make ions reach the wall faster than electrons.
Although not explicitly stated by Stangeby, both $\alpha_{\star 1}$ and $\alpha_{\star 1}$ are of order $\sim M^{-1/2}$ and differ by order unity constants.
These results were complemented by cautious statements that they could differ substantially if the fluid-ion and adiabatic-electron assumptions employed to derive these critical angles were not satisfied.
By including kinetic ions and using a kinetic electron model which neglects finite electron gyro-orbits (i.e. taking $\gamma = 0$), Ewart \emph{et al.} \cite{Ewart-2021} found the critical magnetic field angle for sheath disappearance to be much smaller, $\alpha_{\star 1} \sim M^{-1}$, due to a \emph{gradual weakening} of the Bohm condition, as in (\ref{Bohmint-ordered}).
This weakening allows the flow velocity of ions at the entrance of the Debye sheath to become smaller when the potential drop across the Debye sheath is small (as was also found by Loizu \cite{Loizu-2011} for unmagnetised sheaths next to biased walls).
Ewart \emph{et al.} \cite{Ewart-2021} found that the small ion velocity at the entrance of the Debye sheath is determined by intrinsically kinetic physics. The ion gyration (second term in the square root in (\ref{vx-dse})) sets a lower limit to the ion flow velocity into the Debye sheath even when the sheath collapses ($\phi_{\rm ds}(0) = 0$) and the Bohm condition disappears.
The intrinsic ion flow towards the target is not the velocity of the ion within its orbit, $\sim v_{\rm t,i}$, but the projection of the velocity of the ion within its orbit when it hits the wall, $\alpha^{1/2} v_{\rm t,i}$ \cite{Geraldini-2021}.
Hence, in the Debye sheath for angles $\alpha \lesssim \alpha_{\star 2} \sim M^{-1}$ the ions reach the wall intrinsically faster than electrons, whose motion is constrained along the magnetic field such that their velocity component to the wall is $\sim \alpha v_{\rm t,e}$.
The critical angle below which the Debye sheath disappears, $\alpha_{\star 1}$, and the critical angle below which a monotonic electron-repelling sheath cannot exist, $\alpha_{\star 2}$, coincide and are much smaller than predicted by Stangeby, $\alpha_{\star 1} = \alpha_{\star 2} \approx 0.3^{\circ}$ for $M = 3600$.

Here, it is found that the Debye sheath cannot disappear when finite electron gyro-orbits are included such that $\gamma = \rho_{\rm e} / \lambda_{\rm D} \neq 0$.
Hence, the critical angle $\alpha_{1 \star}$ does not exist at all.
Yet, there is still a critical angle $\alpha_{\star} = \alpha_{2 \star}$ below which a monotonic and electron-repelling Debye sheath solution cannot exist.
Considering $\tau \sim 1$ and $M^{-1} \lesssim \gamma \lesssim 1$, the critical angle is found to follow the scaling $\alpha_{\star} \sim \gamma^{1/2} M^{-1/2}$, shown to be consistent with numerical results (figure~\ref{fig-alphacrit}).
The critical angle is numerically found to further increase with $\gamma$ for $\gamma > 1$.
The enhancement of the critical angle for finite $\gamma$ can be understood by realising that a collapsed Debye sheath (with a flat potential profile, and therefore no electric field) is inconsistent with the presence of finite electron gyro-orbits.
The electron density has an intrinsic (i.e., not requiring an electric field) drop near the wall due to the electron gyro-orbits intersecting the wall, causing a charge separation which must be balanced by a gradient of the electric field.
Therefore, the potential drop across the Debye sheath cannot be reduced to zero while preserving monotonicity.
Below the critical angle, a steady-state solution would require the electrostatic potential to reverse at least partially in the Debye sheath or the magnetic presheath, thus becoming non-monotonic in the magnetised sheath as a whole.
A non-monotonic or a non-steady-state solution could significantly affect the sheath structure and its characteristics, and consequently also the ion velocity distribution and the electron energy flux reaching the target, as well as the boundary conditions at the magnetised sheath entrance.

The code developed and used in this work builds onto, and extends, the code developed in references \cite{Geraldini-2018} and \cite{Ewart-2021} for the magnetic presheath.
The code runs in \textit{seconds to hundreds of seconds on a laptop}, depending on both physical parameters and numerical (resolution) parameters.
Although this is already very fast, we plan to significantly reduce the upper bound of the computational time by developing an accurate analytical model for the electron parallel velocity cutoff for small finite values of $\gamma$, for which the code takes longest, and more refined initial electrostatic potential guesses.
The full code has been made available at \url{https://github.com/alessandrogeraldini/GYRAZE}.

~

This work has been carried out within the framework of the EUROfusion Consortium, partially funded by the European Union via the Euratom Research and Training Programme (Grant Agreement No 101052200 — EUROfusion). The Swiss contribution to this work has been funded by the Swiss State Secretariat for Education, Research and Innovation (SERI). Views and opinions expressed are however those of the author(s) only and do not necessarily reflect those of the European Union, the European Commission or SERI. Neither the European Union nor the European Commission nor SERI can be held responsible for them. 
This work was supported by the U.S. Department of Energy under contract number DE-AC02-09CH11466. The United States Government retains a non-exclusive, paid-up, irrevocable, world-wide license to publish or reproduce the published form of this manuscript, or allow others to do so, for United States Government purposes.
RJE was supported by a UK EPSRC studentship.

\appendix

\section{Gyrokinetic particle density far from target} \label{app-densfinorb}

We proceed to analyse the ion density near the entrance of the magnetic presheath and the electron density near the entrance of the Debye sheath, where the electrostatic potential drop is still small, $\hat \phi_R = e\phi_{R} / T_{\rm e} \ll 1$ and $\kappa_R = \rho_{s} \left| \phi'_{R} /\phi_R \right| \ll 1$ (recall that $\rho_{s} = v_{\rm t,s} / |\Omega_s|$ is the thermal ion Larmor radius of species $s$).
By expanding the potential $\phi(x)$ about $\phi (\bar x )$ in equation (\ref{mugk-def}) for the adiabatic invariant, and recalling that the effective potential $\chi_R$ is defined in (\ref{chis}), we obtain
\begin{align}
\mu \simeq & \frac{1}{\pi} \int_{x_-}^{x_+} dx \left( 2U_{\perp} - \Omega_{s}^2 (x - \bar{x})^2 - \frac{2\Omega_{ s} \phi_R(\bar{x}) }{B}   \right. 
\\ &  \left.   - \frac{2\Omega_{s} \phi'_{R}(\bar{x})  }{B} \left(x - \bar{x} \right) -  \frac{\Omega_{s} \phi''_{R}(\bar{x})  }{B} (x - \bar{x})^2 \right)^{1/2} \text{.}
\end{align}
Completing the square gives
\begin{align} \label{mus-int}
\mu \simeq & \frac{1}{\pi} \int_{x_-}^{x_+} dx \left[ 2U_{\perp} - \frac{2\Omega_{s} \phi_{R}(\bar{x})  }{B} + \frac{[\phi_{R}'(\bar{x})]^2}{B^2 \left[ 1+\frac{\phi''_{R}(\bar{x})}{\Omega_s B} \right]}   \right. \nonumber \\
& \left. -  \left[ \Omega_{s}^2 + \frac{\Omega_{s} \phi_{R}''(\bar x)}{B} \right] \left( x - \bar{x} + \frac{ \phi_{R}'(\bar{x})}{B\left[ \Omega_{s} + \frac{ \phi_{R}''(\bar x)}{B} \right] } \right)^2   \right]^{1/2} \rm .
\end{align} 
Evaluating the integral in (\ref{mus-int}) gives
\begin{align} \label{mus-corr1}
\mu \simeq  & \frac{ 1 }{|\Omega_{s}|}  \left( U_{\perp} - \frac{ \Omega_{ s} \phi_{R} (\bar{x})}{B} + \frac{1}{2}  \frac{[\phi_{R}'(\bar{x})]^2}{B^2 \left[ 1+\frac{\phi''_{R}(\bar{x})}{\Omega_s B} \right]}  \right) \left( 1 + \frac{\phi_{R}''(\bar{x})}{\Omega_{s} B} \right)^{-1/2}  \rm .
\end{align}
The effect on $\mu$ of the electric field $\phi_{R}'(x)$ is quadratic and of order $\hat{\phi}^2 \kappa^2$.
Since we retain only terms that are quadratic in $\hat{\phi}$ but lowest order in $\kappa$, we neglect the term $[\phi_{R}'(\bar x)]^2$ in (\ref{mus-corr1}).
Expanding (\ref{mus-corr1}) in $\hat \phi_R \ll 1$ thus gives 
\begin{align} \label{mus-corr}
\mu \simeq  \frac{ 1 }{|\Omega_{s}|} \left(  U_{\perp}  - \frac{ \Omega_{s} \phi_{\rm R} (\bar{x})}{B}  -  \frac{\phi_{R}''(\bar{x})}{2\Omega_{s} B} U_{\perp}  \right)   \rm .
\end{align}

In the expression for the particle density (\ref{n-finorb}), the second term owing to the contribution of open orbits is negligible far from the target.
Hence, we keep only the first term in equation (\ref{n-finorb}) and use (\ref{mus-corr}) to obtain
\begin{align} \label{nsclosed-app1}
n_{\text{s},R} \left(x \right) \simeq  & \sum_{\sigma_{\parallel} = \pm 1} \int_{-\infty}^{\infty} |\Omega_{s}| d\bar{x}  
\int_{ \frac{\Omega_{s}\phi_{R} (x)}{B} + \frac{\Omega_{s}^2}{2} (x - \bar{x})^2 }^{\infty}  \frac{ 2dU_{\perp}}{\sqrt{2U_{\perp} - \frac{2\Omega_{s} \phi_{R}(x)}{B}- \Omega_{s}^2 (x - \bar{x})^2 }} \times
\nonumber \\
& \int_{U_{\perp}}^{\infty} \frac{ dU_{R}  }{\sqrt{2\left( U_{R} - U_{\perp} \right)}}   \bar{F}_{s,R} \left( \frac{U_{\perp}}{|\Omega_{s}|} - \frac{ \Omega_{s} \phi_{R} (\bar{x})}{|\Omega_{s}|B}  -  \frac{\phi_{R}''(\bar{x})}{2 \Omega_s B}  \frac{ U_{\perp} }{|\Omega_{s}|}  , U_{R}, \sigma_{\parallel} \right)
 \text{.}
\end{align}
To be precise, the lower limit of integration in $\bar{x}$ of (\ref{nsclosed-app1}) should be $\bar{x}_{\text{m}}(x) \simeq x/2$, obtained from (\ref{xbarm-general}). 
However, the integrand is small for $\bar x \lesssim x/2$ due to the constraint $U_{\perp} > \Omega_{\rm s}^2 (\bar x - x)^2 / 2 \gtrsim \Omega_{\rm s}^2 x^2 / 4$ arising from the lower limit of the integral in $U_{\perp}$, and the fact that the distribution function is assumed to be exponentially small at large energies.
Hence, the lower limit of the integration in $\bar x$ has been extended to $-\infty$ without it causing a significant change in the integral.
Changing variables from $U_{\perp}$ to $U_{\perp}' = U_{\perp} - \Omega_{s} \phi_{R}(x) / B$ and from $U_{R}$ to $U_{R}' = U_{R} - \Omega_{s} \phi_{R}(x) / B$, we obtain
\begin{align} \label{nsclosed-app2}
n_{s,R} \left(x \right) \simeq  & \sum_{\sigma_{\parallel} = \pm 1}  \int_{-\infty}^{\infty} |\Omega_{s}| d\bar{x}  
\int_{ \frac{\Omega_{s}^2}{2} (x - \bar{x})^2 }^{\infty}  \frac{ 2dU'_{\perp}}{\sqrt{2U'_{\perp} - \Omega_{s}^2 (x - \bar{x})^2 }} \int_{U'_{\perp}}^{\infty} \frac{ dU'_{R}  }{\sqrt{2\left( U'_{R} - U'_{\perp} \right)}}   \times
\nonumber \\
&  \bar{F}_{s, R} \left( \frac{U'_{\perp}}{|\Omega_{s}|} - \frac{ \Omega_s }{|\Omega_s|B} \left( \phi_{R} (\bar{x}) - \phi_{R} (x) \right) -  \frac{\phi_{R}''(\bar{x})}{2\Omega_{s} B}  \frac{ U'_{\perp} }{|\Omega_{s}|}  , U'_R +  \frac{\Omega_{s} \phi_{R} (x)}{B}, \sigma_{\parallel} \right) \times \nonumber \\
& \Theta\left( U_{R}' - U'_{\perp} + \frac{ \Omega_{s} }{B}  \phi_{R} (\bar{x}) + \frac{\phi_{R}''(\bar{x})}{2\Omega_{s} B} U'_{\perp}  \right)
 \text{.}
\end{align}
The distribution function $\bar F_{s, R}$ appearing in (\ref{nsclosed-app2}) can be expanded about $U'_{\perp} / |\Omega_{s}| $ and $U_{R}'$\footnote{This assumes the distribution function to be Taylor expandable in energy, following Riemann \cite{Riemann-review}. Since this assumption has been contentious \cite{Baalrud-Hegna-2011-Bohm, Riemann-2012-comment, Baalrud-Hegna-2012-reply}, a more general expansion of the ion density near the magnetic presheath entrance has been carried out in the appendix of reference \cite{Geraldini-2024-Chodura}, where the distribution function at the magnetic presheath entrance is expanded in $v_{\parallel}$ instead of $v_{\parallel}^2$ (in addition, the more general orderings $\alpha \sim 1$ and $\kappa_{\rm mp} \sim 1$ are taken, and spatial fluctuations tangential to the wall and perpendicular to $\vec{B}$ are kept). In reference \cite{Geraldini-2024-Bohm}, an expansion of the ion density near the unmagnetised Debye sheath entrance which only assumes that the ion distribution function may be expanded in \textit{arbitrary} powers (even fractional) of velocity is performed to put the kinetic Bohm criterion on firmer ground following the criticisms in \cite{Baalrud-Hegna-2011-Bohm, Baalrud-Hegna-2012-reply}.}.
Before doing so, it is important to recall that $\bar F_{s, R}(\mu, U_{\rm R}, \sigma_{\parallel}) = 0$ for $U_{R} < |\Omega_{\rm s}| \mu$, which are values that cannot correspond to particles in the system, since such particles would have entered with $v_z^2 = 2\left( U_R - |\Omega_{s}|\mu \right) < 0$ at $x/l_R \rightarrow \infty$. 
In the new integration variables of (\ref{nsclosed-app2}), the distribution function is zero for $U_{R}' < U'_{\perp} - \Omega_{s} \phi_{R} (\bar{x}) / B -  U'_{\perp}  \phi_{R}''(\bar{x}) / (2\Omega_{s} B )  $.
In view of the imminent expansion, and in order to make explicit the fact that $\bar F_{s, R}(\mu, U_R, \sigma_{\parallel}) = 0$ for $U_R < |\Omega_{\rm s}|\mu$, the Heaviside step function (defined in (\ref{Heaviside-def})) $\Theta\left( U_{R}' - U'_{\perp} + \Omega_{s} \phi_{R} (\bar{x}) / B + U'_{\perp} \phi_{R}''(\bar{x}) / (2\Omega_{s} B)  \right)$ has been introduced in the integrand.
We proceed by expanding the distribution function $\bar F_{s, R}$ in (\ref{nsclosed-app2}) about $\mu' = U'_{\perp} / |\Omega_{s}| $ and $U_{R}'$ and exchanging the order of integration such that the integral in $\bar{x}$ is performed first with the modified integration limits $\bar x_{\pm} = x \pm \sqrt{2\mu'/|\Omega_{\rm s}|}$ (obtained by requiring that $\sqrt{2\left( \Omega_{\rm s} \mu - \Omega^2 (x-\bar x)^2 / 2 \right)}$ be real). 
For ions in the magnetic presheath, the density at large $x / \rho_{\rm i}$ thus becomes, upon using also (\ref{Fi-mp}),
\begin{align} \label{niclosed-far-app}
n_{\rm i,mp} \left(x \right) & \simeq  \int_{ 0 }^{\infty}  \Omega_{\rm i} d\mu'  \int_{\Omega_{\rm i} \mu'}^{\infty} \frac{ dU_{\rm mp}'  }{\sqrt{2\left( U_{\rm mp}' - \Omega_{\rm i} \mu' \right)}}   \int_{\bar{x}_{-}}^{\bar{x}_+} \frac{ 2 \Omega_{\rm i} d\bar{x} }{\sqrt{2 \Omega_{\rm i} \mu' - \Omega_{\rm i}^2 (x - \bar{x})^2 }}  \times
\nonumber \\
& \left[  F_{\rm i \infty} \left( \mu', U_R'\right) - \left( \frac{ \phi_{\rm  mp} (\bar{x})}{B} -  \frac{ \phi_{\rm mp} (x)}{B} +  \frac{\phi_{\rm mp}''(\bar{x}) \mu' }{2\Omega_{\rm i} B}  \right)  \partial_{\mu}  F_{\rm i\infty} \left( \mu', U_{\rm mp}'\right) \phantom{\left(\frac{1}{1} \right)^2} \right.  \nonumber \\  
& \left.  + \frac{\Omega_{\rm i} \phi_{\rm mp} (x)}{B} \partial_{U_{\rm mp}}  F_{\rm i \infty} \left( \mu', U_{\rm mp}', \right)  + \frac{1}{2} \left( \frac{\Omega_{\rm i} \phi_{\rm mp} (x)}{B} \right)^2 \partial_{U_{\rm mp}}^2  F_{\rm i \infty} \left( \mu', U_{\rm mp}' \right)    \right]
\nonumber \\
& - \int_{ 0 }^{\infty}  \Omega_{\rm i} d\mu'   \int_{\bar{x}_{-}}^{\bar{x}_+} 2 \Omega_{\rm i} d\bar{x} \frac{  \sqrt{- 2\Omega_{\rm i} \phi_{\rm mp} (\bar{x}) / B -  \mu' \phi_{\rm mp}''(\bar{x}) / B }  }{\sqrt{2 \Omega_{\rm i} \mu' - \Omega_{\rm i}^2 (x - \bar{x})^2  }} \times \nonumber \\
&  \left[ F_{\rm i \infty} \left( \mu', \Omega_{\rm i} \mu' \right)  \phantom{\frac{1}{1}}   - \frac{1}{3} \left( - \frac{ 2\Omega_{\rm i} \phi_{\rm mp} (\bar{x})  }{B}  \right) \partial_{U_{\rm mp}}  F_{\rm i \infty} \left( \mu', |\Omega_{\rm i}| \mu' \right)     \right]
 \text{,}
\end{align}
where terms of order $O\left( \kappa^2 \hat \phi^{3/2} n_{\rm i,mp}(\infty), \hat \phi^{5/2} n_{\rm i,mp}(\infty) \right)$ have been neglected.
For electrons in the Debye sheath at large $x/\rho_{\rm e}$, the density correspondingly becomes
\begin{align} \label{neclosed-far-app}
n_{\rm e, ds} & \left(x \right)  \simeq   \sum_{\sigma_{\parallel} = \pm 1} \int_{ 0 }^{\infty}  |\Omega_{\rm e}| d\mu'  \int_{\Omega_{\rm e} \mu'}^{\infty} \frac{ dU_{\rm ds}'  }{\sqrt{2\left( U_{\rm ds}' - |\Omega_{\rm e}| \mu' \right)}}   \int_{\bar{x}_{-}}^{\bar{x}_+} \frac{ 2 |\Omega_{\rm e}| d\bar{x} }{\sqrt{2 |\Omega_{\rm e}| \mu' - \Omega_{\rm e}^2 (x - \bar{x})^2 }}  \times
\nonumber \\
& \left[  \bar{F}_{\rm e, ds} \left( \mu', U_{\rm ds}', \sigma_{\parallel} \right) - \frac{\Omega_{\rm e}}{|\Omega_{\rm e}|} \left( \frac{ \phi_{\rm ds} (\bar{x})}{B} -  \frac{ \phi_{\rm ds} (x)}{B} +  \frac{\phi_{\rm ds}''(\bar{x}) \mu' }{2|\Omega_{\rm e}| B}  \right)  \partial_{\mu}  \bar{F}_{\rm e,ds} \left( \mu', U_{\rm ds}' , \sigma_{\parallel}\right)   \right.  \nonumber \\  
& \left.  + \frac{\Omega_{\rm e} \phi_{\rm ds} (x)}{B} \partial_{U_{\rm ds}}  \bar{F}_{\rm e, ds} \left( \mu', U_{\rm ds}', \sigma_{\parallel} \right)  + \frac{1}{2} \left( \frac{|\Omega_{\rm e}| \phi_{\rm ds} (x)}{B} \right)^2 \partial_{U_{\rm ds}}^2  \bar{F}_{\rm e, ds} \left( \mu', U_{\rm ds}', \sigma_{\parallel} \right)    \right]
 \text{.}
\end{align}
The second set of terms in the ion density expansion (\ref{niclosed-far-app}) comes from subtracting off the portion of the integral in the domain $0 \leqslant U_{\rm mp}' - \Omega_{\rm i} \mu'  <  -\Omega_{\rm i} \phi_{\rm mp} (\bar{x}) / B - \mu'   \phi_{\rm mp}''(\bar{x}) / (2 B) $ which has been included in the integrals of the first set of terms in (\ref{niclosed-far-app}) for convenience, but which should not have been included due to the Heaviside function in (\ref{nsclosed-app2}).
Physically, ions with $v_z < \sqrt{ - 2\Omega_{\rm i} \phi_{\rm mp} (\bar{x}) / B - \mu'   \phi_{\rm mp}''(\bar{x}) / B  }$, which should be absent because all ions have been accelerated to larger velocities, have been artificially included in the first set of terms of (\ref{niclosed-far-app}) and must hence be subtracted.
To calculate the subtracted piece, the distribution function has been expanded about $U_{\rm mp}' = \Omega_{\rm i} \mu' - \Omega_{\rm i} \phi_{\rm mp} (\bar x) / B - \phi_{\rm mp}''(\bar x) \mu / (2B)$ (about $v_z^2 = 0$) and the integrals in $U_{\rm mp}'$ have thus been carried out explicitly (carefully accounting for the two different contributions to the factor of $\frac{1}{3}$ in the last term).
The additional terms present in the ion density (\ref{niclosed-far-app}) are absent in the electron density (\ref{neclosed-far-app}) because the Heaviside function is unity over the entire integration domain in (\ref{nsclosed-app2}) for electrons, since a domain in which $0 \leqslant U_{\rm ds}' - |\Omega_{\rm e}| \mu'  <  -\Omega_{\rm \rm e} \phi_{\rm ds} (\bar{x}) / B - \mu'  |\Omega_{\rm e}| \phi_{\rm ds}''(\bar{x}) / (2|\Omega_{\rm e}| B) $ does not exist due to the fact that $\Omega_{\rm e} \phi_{\rm ds} > 0$ ($\phi_{\rm ds}''/\Omega_{\rm e} > 0$ is anyway smaller due to $\kappa_{\rm ds} \ll 1$). 
That is, electrons are being repelled away from the wall in the Debye sheath, such that their $v_z \simeq v_{\parallel}$ always decreases, and so the contribution from electrons with $v_z$ close to zero is physical (coming from electrons that are about to reflect or have just reflected) and need not be subtracted.

Upon inserting $\phi_{R}(\bar{x}) - \phi_{R}(x) \simeq \phi'(x)(\bar{x} - x) + \frac{1}{2} \phi_{R}''(x) (\bar{x} -x)^2 $ and $\phi_R''(\bar x ) \simeq \phi_R'' (x)$ in (\ref{niclosed-far-app}), evaluating the integrals over $\bar{x}$ appearing in (\ref{niclosed-far-app}) and (\ref{neclosed-far-app}) using the results
 \begin{align} \label{id1}
 \int_{\bar{x}_{-}}^{\bar{x}_{+}} \frac{ |\Omega_{s}| d\bar{x}}{\sqrt{2|\Omega_{s}| \mu - \Omega_{s}^2 (x - \bar{x})^2 }} =  \int_{-1}^1 \frac{d\hat x'}{\sqrt{1-\hat x'^2}} = \pi  \rm ,
\end{align}
\begin{align}
 \sqrt{\frac{|\Omega_{\rm s}|}{2\mu}} \int_{\bar{x}_-}^{\bar{x}_+} \frac{  ( \bar{x} - x)  |\Omega_s|  d\bar{x}}{\sqrt{2|\Omega_s| \mu - \Omega_s^2 (x - \bar{x})^2 }} 
=  \int_{-1}^1 \frac{\hat x' d\hat x'}{\sqrt{1-\hat x'^2}} = 0 \rm ,
\end{align}
\begin{align} \label{id3}
 \frac{|\Omega_{\rm s}|}{2\mu} \int_{\bar{x}_-}^{\bar{x}_+} \frac{  ( \bar{x} - x)^2  |\Omega_s|  d\bar{x}}{\sqrt{2|\Omega_s| \mu - \Omega_s^2 (x - \bar{x})^2 }} 
=  \int_{-1}^1 \frac{\hat x'^2 d\hat x'}{\sqrt{1-\hat x'^2}} = \frac{1}{2} \pi   \rm ,
\end{align}
and relabelling the integration variables such that $U_R' \rightarrow U_R$ and $\mu' \rightarrow \mu$, we re-express the ion density as (\ref{niclosed-far}) and the electron density as (\ref{neclosed-dse-ds-1}).

\section{Analytical evaluation of the integral in (\ref{phids-min-gammasmall-exact})} \label{app-integral}

We proceed to prove the result used in (\ref{phids-min-gammasmall-exact}): $I = 3/2$, with $I$ given by
\begin{align} \label{I-def}
I = \int_{0}^{\infty} d\hat{x} \int_{\hat{x}}^{\infty}   \left[ 1 - \eta_{\rm flat}\left( \hat{x}'  \right) \right] d\hat{x}'  \rm .
\end{align}
Inserting $\eta_{\rm flat}$ from (\ref{n-closed-cold}), and using the expression for the error function (\ref{erf}), gives the full expression for $I$:
\begin{align}
I = \int_{0}^{\infty} d\hat{x} \int_{\hat{x}}^{\infty} d\hat{x}'  \left[ 1 - \frac{1}{\pi}  \int_{-\frac{\hat x'}{2}}^{\infty} dw   \int_0^{\sqrt{ 2\hat x' \left( w + \frac{\hat x'}{2} \right) }} du \exp \left(- \frac{w^2+u^2}{2} \right) \right]  \rm .
\end{align}
Integrating by parts the outermost integral with respect to $\hat{x}$ eliminates the integral with respect to $\hat x'$,
\begin{align}
I = \int_{0}^{\infty} d\hat{x} \hat{x}  \left[ 1 - \frac{1}{\pi}  \int_{-\frac{\hat x}{2}}^{\infty} dw   \int_0^{\sqrt{ 2\hat x \left( w + \frac{\hat x}{2} \right) }} du \exp \left(- \frac{w^2+u^2}{2} \right) \right]  \rm .
\end{align}
Integrating by parts again with respect to $\hat{x}$ eliminates the integral with respect to $u$ (the contribution from differentiating with respect to the lower limit of the integral in $w$ vanishes, since the integrand vanishes at its lower limit),
\begin{align}
I = \frac{1}{2\pi} \int_{0}^{\infty} d\hat{x} \hat{x}^2  \int_{-\frac{\hat x}{2}}^{\infty} dw   \exp \left(- \frac{w^2+2 \hat x w + \hat x^2}{2} \right)  \frac{w + \hat x}{\sqrt{ 2\hat x \left( w + \frac{\hat x}{2} \right) }} \rm . 
\end{align}
Changing integration variable to $w' = w + \hat x$ gives
\begin{align}
I = \frac{1}{2\pi} \int_{0}^{\infty} d\hat{x} \hat{x}^2  \int_{\frac{\hat x}{2}}^{\infty} dw   \exp \left(- \frac{w'^2}{2} \right)  \frac{w'}{\sqrt{ 2\hat x \left( w' - \frac{\hat x}{2} \right) }} \rm .
\end{align}
Reversing the order of integration gives
\begin{align}
I = \frac{1}{2\pi} \int_{0}^{\infty} dw' w'  \exp \left(- \frac{w'^2}{2} \right) \int_{0}^{2w'} d\hat{x} \frac{\hat{x}^2 }{\sqrt{ 2\hat x w' - \hat x^2 }} \rm .
\end{align}
Changing integration variable from $\hat x$ to $\hat x' = (\hat{x}-w')/w'$ finally turns the expression into a product of two integrals,
\begin{align} \label{integral-laststep}
I = \frac{1}{2\pi} \int_{0}^{\infty} dw' w'^3  \exp \left(- \frac{w'^2}{2} \right) \int_{-1}^{1} d\hat{x}' \frac{(\hat{x}'+1)^2 }{\sqrt{ 1-\hat x'^2 }} \rm .
\end{align}
Expanding $(\hat{x}'+1)^2 = \hat{x}'^2 + 2 \hat{x}' + 1$ and using (\ref{id1})-(\ref{id3}) 
leads to $\int_{-1}^{1} d\hat{x}' (\hat{x}'+1)^2 / \sqrt{ 1-\hat x'^2 } = 3\pi / 2$.
This result can be substituted into (\ref{integral-laststep}), together with $\int_{0}^{\infty} dw' w'^3  \exp \left(- w'^2 / 2 \right) = 2$, to obtain $I = [1/(2\pi)] \times 2 \times (3\pi/2) = 3/2$, which recovers the numerical prefactor in (\ref{phids-min-gammasmall-exact}).

\section{Electron reflection cutoff model for $\gamma \rightarrow \infty$} \label{app-gammainf}

In the asymptotic limit $\gamma \rightarrow \infty$, the potential drop across the Debye sheath occurs over a scale that is infinitely smaller than the size of the electron orbit, $\lambda_{\rm D} / \rho_{\rm e} \rightarrow 0$. 
The electrostatic potential in the Debye sheath can thus be treated as a stepwise barrier at $x / \rho_{\rm e} = 0$, with a negligible potential drop on the scale $x \sim \rho_{\rm e}$ \cite{Cohen-Ryutov-1995-spreading}. 
This can only occur self-consistently if the potential barrier is sufficiently large to reflect most electrons, 
and thus requires $\alpha \gg M^{-1/2}$ \cite{Cohen-Ryutov-1995-spreading}. 
This ordering is not satisfied for the angles considered in this paper, but this limit can still serve as an analytical model which includes the effect of finite electron orbits on the electron reflection cutoff. 

\begin{figure} 
\centering
\includegraphics[width=1.0\textwidth]{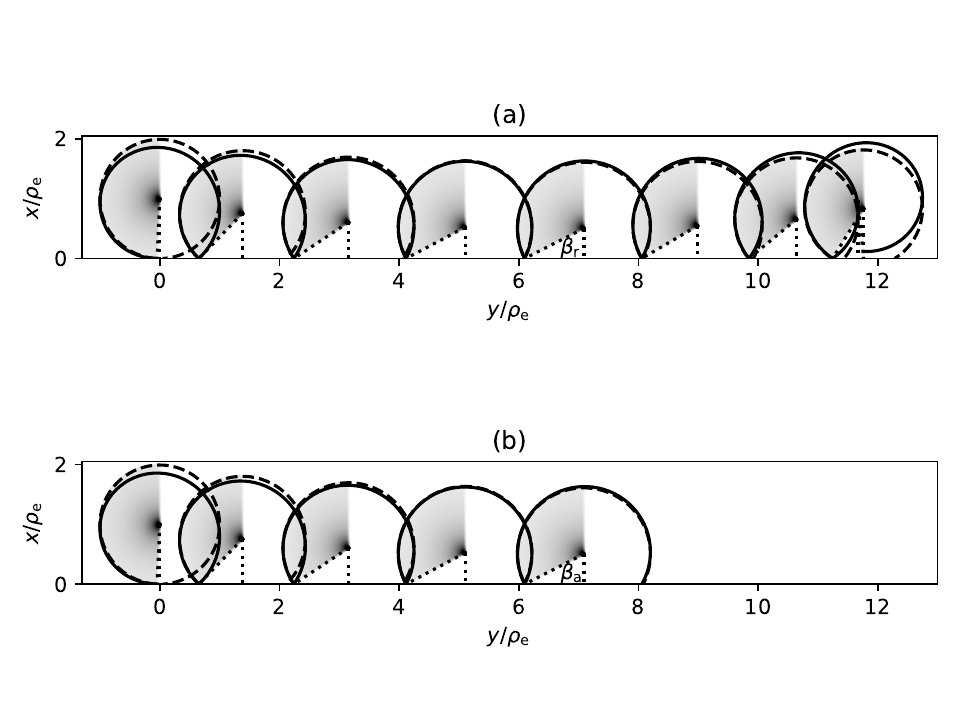} 
\caption{Electron trajectories bouncing off the Debye sheath in the limit $\rho_{\rm e} / \lambda_{\rm D} \rightarrow \infty$ are shown in the $xy$ plane, calculated to lowest order (dashed lines) and exactly (solid lines), starting from a first reflection from $x/\rho_{\rm e} = 0$, $y/\rho_{\rm e} = 0$ (the initial $y$ coordinate is arbitrary and unimportant), $v_{\perp} / v_{\rm t,e} = \sqrt{2|\Omega_{\rm e}| \mu} / v_{\rm t,e} = 1$, $v_{\parallel} / v_{\rm t,e} = 0.5 $.
The magnetic field angle is $\alpha = 5^{\circ}$.
The gyrophase (defined in (\ref{varphi-def}) and (\ref{vxtil-varphi})-(\ref{vytil-varphi})) of a particle at a point in its lowest-order orbit is the angle, measured clockwise, between the dashed line going vertically downwards from the centre of its gyro-orbit and another line from the centre of the gyro-orbit to the particle position.
The value of $\varphi$ with which the electron bounces back from the Debye sheath entrance is $\beta$.
The gyrophase domain $\varphi \in [\beta, \pi]$ is shaded for each lowest-order orbit.
In (a) the electron is reflected when $\beta = \beta_{\rm r}$, while in (b) the electron is absorbed when $\beta = \beta_{\rm a}$.
Here, $v_{\perp} \sin \beta_{\rm r} / v_{\rm t,e} = v_{\perp} \sin \beta_{\rm a} / v_{\rm t,e} = \hat v \approx 0.98 $, and thus the potential drop $e|\phi_{\rm ds}(0)|/T_{\rm e}$ must be larger (smaller) than $\frac{1}{2} \hat v^2 \approx 0.48$ for the reflected (absorbed) trajectory.}
\label{fig-electron-bouncing}
\end{figure}

The electron gyro-orbit is circular at distances from the wall $x \sim \rho_{\rm e}$, since the Debye sheath potential $\phi_{\rm ds}$ decays to zero at $x/\lambda_{\rm D} \rightarrow \infty$.
An electron reaches the Debye sheath ($x \sim \lambda_{\rm D}$) for the first time when its circular gyro-orbit reaches the guiding center position $\bar{x} \leqslant v_{\perp} / |\Omega_{\rm e}| \sim \rho_{\rm e}$.
Here, we have introduced the magnitude of the velocity in the $xy$ plane, approximately coinciding with the plane perpendicular to the magnetic field,
\begin{align}
v_{\perp} = \sqrt{v_x^2 + v_y^2} \rm .
\end{align}
When an electron first reaches the Debye sheath, the component of its velocity normal to the target is small, $|v_x | \sim \alpha^{1/2} v_{\rm t,e}$.
This is because the displacement of an electron gyro-orbit towards the wall during one gyro-period is small, $\sim \alpha \rho_{\rm e}$ \cite{Cohen-Ryutov-1995-spreading, Cohen-Ryutov-1998, Geraldini-2018, Geraldini-2021}.
Since the potential drop in the Debye sheath is large, the electron reflects while its orbit position, satisfying equation (\ref{xbardot}), continues to slowly move closer to the wall.
At the subsequent approaches to the wall, the component of the velocity at $x=0$ towards the wall is larger.
To lowest order in $\alpha v_{\parallel} / v_{\perp} \ll 1$, it satisfies $v_x = - v_{\perp} \sin \beta$ with $\cos \beta = \bar{x} |\Omega_{\rm e}| / v_{\perp}$.
The angle $\beta$ is shown in figure~\ref{fig-electron-bouncing} in the particular cases of a particle which is about to reflect ($v_{\parallel}$ changes sign) and one which is about to be absorbed (such that $|v_x|$ is sufficiently large to overcome the Debye sheath potential barrier and reach the target).

We proceed to calculate the relationship between $v_{\perp}$ and $\beta$, as the gyromotion of an electron causes it to bounce in and out of the Debye sheath.
This relation was derived in reference \cite{Cohen-Ryutov-1995-spreading}, although we propose here a different derivation exploiting the adiabatic invariant first derived by the same authors in a subsequent paper \cite{Cohen-Ryutov-1998}.
The adiabatic invariant $\mu$ of the electron motion illustrated in figure~\ref{fig-electron-bouncing} is
\begin{align} \label{mu(beta, vperp)}
\mu = \frac{v_{\perp}^2}{2|\Omega_{\rm e}|} \left( 1 - \frac{\beta}{\pi} + \frac{1}{2\pi} \sin \left(2\beta \right) \right) \rm .
\end{align}
To obtain equation (\ref{mu(beta, vperp)}), we have made use of equation (\ref{mugk-def}) with $v_{x} =  |\Omega_{\rm e}| \partial x / \partial \varphi = \sigma_{x} \sqrt{2\left(U_{\perp} - \chi_R (x, \bar{x}) \right)} = v_{\perp} \sin \varphi$ and with the limits of integration in $\varphi$ from $\beta$ to $\pi$, \emph{viz} figure \ref{fig-electron-bouncing}.
The electron bounces back and forth from the stepwise barrier conserving the adiabatic invariant, 
\begin{align} \label{dmudt(beta,vperp)}
\dot{\mu} = \frac{d}{dt} \left[ \frac{v_{\perp}^2}{2|\Omega_{\rm e}|} \left( 1 - \frac{\beta}{\pi} + \frac{1}{2\pi} \sin \left(2\beta \right) \right)  \right] = 0 \rm .
\end{align}
From (\ref{dmudt(beta,vperp)}), we obtain
\begin{align} \label{dvperpdbeta}
\frac{ d v_{\perp} }{d\beta} = \frac{v_{\perp} \sin^2 \beta }{ \pi - \beta + \sin \beta \cos \beta } \rm .
\end{align}
Upon using the boundary condition $v_{\perp}^2 = 2|\Omega_{\rm e}| \mu$ at $\beta = 0$, corresponding to a fully circular gyro-orbit not interrupted by the wall, the integration of (\ref{dvperpdbeta}) leads to the integral of motion found in \cite{Cohen-Ryutov-1995-spreading}, 
\begin{align} \label{F-beta-def-app}
\frac{1}{2} \ln \left( \frac{v_{\perp}^2}{2|\Omega_{\rm e}| \mu} \right) = \int_0^{\beta} \frac{ \sin^2 \beta' }{\pi - \beta' + \sin \beta' \cos \beta'} d\beta' \equiv F (\beta) \rm .
\end{align}

This integral of motion can be used to calculate the cutoff in phase space between absorbed and reflected electrons, $U_{\rm cut, ds} (\mu)$.
If the perpendicular velocity of the electron and its gyrophase at the target finally (after multiple bounces) satisfy 
\begin{align} \label{1}
v_{\perp} \sin \beta_{\rm a} = \sqrt{\frac{ - 2e\phi_{\rm ds}(0)}{m_{\rm e} } } \rm ,
\end{align}
then the electron overcomes the potential barrier of the Debye sheath, and to lowest order in $\alpha \ll 1$, it enters it with $|v_x| = \sqrt{- 2e\phi_{\rm ds}(0)/m_{\rm e} }$ and touches the target with $|v_{x}| = 0$.
The electron is thus absorbed when it reaches $\beta = \beta_{\rm a} (\mu)$, defined implicitly by
\begin{align} \label{absorbed}
\frac{1}{2} \ln \left( \frac{- \phi_{\rm ds}(0) }{ \mu B \sin^2 \beta_{\rm a} } \right) = F (\beta_{\rm a}) \rm .
\end{align}
If, however, the electron reaches a state where all of its kinetic energy is in the velocity components perpendicular to the magnetic field,
\begin{align} \label{2}
\frac{1}{2} v_{\perp}^2  = U_{\rm ds} + \frac{ e\phi_{\rm mp} (0) }{ m_e } \rm ,
\end{align} 
before reaching (\ref{1}), the electron is reflected as $v_z $ crosses through zero and changes sign, which occurs at $\beta = \beta_{\rm r}(\mu, U_{\rm ds})$, defined by
\begin{align} \label{reflected}
\frac{1}{2} \ln \left( \frac{m_{\rm e} U_{\rm ds} + e \phi_{\rm mp} (0) }{ e B \mu} \right) = F (\beta_{\rm r}) \rm .
\end{align}
In the limit in which conditions (\ref{1}) and (\ref{2}) are simultaneously satisfied, the electron parallel velocity changes sign as the electron just touches the target.
In this limit, $\beta_{\rm r}(\mu, U_{\rm ds}) = \beta_{\rm a} (\mu)$ (see (\ref{absorbed}) and (\ref{reflected}) for the definition of these angles), which gives the phase space boundary between reflected and absorbed electrons, that is, the cutoff function,
\begin{align} \label{U-cut-infty}
U_{\rm cut,ds, \infty} (\mu)  =  \frac{|\Omega_{\rm e}| \phi_{\rm ds}(0) }{B  \sin^2 \left[ \beta_{\rm a}(\mu) \right] } - \frac{|\Omega_{\rm e}|\phi_{\rm mp}(0) }{B}  \rm  .
\end{align}
Note that in $U_{\text{cut,ds,}\infty}$ the subscript $\infty$ refers to $\gamma \rightarrow \infty$.

\section*{References}

\bibliography{gyrokineticsbibliography}{}

\begin{thebibliography}{10}

\bibitem{Geraldini-2017}
A.~Geraldini, F.~I. Parra, and F.~Militello.
\newblock Gyrokinetic treatment of a grazing angle magnetic presheath.
\newblock {\em Plasma Physics and Controlled Fusion}, 59(2):025015, 2017.

\bibitem{Geraldini-2018}
A.~Geraldini, F.~I. Parra, and F.~Militello.
\newblock Solution to a collisionless shallow-angle magnetic presheath with
  kinetic ions.
\newblock {\em Plasma Physics and Controlled Fusion}, 60(12):125002, 2018.

\bibitem{Ewart-2021}
R.~J Ewart, F.~I. Parra, and A.~Geraldini.
\newblock Sheath collapse at critical shallow angle due to kinetic effects.
\newblock {\em Plasma Physics and Controlled Fusion}, 64(1):015010, 2021.

\bibitem{Stangeby-book}
P.~C. Stangeby.
\newblock The plasma boundary of magnetic fusion devices ({IOP} publishing,
  {B}ristol, {UK}).
\newblock 2000.

\bibitem{Martinez-1998}
M.~Martinez-Sanchez and J.~E. Pollard.
\newblock Spacecraft electric propulsion - an overview.
\newblock {\em Journal of Propulsion and Power}, 14(5):688--699, 1998.

\bibitem{Anders-1995-filters}
A.~Anders, S.~Anders, and I.~G. Brown.
\newblock Transport of vacuum arc plasmas through magnetic macroparticle
  filters.
\newblock {\em Plasma Sources Science and Technology}, 4(1):1, 1995.

\bibitem{Hutchinson-book}
I~H Hutchinson.
\newblock Principles of plasma diagnostics.
\newblock {\em Plasma Physics and Controlled Fusion}, 44(12):2603, 2002.

\bibitem{Chodura-1982}
R.~Chodura.
\newblock Plasma--wall transition in an oblique magnetic field.
\newblock {\em Physics of Fluids (1958-1988)}, 25(9):1628--1633, 1982.

\bibitem{Geraldini-2024-Chodura}
A.~Geraldini, S.~Brunner, and F.~Parra.
\newblock Sheath constraints in turbulent magnetised plasmas.
\newblock {\em Plasma Physics and Controlled Fusion}, 66:105021, 2024.

\bibitem{Stangeby-2012}
P.~C. Stangeby.
\newblock The {C}hodura sheath for angles of a few degrees between the magnetic
  field and the surface of divertor targets and limiters.
\newblock {\em Nuclear Fusion}, 52(8):083012, 2012.

\bibitem{Coulette-Manfredi-2016}
D.~Coulette and G.~Manfredi.
\newblock Kinetic simulations of the {C}hodura and {D}ebye sheaths for magnetic
  fields with grazing incidence.
\newblock {\em Plasma Physics and Controlled Fusion}, 58(2):025008, 2016.

\bibitem{Riemann-1994}
K.-U. Riemann.
\newblock Theory of the collisional presheath in an oblique magnetic field.
\newblock {\em Physics of Plasmas (1994-present)}, 1(3):552--558, 1994.

\bibitem{Tskhakaya-2017}
D.~Tskhakaya.
\newblock One-dimensional plasma sheath model in front of the divertor plates.
\newblock {\em Plasma Physics and Controlled Fusion}, 59(11):114001, 2017.

\bibitem{Cohen-Ryutov-2004-sheath-boundary-conditions}
R.~H. Cohen and D.~D. Ryutov.
\newblock Sheath physics and boundary conditions for edge plasmas.
\newblock {\em Contributions to Plasma Physics}, 44(1-3):111--125, 2004.

\bibitem{Geraldini-2021}
A.~Geraldini.
\newblock Large gyro-orbit model of ion velocity distribution in plasma near a
  wall in a grazing-angle magnetic field.
\newblock {\em Journal of Plasma Physics}, 87(1), 2021.

\bibitem{Mandell-2022}
N.~R. Mandell, G.~W. Hammett, A.~Hakim, and M.~Francisquez.
\newblock Turbulent broadening of electron heat-flux width in electromagnetic
  gyrokinetic simulations of a helical scrape-off layer model.
\newblock {\em Physics of Plasmas}, 29(4):042504, 2022.

\bibitem{Chang-2024}
C.~S. Chang, S.~Ku, R.~Hager, J.~Choi, D.~Pugmire, S.~Klasky, A.~Loarte, and
  R.~A. Pitts.
\newblock Role of turbulent separatrix tangle in the improvement of the
  integrated pedestal and heat exhaust issue for stationary-operation tokamak
  fusion reactors.
\newblock {\em Nuclear Fusion}, 64(5):056041, 2024.

\bibitem{Stangeby-Chankin-1995}
P.~C. Stangeby and A.~V. Chankin.
\newblock The ion velocity ({B}ohm--{C}hodura) boundary condition at the
  entrance to the magnetic presheath in the presence of diamagnetic and
  {E}$\times${B} drifts in the scrape-off layer.
\newblock {\em Physics of Plasmas (1994-present)}, 2(3):707--715, 1995.

\bibitem{Loizu-2012}
J.~Loizu, P.~Ricci, F.~D. Halpern, and S.~Jolliet.
\newblock Boundary conditions for plasma fluid models at the magnetic presheath
  entrance.
\newblock {\em Physics of Plasmas (1994-present)}, 19(12):122307, 2012.

\bibitem{Mosetto-2015}
A.~Mosetto, F.~D. Halpern, S.~Jolliet, J.~Loizu, and P.~Ricci.
\newblock Finite ion temperature effects on scrape-off layer turbulence.
\newblock {\em Physics of Plasmas}, 22(1):012308, 2015.

\bibitem{Castillo-2024}
A.~M. Castillo and K.~Hara.
\newblock Loss cone effects and monotonic sheath conditions of a partially
  magnetized plasma sheath.
\newblock {\em Physics of Plasmas}, 31(3), 2024.

\bibitem{Gerver-Parker-Theilhaber-1990}
M.~J. Gerver, S.~E. Parker, and K.~Theilhaber.
\newblock Analytic solutions and particle simulations of cross--field plasma
  sheaths.
\newblock {\em Physics of Fluids B}, 2(5):1069--1082, 1990.

\bibitem{Cohen-Ryutov-1998}
R.~H. Cohen and D.~D. Ryutov.
\newblock Particle trajectories in a sheath in a strongly tilted magnetic
  field.
\newblock {\em Physics of Plasmas (1994-present)}, 5(3):808--817, 1998.

\bibitem{Cohen-Ryutov-1995-spreading}
R.~H. Cohen and D.~D. Ryutov.
\newblock Spreading particle trajectories near a perfectly reflecting surface
  in a tilted magnetic field.
\newblock {\em Physics of Plasmas}, 2(11):4118--4121, 1995.

\bibitem{Geraldini-2019}
A.~Geraldini, F.~I. Parra, and F.~Militello.
\newblock Dependence on ion temperature of shallow-angle magnetic presheaths
  with adiabatic electrons.
\newblock {\em Journal of Plasma Physics}, 85(6):795850601, 2019.

\bibitem{Parker-1993}
S.~E. Parker, R.~J. Procassini, C.~K. Birdsall, and B.~I. Cohen.
\newblock A suitable boundary condition for bounded plasma simulation without
  sheath resolution.
\newblock {\em Journal of Computational Physics}, 104(1):41--49, 1993.

\bibitem{Shi-2017}
E.~L. Shi, G.~W. Hammett, T.~Stoltzfus-Dueck, and A.~Hakim.
\newblock Gyrokinetic continuum simulation of turbulence in a straight
  open-field-line plasma.
\newblock {\em Journal of Plasma Physics}, 83(3), 2017.

\bibitem{Krasheninnikova-2010}
N.~S. Krasheninnikova, X.~Tang, and V.~S. Roytershteyn.
\newblock Scaling of the plasma sheath in a magnetic field parallel to the
  wall.
\newblock {\em Physics of Plasmas}, 17(5), 2010.

\bibitem{Militello-Fundamenski-2011}
F.~Militello and W.~Fundamenski.
\newblock Multi-machine comparison of drift fluid dimensionless parameters.
\newblock {\em Plasma Physics and Controlled Fusion}, 53(9):095002, 2011.

\bibitem{Loizu-2011}
J.~Loizu, P.~Ricci, and C.~Theiler.
\newblock Existence of subsonic plasma sheaths.
\newblock {\em Physical Review E}, 83(1):016406, 2011.

\bibitem{Riemann-review}
K.-U. Riemann.
\newblock The {B}ohm criterion and sheath formation.
\newblock {\em Journal of Physics D: Applied Physics}, 24(4):493, 1991.

\bibitem{Baalrud-Hegna-2011-Bohm}
S.~D. Baalrud and C.~C. Hegna.
\newblock Kinetic theory of the presheath and the {B}ohm criterion.
\newblock {\em Plasma Sources Science and Technology}, 20(2):025013, 2011.

\bibitem{Riemann-2012-comment}
K.-U. Riemann.
\newblock Comment on ‘{K}inetic theory of the presheath and the {B}ohm
  criterion’.
\newblock {\em Plasma Sources Science and Technology}, 21(6):068001, 2012.

\bibitem{Baalrud-Hegna-2012-reply}
S.~D. Baalrud and C.~C. Hegna.
\newblock Reply to comment on ‘{K}inetic theory of the presheath and the
  {B}ohm criterion’.
\newblock {\em Plasma Sources Science and Technology}, 21(6):068002, 2012.

\bibitem{Geraldini-2024-Bohm}
A.~Geraldini and S.~Brunner.
\newblock On the importance of slow ions in the kinetic {B}ohm criterion.
\newblock {\em Journal of Plasma Physics}, 90(6):905900604, 2024.

\end{thebibliography}
\bibliographystyle{unsrt}

\end{document}